\pdfoutput=1

\documentclass[
	sn-nature,
	pdflatex
	]{sn-jnl}

\usepackage{graphicx}%
\usepackage{multirow}%
\usepackage{amsmath,amssymb,amsfonts}%
\usepackage{amsthm}%
\usepackage{mathrsfs}%
\usepackage[title]{appendix}%
\usepackage{xcolor}%
\usepackage{textcomp}%
\usepackage{manyfoot}%
\usepackage{booktabs}%
\usepackage{listings}%
\usepackage[range-units = single,
            separate-uncertainty-units = single,
            separate-uncertainty = true,
            multi-part-units=single]
            {siunitx}%
\usepackage{natbib}
\usepackage[utf8]{inputenc}
\usepackage[T1]{fontenc}

\unnumbered

\begin{document}

\title[Solid-State Oxide-Ion Synaptic Transistor]{Solid-State Oxide-Ion Synaptic Transistor for Neuromorphic Computing}

\author[1]{\fnm{Philipp} \sur{Langner}}

\author*[1]{\fnm{Francesco} \sur{Chiabrera}}\email{fchiabrera@irec.cat}

\author[1]{\fnm{Paul} \sur{Nizet}}

\author[1]{\fnm{Nerea} \sur{Alayo}}

\author[2]{\fnm{Luigi} \sur{Morrone}}

\author[1]{\fnm{Carlota} \sur{Bozal-Ginesta}}

\author[1]{\fnm{Alex} \sur{Morata}}

\author*[1,3]{\fnm{Albert} \sur{Taranc\`on}}\email{atarancon@irec.cat}

\affil[1]{\orgdiv{Department of Advanced Materials for Energy}, \orgname{Catalonia Institute for Energy Research (IREC)}, \orgaddress{\street{Jardins de les Dones de Negre 1, 2}, \city{Sant Adrià de Besós}, \postcode{08930}, \state{Barcelona}, \country{Spain}}}

\affil[2]{\orgdiv{Scientific Services NANOQUIM}, \orgname{Institut de Ci\`encia de Materials de Barcelona (CSIC-ICMAB)}, \orgaddress{\street{Campus UAB}, \city{Bellaterra}, \postcode{08193}, \state{Barcelona}, \country{Spain}}}

\affil[3]{\orgname{Catalan Institution for Research and Advanced Studies (ICREA)}, \orgaddress{\street{Passeig Lluis Companys 23}, \postcode{08010}, \city{Barcelona}, \country{Spain}}}


\abstract{Neuromorphic hardware facilitates rapid and energy-efficient training and operation of neural network models for artificial intelligence. 
However, existing analog in-memory computing devices, like memristors, continue to face significant challenges that impede their commercialization. 
These challenges include high variability due to their stochastic nature. 
Microfabricated electrochemical synapses offer a promising approach by functioning as an analog programmable resistor based on deterministic ion-insertion mechanisms.
Here, we developed an all-solid-state oxide-ion synaptic transistor employing $\text{Bi}_\text{2}\text{V}_\text{0.9}\text{Cu}_{0.1}\text{O}_\text{5.35}$ as a superior oxide-ion conductor electrolyte and $\text{La}_\text{0.5}\text{Sr}_\text{0.5}\text{F}\text{O}_\text{3-$\delta$}$ as a variable-resistance channel able to efficiently operate at temperatures compatible with conventional electronics. 
Our transistor exhibits essential synaptic behaviors such as long- and short-term potentiation, paired-pulse facilitation, and post-tetanic potentiation, mimicking fundamental properties of biological neural networks. 
Key criteria for efficient neuromorphic computing are satisfied, including excellent linear and symmetric synaptic plasticity, low energy consumption per programming pulse, and high endurance with minimal cycle-to-cycle variation. 
Integrated into an artificial neural network (ANN) simulation for handwritten digit recognition, the presented synaptic transistor achieved a \qty{96}{\percent} accuracy on the MNIST dataset, illustrating the effective implementation of our device in ANNs. 
These findings demonstrate the potential of oxide-ion based synaptic transistors for effective implementation in analog neuromorphic computing based on iontronics.}



\maketitle

\subsection{Introduction}\label{sec: introduction}
Neuromorphic computing seeks to emulate the complex functionality of biological neural networks, which rely on the dynamic establishment of trillions of synaptic connections between neurons. 
This intricate network continuously evolves in response to stimuli, strengthening or weakening connections through a mechanism known as “synaptic plasticity” crucial for learning and memory processes \cite{dan2006spike,harvey2007locally,kuzum2013synaptic,atluri1996determinants,zucker2002short,debanne1996paired}. 
Individual digital transistors are unable to replicate this dynamic behavior, requiring tens of devices to mimic a single synapse, which involves huge energy consumption and a high architecture complexity. 
To overcome this issue, a collection of energy-efficient synaptic transistors has been developed in the last years \cite{kuzum2013synaptic,aguirre2024hardware,shi2013correlated,fuller2016li}. 
In particular, memristor devices based on existing non-volatile material properties (ferroelectricity, ferromagnetism, phase changes or resistivity) that change after applying a voltage have been object of study showing a great potential \cite{song2023recent,bauer2015magneto}.
Crossbar arrays based on these memristor devices are expected to enable ultra-low power consumption for massive parallel computing \cite{mehonic2022brain,li2018efficient,li2018analogue}.
However, memristors still suffer from poor device-to-device reproducibility, non-linear switching behavior, require high write/read currents and present excessive power dissipation, which constrain the performance and energy efficiency of the derived neuromorphic computing network\cite{snider2007self,burr2015experimental,li2018review,islam2019device}. 
Alternatively, recent advancements have introduced electrochemical ionic synapses, also known as electrochemical random-access memory (ECRAM) devices, as a novel type of programmable synaptic transistor \cite{van2017non,fuller2016li,zhang2022reconfigurable,yang2018all,yang2018artificial,tsuchiya2013all,schwacke2024electrochemical,ling2020electrolyte,tang2018ecram,fuller2019parallel,nikam2019near,cui2023cmos}. 
Operating with a three-terminal configuration, these devices enable power-efficient and fast analog in-memory computing, effectively addressing the von Neumann bottleneck \cite{shi2013correlated,schwacke2024electrochemical,ling2020electrolyte}. 
By this, they hold significant potential as a disruptive technology for applications in neuromorphic computing and machine learning, consumer electronics in smart devices, medical devices, and space applications, particularly in deep neural networks and inference systems for IoT applications. 
These synaptic transistors are based on the control of the electronic resistance of a channel material by ionic insertion/extraction, mediated by the application of an electrochemical potential to an ionic reservoir gate through a pure ionic electrolyte, in a battery-like configuration \cite{kuzum2013synaptic,shi2013correlated,zhu2020comprehensive}. 
Promising complementary metal-oxide-semiconductor (CMOS) compatible oxide films, such as perovskite and pseudo-perovskite structures, are utilized as ion reservoirs and channel electrodes in synaptic transistors. 
These films possess high ion mobility, enabling efficient ion insertion, electron doping, and resistance modulation of the oxide films\cite{shi2013correlated,nizet2024optoionic,shi2021solid,talin2023ecram,huang2023electrochemical}. 
Synaptic transistors based on different types of ions have already been explored. 
As presented by Fuller \textit{et al.} (2016), lithium-ion synaptic transistors offer advantages such as fast diffusion and good retention\cite{fuller2016li}.
However, their integration in silicon presents significant challenges, including limited compatibility with CMOS technology and reduced stability at lower temperatures and ambient atmospheres
\cite{li2019low,nguyen2022ultralow,fuller2016li,tang2018ecram,talin2023ecram}. 
Proton-based synaptic transistors, on the other hand, show rapid ion transport, linear and symmetric switching but suffer from environmental instability, and restricted choice of oxide hosting materials\cite{kim2019metal,onen2021cmos,onen2022nanosecond}. 
Conversely, utilizing oxygen as the mobile ion in synaptic transistors provides access to a broad range of oxide materials with orders of magnitude of change in conductance and excellent stability under varying environmental conditions, such as temperature, humidity, and time. 
This stability is crucial for accurately emulating dynamic synaptic networks in neuromorphic computing architectures.
 In particular, the capability to operate at intermediate and high temperatures makes them suitable for extreme environments targeting edge computing in the industrial or space sectors together with recently developed high-temperature electronics \cite{pradhan2024scalable,melianas2020temperature}. 
Moreover, the well-known compatibility of oxides with CMOS technology makes oxygen-based synaptic transistors promising candidates for seamless CMOS- or BEOL-integration in conventional silicon chips\cite{bauer2015magneto,talin2023ecram}. 
However, despite the potential of oxygen-based synaptic transistors, their development is still in its infancy, primarily due to the use of solid-state electrolytes like yttria-stabilized zirconia (YSZ), HfO$_\text{2-$\delta$}$, or gadolinium-doped ceria (GDC). 
These materials exhibit insufficient oxide-ionic conductivity at low temperatures, necessitating operation at temperatures incompatible with commercial electronics\cite{talin2023ecram,nikam2021all,lee2022strategies,jeong2021room}. 
In this work, we develop a novel synaptic transistor able to operate at moderate temperatures based on a superior oxide-ion conductor such as stabilized-bismuth vanadate ($\text{Bi}_\text{2}\text{V}_\text{0.9}\text{Cu}_{0.1}\text{O}_\text{5.35}$, BICUVOX). 
This solid-state thin film electrolyte exhibits high ionic conductivity and stability over a broad range of temperatures and oxygen pressures (p$\text{O}_\text{2}$), offering a promising solution to deploy oxygen-based synaptic transistors\cite{garbayo2019thin}. 
As channel material, we employed $\text{La}_\text{0.5}\text{Sr}_\text{0.5}\text{F}\text{O}_\text{3-$\delta$}$ (LSF50) thin films, a mixed ionic-electronic conducting (MIEC) perovskite oxide derived from the parent compound $\text{La}_\text{1-x}\text{Sr}_\text{x}\text{F}\text{O}_\text{3-$\delta$}$  with $x=0.5$.
This material allows precise control of the oxygen content in the range of $\text{$\delta$} = 0 \text{ - } 0.25$, by reversibly applying a voltage enabling precise modulation of its electrical properties over a wide range of conductivities\cite{bae2019investigations,jana2019charge,wang2019hole,tang2021pushing,nizet2024optoionic}. 
The ability to control the oxygen stoichiometry of LSF50 results in non-volatile changes of orders of magnitude in its electronic conductivity, making it an ideal material for a programmable multistate channel\cite{patrakeev2003electron,cheng2005thermochemistry,nizet2024optoionic}.  
Integrating our BICUVOX electrolyte and the LSF50 channel/reservoir into a microfabricated symmetric synaptic transistor, we were able to emulate neurotransmitter transfer between pre- and post-neurons by precisely controlling the channel’s conductance through the application of a small voltage down to $\qty{\pm 0.5}{\volt}$.
Fundamental synaptic functionalities, such as short- and long-term synaptic plasticity, excellent linearity, low asymmetric ratio during programming with low energy consumption, and a broad dynamic range are demonstrated in our oxygen based synaptic transistor. 
Finally, we illustrate the effective implementation of our device in artificial neural networks (ANNs) by simulating the behavior of our transistors into an ANN for handwritten digit recognition.


\subsection{Device Architecture and Operation Principle}
\label{subsec:architecture,principle}

\begin{figure}[t]
\centering
\includegraphics[width=\textwidth]{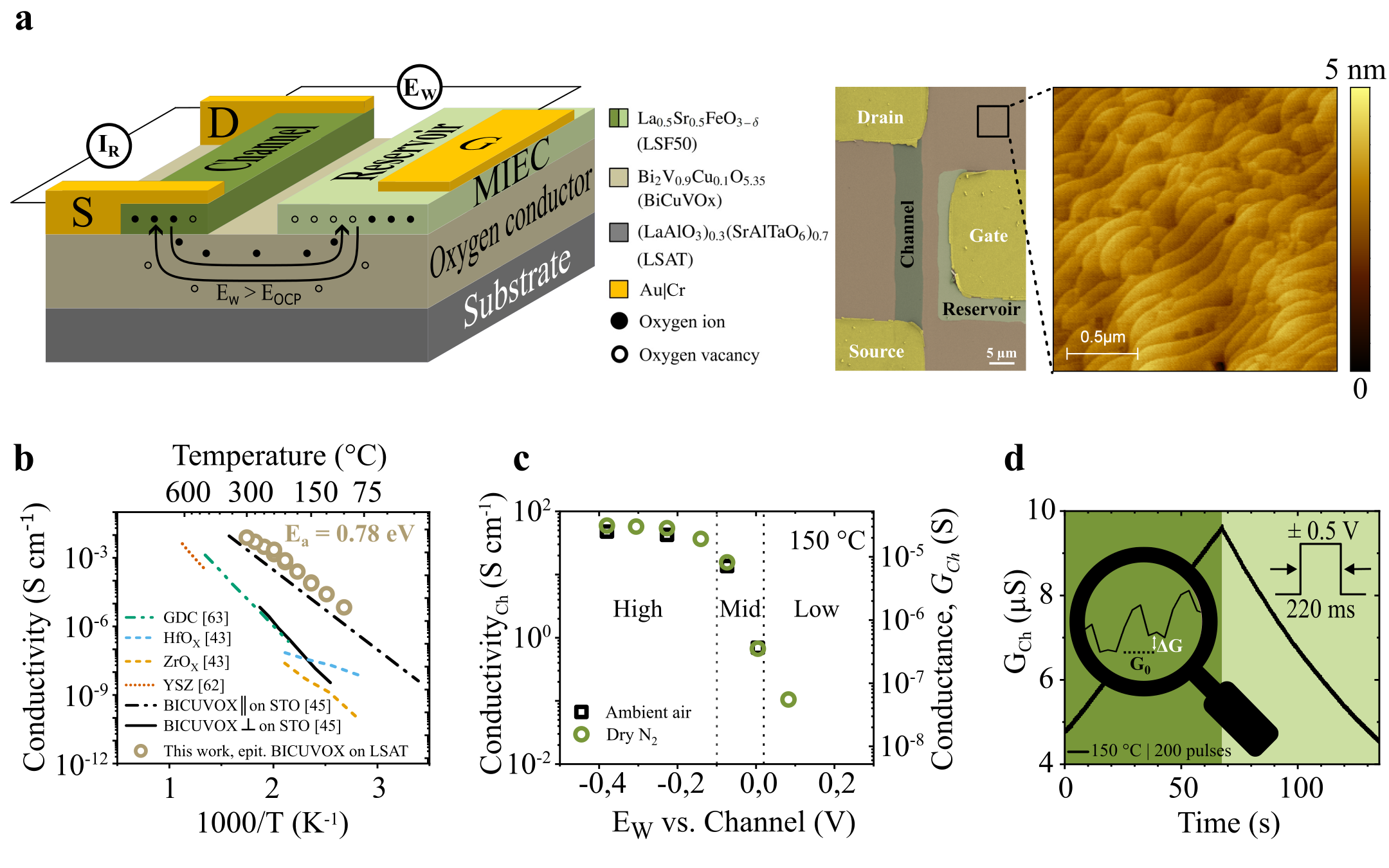}
\caption[width=\linewidth]{Device architecture and operation principle.
\textbf{a} sketch and colorized SEM microscopy of the synaptic transistor with symmetric battery-like architecture. 
\textbf{b} Arrhenius representation of ionic conductivity for [00l] epitaxial BICUVOX thin-film with temperature; in comparison to epitaxial BICUVOX on STO (parallel and perpendicular to the BICUVOX [100]-planes), YSZ, $\text{Zr}\text{O}_\text{x}$, $\text{Hf}\text{O}_\text{x}$ and GDC thin film\cite{garbayo2019thin,joo2006electrical,chiodelli2005synthesis,lee2022strategies}. 
\textbf{c} operational window of MIEC LSF50-channel at \qty{150}{\degreeCelsius} showing high, medium and low conductance regimes. 
\textbf{d} linear and symmetric synaptic plasticity of the synaptic transistor at \qty{150}{\degreeCelsius} during 100 potentiation (dark green) and 100 depression (light green) pulses. }
\label{fig: Fig1}
\end{figure}

Fig.~\ref{fig: Fig1}a sketches the architecture and operation principle of the microfabricated all-solid-state thin film synaptic transistor based on an oxide-ion conductor (BICUVOX) and a mixed ionic-electronic conductor (LSF50) for acting as a channel and gate (ion-reservoir).
The device presents a planar geometry with the channel and the ion-reservoir on top of the BICUVOX layer. 
Additionally, gold electrical contacts are located on top of the channel and the ion-reservoir for reading its conductance state (source and drain) and fixing a homogeneous writing voltage (gate), respectively. The operation principle of the transistor is completely analog to an oxygen battery\cite{schmid2023rechargeable}, \textit{i.e.} the LSF50 channel is a battery electrode charged/discharged by electrochemical ion-pumping through the BICUVOX electrolyte after applying a writing voltage $E_W$. 
The planar configuration was preferred to take advantage of a fastest in-plane ionic conductivity of the BICUVOX electrolyte epitaxially grown on the [00l] direction. 
This preferential conduction was confirmed to be orders of magnitude higher in a previous work by the authors\cite{garbayo2019thin}.
Therefore, epitaxial [00l]-oriented BICUVOX and LSF50 thin films were grown on $(\text{La}\text{Al}\text{O}_\text{3})_\text{0.3}(\text{Sr}\text{Al}\text{Ta}_\text{0.6})_\text{0.7}$ single crystal substrates (LSAT, [100]) using large-area pulsed laser deposition (LA-PLD) (see methods).
Structural characterization of the BICUVOX thin films highlights a very ordered structure, characterized by a single [00l]-orientation (see XRD in S1) and by the presence of well-defined atomic step terraces, see Fig.~\ref{fig: Fig1}a. 
Photolithography and Ar$^\text{+}$-milling was performed (see methods and S2). 
The final microfabricated synaptic transistor consists of a LSF50 channel with width $w_{Ch} = \qty{6}{\micro\metre}$, length $l_{Ch} = \qty{35}{\micro\metre}$, thickness $th_{Ch} = \qty{35}{\nano\metre}$ and BICUVOX electrolyte thickness $th_{Ely} = \qty{140}{\nano\metre}$ with LSF50-channel and -reservoir being separated by a gap of \qty{3}{\micro\metre} (see Fig.~\ref{fig: Fig1}a, S3, and S4). 

Electrochemical impedance spectroscopy (EIS) was first carried out on the final device to confirm a high oxide-ionic conductivity of the electrolyte after the microfabrication flow. 
EIS was performed at different temperatures in dry nitrogen to ensure purely oxide-ion conduction, avoiding the influence of surface’s protonic\cite{gu2023surface}. 
EIS spectra showed the presence of a characteristic high frequency semicircle and a low frequency highly resistant contribution, which were assigned to the ionic conduction of BICUVOX and the oxygen incorporation at the LSF50 electrodes, respectively (see S5).  
Arrhenius representation of the ionic conductivity (see methods for details) shows that BICUVOX thin films in the miniaturized devices present values and activation energy comparable with our previous results, demonstrating low resistance for the BICUVOX$\vert$LSF interface and that the miniaturization did not substantially affect the ionic transport properties of the layer\cite{garbayo2019thin}. 
Moreover, the observed Arrhenius behavior of the BICUVOX conductivity over the whole temperature range, even below \qty{150}{\degreeCelsius} (see Fig.~\ref{fig: Fig1}b), further confirms a negligible protonic conduction, which is typically characterized with non-Arrhenius behavior at lower T\cite{norby2004hydrogen,meng2019recent}. 
In the same figure, we can compare the oxide-ion conductivity of BICUVOX and other conventional electrolytes such as YSZ, $\text{Hf}\text{O}_\text{2-$\delta$}$, $\text{Zr}\text{O}_\text{2-$\delta$}$, and GDC.
Such higher values of conductivity directly impact in a much faster operation of the in-plane transistors based on BICUVOX, which present an enhancement of 3 orders of magnitude at \qty{100}{\degreeCelsius} for $\si{\micro\metre}$-sized channel compared to the out-of-plane YSZ-based counterpart (see discussion in S6 and S20). 

After characterization of the \nolinebreak{BICUVOX}-layer, the modulation of the LSF50 channel conductance ($G_{Ch}$) along the full operational voltage window was evaluated at \qty{150}{\degreeCelsius} in dry nitrogen and ambient air conditions (Fig.~\ref{fig: Fig1}c).
For this, we recorded the readout source-drain current in the channel ($I_R$) using a small voltage ($E_R = \qty{0,05}{\V}$)as a function of the writing voltage between channel and reservoir ($E_W$). 
This measurement showed a wide operational conductance range (high, medium, and low) of the LSF50 channel-electrode (change of $\approx\num{e3}$) within a very low voltage range (\qtyrange{-0,4}{0,1}{\V}).  
To assign the measured conductance solely to oxide ion transport, we first heated up the device to \qty{300}{\degreeCelsius} in ambient air to remove surficial protons followed by cooling-down to \qty{150}{\degreeCelsius} with constant dry $\text{N}_\text{2}$ flow (see S7)\cite{fabbri2010materials}. 
At both temperatures, the operational window does not change when switching from dry $\text{N}_\text{2}$ to ambient air indicating no significant protonic contribution even though the device is exposed to the atmosphere, supporting our previous findings. 
These results also prove the stability of oxygen-based synaptic transistors over ambient conditions.

In the context of electrochemical synaptic transistors, the channel conductance $G_{Ch}$ represents the ”weight” of the synapses within the final neural network \cite{nishitani2012three,yao2020protonic}. 
To ensure the required plasticity of our device, we carried out programming the channel’s synaptic weight $G_{Ch}$ with a train of voltage-controlled depression- and potentiation pulses ($E_W = \qty{+- 0,5}{\V}$ with duration $t_p = \qty{220}{\milli\second}$) at \qty{150}{\degreeCelsius}.
According to Fig.~\ref{fig: Fig1}d, this test resulted in a high level of symmetric and linear analog tuning through 100$\text{+}$ discrete conductance states (equivalent to 7 bit of information). 
This demonstrates a well-defined multi-state synaptic plasticity behavior of our synaptic transistor addressing the sluggish oxygen kinetics by employing epitaxial BICUVOX and precisely tunable LSF50 MIECs. 
It is important to mention here that operation at \qty{150}{\degreeCelsius} was selected to ensure sub-second operation speeds, which sets a significant benchmark for electrochemical solid-state synaptic transistors based on oxide-ion conductors while keeping the potential for application in CMOS-compatible electronics. 
Additional measurements at higher temperature ($T = \qty{300}{\degreeCelsius}$) confirm the thermal stability of the of the device (Fig.~S8) while operability at temperatures as low as \qty{100}{\degreeCelsius} is also proved in this work for the current geometry (see S9).
Furthermore, miniaturization of the devices will enable operation below \qty{90}{\degreeCelsius} in the kHz-range according to our finite element simulations (see S21), opening the door for multiple applications in conventional scenarios.  
In this regard, despite different authors have claimed oxide-ion modulation at room temperature employing synaptic transistors based on $\text{Zr}\text{O}_\text{2}$ or $\text{Hf}\text{O}_\text{2}$ electrolytes\cite{lee2022strategies,kwak2021experimental,nikam2021all,li2020filament,kim2023nonvolatile,kim2019metal}, our simulations based on the ionic conductivity of these materials suggest that these results probably benefited from surface’s protonic contributions or redox changes in the electrolyte composition due to high potentials (\textit{i.e.} memristive behavior), see discussion in S6, with the corresponding limitations.


\subsection{Synaptic Properties of the Synaptic Transistor}
\label{subsec:Synaptic properties of the synaptic transistor}

\begin{figure}[t]
\centering
\includegraphics[width=\textwidth]{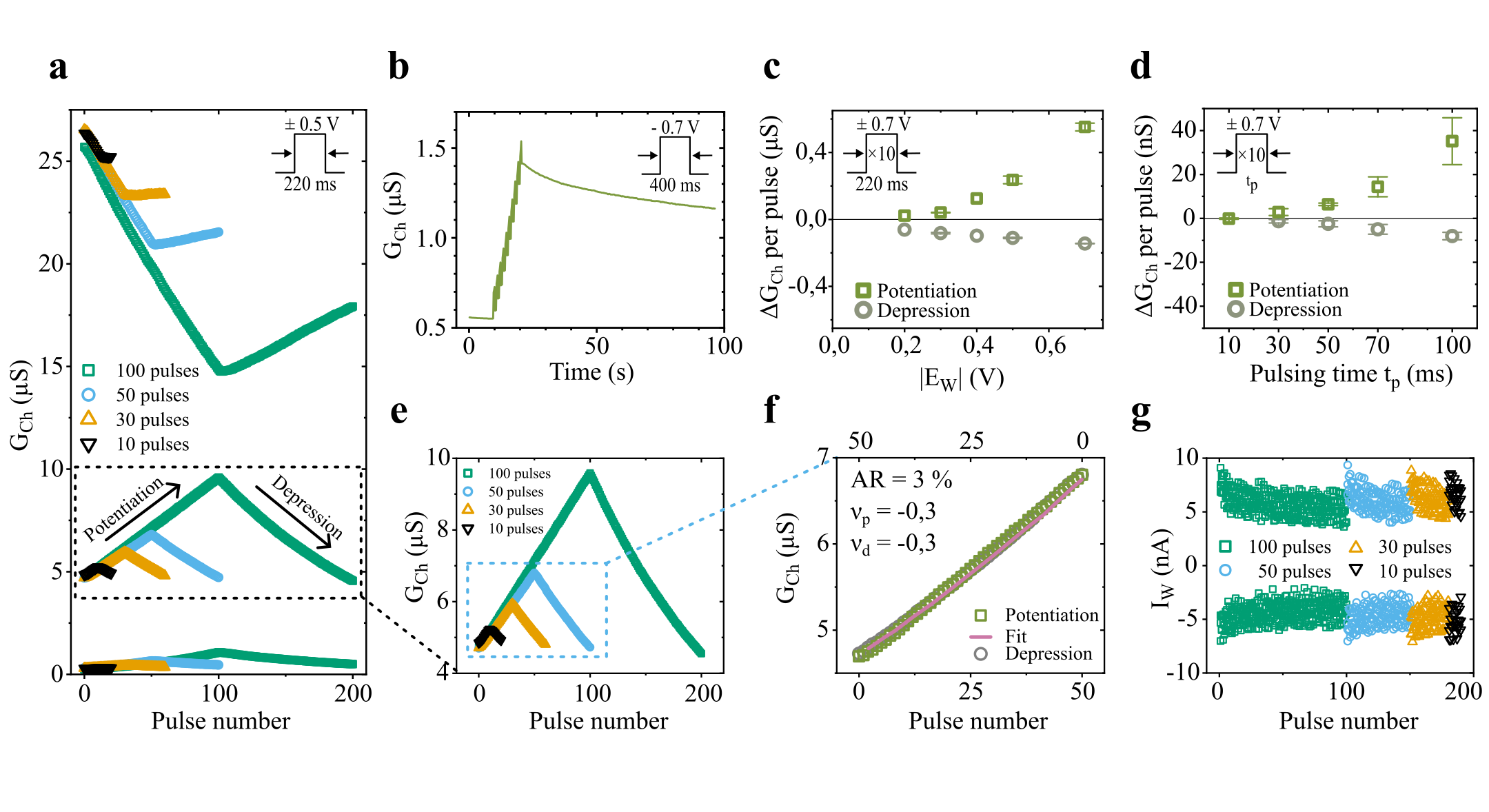}
\caption[width=\linewidth]{Operation speed, long term plasticity and energy consumption at \qty{150}{\degreeCelsius}. 
\textbf{a} Linear, reversible, and symmetric synaptic plasticity across the high, mid and low conductance range of the LSF50 channel. 
\textbf{b} Stimulus of channel conductance demonstrating non-volatility and long-term potentiation. 
\textbf{c-d} Dependence of change in channel conductance $\Delta G_{Ch}$ (“synaptic weight”) when \textbf{c} the applied pulse duration is kept constant ($tp = \qty{220}{\milli\second}$), and \textbf{d} the applied voltage is kept constant ($E_W = \qty{+- 0,7}{\V}$). 
\textbf{e} Zoom into medium operational regime. 
\textbf{f} Linearity- and symmetry comparison of 50 potentiation and depression pulses in medium operational regime. 
\textbf{g} Measured writing current during programming in medium operational regime for energy consumption calculation.}
\label{fig: Fig2}
\end{figure}

Figure \ref{fig: Fig2} shows the main characteristics of the synaptic behavior of our oxygen transistor including the operation speed, long term plasticity and energy consumption at an operation temperature of \qty{150}{\degreeCelsius}.
Application of voltage pulse trains clearly confirms that changes in conductance ($\Delta{G_{Ch}}$), often referred to as “synaptic plasticity”, are essentially proportional to the transferred charge $\Delta{Q}$ ($\Delta{G_{Ch}} \propto \Delta{Q} = I_W \cdot t_p$) independently on the sign of the voltage signal or the conductivity regime studied (see Fig.~\ref{fig: Fig2}a). 
This proportionality is one of the strengths of ECRAM devices, providing a deterministic control of channel’s conductance\cite{chiodelli2005synthesis,bisri2017endeavor}.  
Between pulses of the voltage train (with off-time $t_\textit{off} = \qty{1,4}{\second}$)and after the stimulus signal (with closed circuit but measuring the channel resistance continuously), we observe multi-state characteristics as well as a slow gradual decay of the conductance after application of the last pulse (see Fig.~\ref{fig: Fig2}b).
This behavior is known as long-term plasticity (LTP)\cite{van2017non} and is in accordance with a second device of the same microfabricated chip (see S10)
LTP is a persistent change in synaptic weight, spans from minutes over hours to years and is crucial for long-term memory (LTM) formation, supporting learning and memory functions\cite{dan2006spike,harvey2007locally,lynch2004long}. 
To enhance the learning accuracy and energy efficiency of an artificial neural network (ANN), LTP needs to ensure symmetry and linearity in synaptic plasticity, as well as provide an adequate number of states in conductance\cite{merolla2014million,burr2015experimental,kuzum2013synaptic,seo2020recent}. 
Thus, the LTP pulse-schematic from Fig.~\ref{fig: Fig2}b was then applied with ten potentiation/depression pulses for three cycles in a range of amplitude and duration for both, current and voltage-controlled pulse trains at \qty{150}{\degreeCelsius}.
Fig.~\ref{fig: Fig2}c and Fig.\ref{fig: Fig2}d illustrate the change in conductance $\Delta{G}$ per pulse, averaged over the 10 potentiation and depression pulses of the train. Fig.~\ref{fig: Fig2}d indicates that a pulse of amplitude \qty{0.7}{\V}, lasting for \qty{70}{\ms}, is sufficient to induce a discernible change in $\Delta{G}$. 
We expect faster pulsing times with further miniaturization of the synaptic transistor dimensions.
Error bars in these measurements are derived from three cycles for each $\vert E_W \vert$ and $t_p$.
As outlined in Schwacke et al. (2024), it is anticipated that $\Delta{G}$ is dependent upon the initial conductance state, $G_0$\cite{schwacke2024electrochemical}.
Consequently, the error in $\Delta{G}$ reflects the change in $G_0$ throughout the pulse train, especially when increasing $|E_W|$ or $t_p$. 
The current controlled measurements ($\Delta{G}\propto I_W$, $t_p$) are shown in S11. 
For both dependencies, $\Delta{G}$ on pulse amplitude and duration, potentiation and depression demonstrate linear behavior, although a slight asymmetry is noticeable between the two programming directions.
The synaptic plasticity within the full operational window of our synaptic transistor was additionally explored in the different conductivity regimes discussed before (high, medium and low in Fig.~\ref{fig: Fig1}c) to find the best suited region for operation.
Different constant voltages were applied to the gate to stabilize three initial conductance values for imposing subsequent pulse trains with constant voltage ($E_W = \qty{\pm{0,5}}{\V}$) and symmetric duration ($t_p = t_\textit{off} = \qty{220}{ms}$) in a range of pulse numbers ($\numrange{10}{100}$ pulses), see Fig.~\ref{fig: Fig2}a.
The total variation of conductance is observed to be higher for the high conductance states but in the mid-conductance range, the synaptic plasticity shows a highly symmetric and linear electric response, which was therefore chosen for the next experiments. 
The linearity of synaptic weight update during potentiation and depression can be determined by the non-linearity factor $\nu$ referring to either potentiation ($\nu_p$) or depression ($\nu_d$) (see S12)\cite{jang2015optimization,nikam2019near} 
In a perfectly symmetric and linear device, $\nu$ equals zero. 
The obtained linearity for potentiation and depression in our device is $\nu_p = \nu_d = \num{0.3}$ for \num{50} potentiation and depression pulses (Fig.~\ref{fig: Fig2}f).  
Moreover, the symmetry in synaptic weight update during potentiation and depression can be quantified by the asymmetric ratio (AR, see S12)\cite{nikam2019near}. 
AR is zero for an ideal symmetric case. For the mid conductance range, a remarkable low asymmetric ratio of $AR = \num{0.03}$ and \newline $AR= \num{0.1}$ between $n=\num{50}$ and $n=\num{100}$ potentiation and depression pulses, respectively, is observed (data for $n=\num{100}$ see S13).  
Accordingly, we can conclude that our synaptic transistor shows a highly symmetric and linear multi-state synaptic plasticity, fulfilling previously mentioned criteria of LTP for efficient ANNs.

Another important property is the large dynamic range (DR) strongly related to the synaptic devices’ learning capability and energy consumption\cite{nikam2021all,merolla2014million,burr2015experimental}. 
DR is determined by the difference between maximum and minimum conductance ($G_\text{max}$, $G_\text{min}$) state during programming. 
For the \numlist{50;100} multi-state programming cycles, respective DR values of \numlist{1.4;2} are observed.  
Here, the programming currents exhibit an average value of $I_W = \qty{4,6(0,1)}{\nano\ampere}$ for all the potentiation pulses and $I_W = \qty{6,2(0,1)}{\nano\ampere}$ for all the depression pulses, demonstrating the high stability of the BICUVOX electrolyte (see Fig.~\ref{fig: Fig2}g and S14).
This results in an averaged energy efficiency per pulse of $\qty{1.1(0.2)}{\milli\joule\per\siemens}$ for potentiation and to $\qty{1.5(0.3)}{\milli\joule\per\siemens}$ for depression (see S15). 
This energy efficiency for our in-plane synaptic transistor is comparable with protonic devices and increase the energy efficiency of previously reported oxide-ion intercalation devices by $\sim \num{e2}$ \cite{kim2019metal}. 
Similar to the calculations of Onen \textit{et al.}~\cite{onen2021cmos}, we estimate the energy efficiency for our synaptic transistor with a 100~$\times$~100~\si{\nm\squared} channel-architecture to $\sim \qty{10}{\femto\joule}$.
Overall, our device offers low energy consumption during weight updates fulfilling another crucial criterion for efficient neuromorphic computing\cite{merolla2014million,burr2015experimental,kuzum2013synaptic,seo2020recent}.

\begin{figure}[t]
\centering
\includegraphics[width=\textwidth]{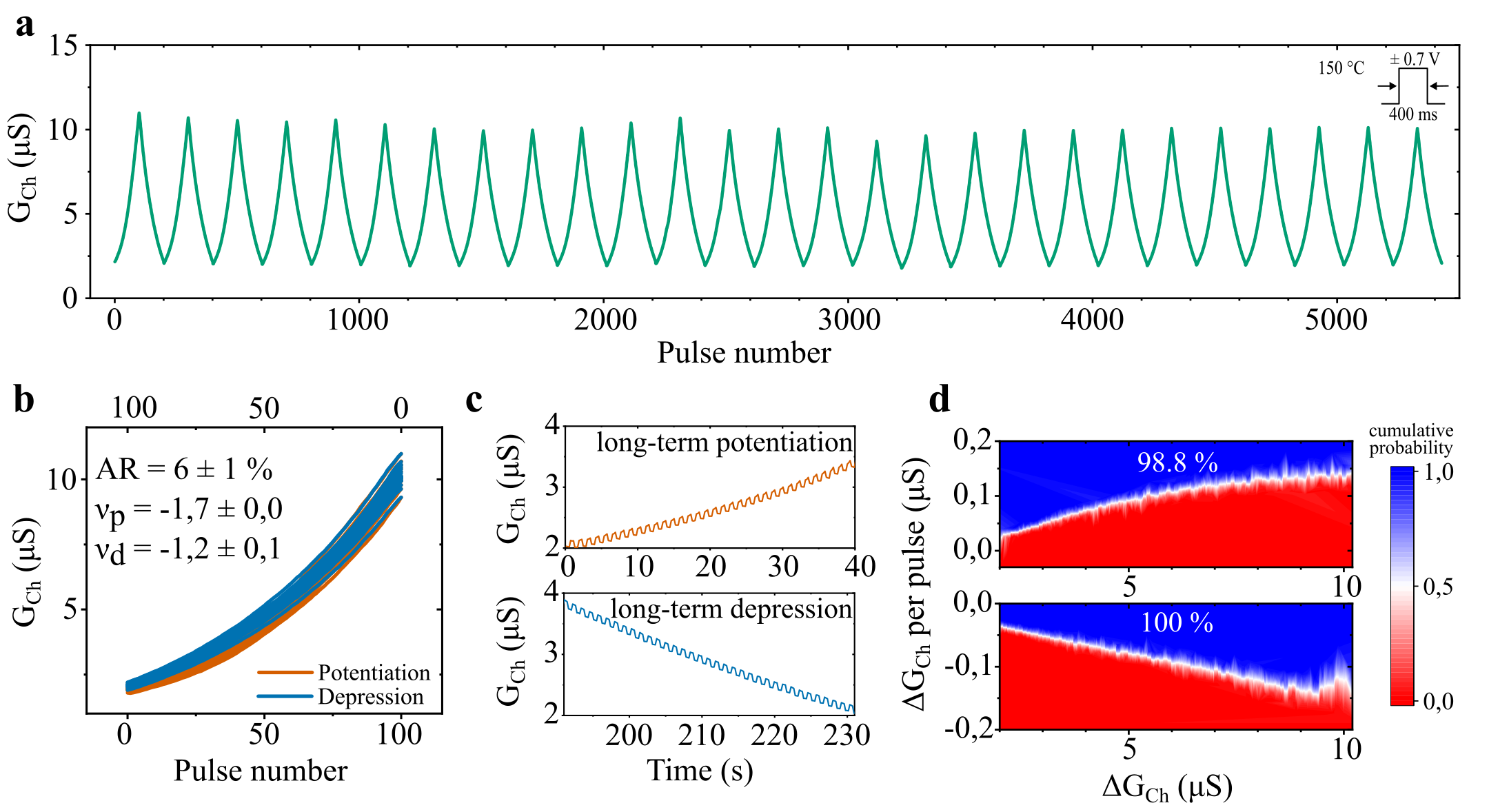}
\caption{Device endurance and short-term plasticity. 
\textbf{a} Reversible voltage controlled programming at \qty{150}{\degreeCelsius} across 100 states for \num{> 5000} operations with $tp = \qty{400}{\milli\second}$ and $E_W = \qty{+- 0,7}{\V}$ in the medium conductance regime demonstrating no significant change in $G_{max}/G_{min}$ for each cycle (see~S16). 
\textbf{b} Linearity- and symmetry comparison of potentiation and depression pulses for all cycles. 
\textbf{c} Characteristic saw-tooth-like long-term potentiation (LTP) and -depression (LTD) behavior during conductance programming. 
\textbf{d} Cumulative probability maps for potentiation (top panel) and depression (bottom panel) during programming for \num{5000} states demonstrating excellent switching-accuracy. 
}
\label{fig: Fig3}
\end{figure}

To investigate endurance, retention, and cycle-to-cycle variations, cycling across 100 states in the mid-conductance range ($\qtyrange{5}{10}{\micro\siemens}$) at a temperature of \qty{150}{\degreeCelsius} was conducted ($t_p = t_\textit{off} = \qty{400}{\milli\second}$, $E_W = \qty{\pm 0.7}{\V}$).
Each cycle involves 100 potentiation and depression pulses. 
Fig.~\ref{fig: Fig3}a illustrates predictable and great endurance of synaptic plasticity without degradation in DR ($G_{max}\text{/}G_{min} = \num{5.1 +- 0.1}$) across the cycles over 5000 operations (see also S16), matching the requirements for implementation into ANNs\cite{fuller2019parallel}. 
Fig.~\ref{fig: Fig3}b displays weight update characteristics of all cycles in dependance on applied pulse number, showing high symmetry ($AR = \num{0.06+-0.01}$) and sufficient linearity ($\nu_p = \num{-1,7 +- 0,0}$ and $\nu_p = \num{-1,2 +- 0,1}$) during cycling. 
Retention is determined from a voltage-controlled $\Delta G$ on $t_p$ dependence experiment (see S17), revealing a retention rate of $\qty{2}{\percent}$. 
At this point it shall be noted that a small amount of weight decay can be beneficial to prevent overfitting in neural network simulations\cite{krogh1991simple}. 
During the cycling, we observed the distinctive saw-tooth-like pattern in conductance programming, which is characteristic of long-term plasticity (see Fig.~\ref{fig: Fig3}c)\cite{van2017non}. 
To fully account for noise and accuracy in synaptic weight programming we acquired a statistical distribution of conductance levels during the cycling from Fig.~\ref{fig: Fig3}a. 
The resulting cumulative probability maps for potentiation (Fig.~\ref{fig: Fig3}d top panel) and depression (Fig.~\ref{fig: Fig3}d bottom panel) show excellent accuracies of \qty{98,8}{\percent} and \qty{100}{\percent}, respectively. 
The mean change in conductance for potentiation ($\Delta G_p = \qty{81 +- 39}{\nano\siemens}$) and depression ($\Delta G_d = \qty{80 +- 32}{\nano\siemens}$) resulted in a signal-to-noise ratio $\Delta G^2\text{/}\sigma^2$ of \num{4.3} and \num{6.3}, respectively (see S18). 
These findings demonstrate that the here developed ElSys fulfills essential parameters for future implementation in efficient neural networks such as great endurance, retention, and negligible cycle-to-cycle variations \cite{merolla2014million,burr2015experimental,kuzum2013synaptic,seo2020recent}. 

\begin{figure}[t]
\centering
\includegraphics[width=0.8\textwidth]{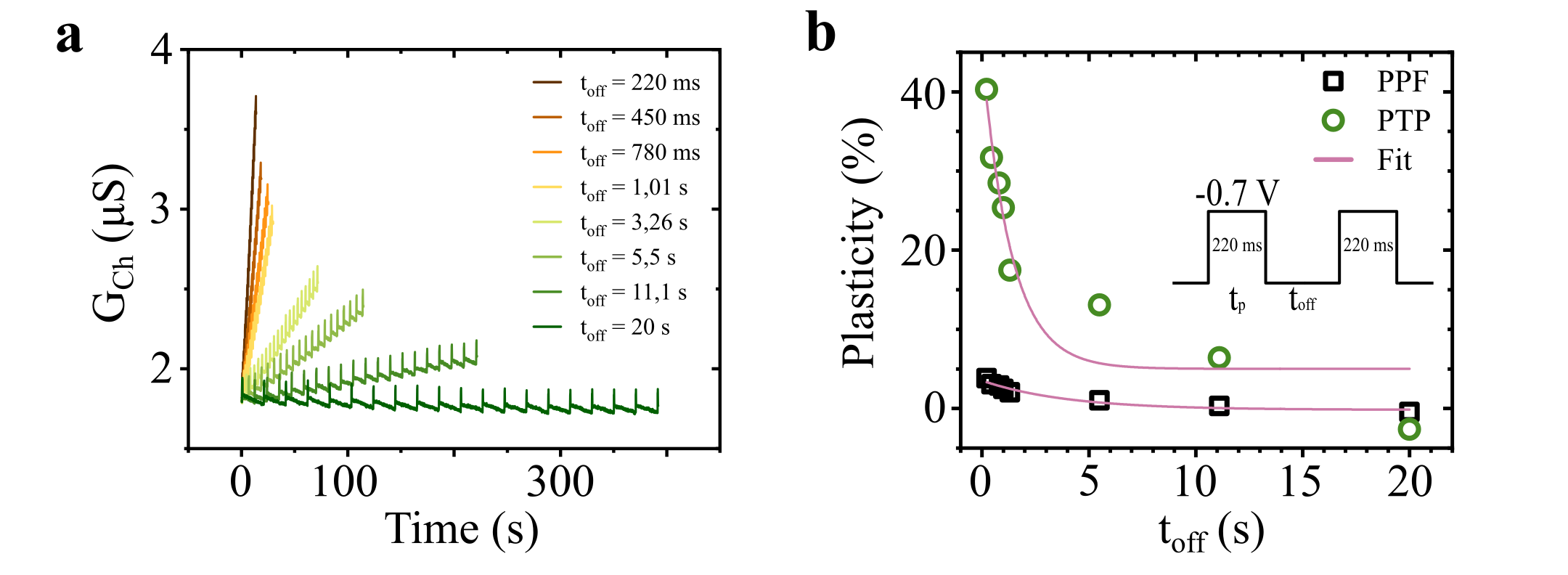}
\caption{Synaptic transistor STP dynamics under the application of voltage trains of 20 pulses with duration $t_p = \qty{220}{\milli\second}$ and off-time $t_\textit{off}$ varying from $\qtyrange{0.22}{20}{\second}$. Change in synaptic strenght ("Plasticity") reported 
\textbf{a} vs. measurement time and 
\textbf{b} vs. dependance of $t_\textit{off}$.
} 
\label{fig: Fig4}
\end{figure}

In addition to LTP, short-term plasticity (STP), \textit{i.e.} the temporary change of synaptic strength in response to stimulation\cite{zucker2002short,atluri1996determinants,ling2020electrolyte}, plays a crucial role in decoding temporal information and enables synapses to adapt based on the timing and frequency of stimuli\cite{zucker2002short,atluri1996determinants,debanne1996paired,dan2006spike}. 
To investigate the presence of STP in our device, we tested its dynamic response under the application of voltage-controlled 20-pulse trains with constant duration ($t_p = \qty{220}{\milli\second}$) while varying pulse interval times ($t_\textit{off}=\qty{220}{\milli\second} \text{-} \qty{20}{\second}$) as presented in Fig.~\ref{fig: Fig4}a. 
Complementary, Fig.~\ref{fig: Fig4}b shows the relative variation of conductance in relation to the first pulse obtained after the second (paired pulse facilitation, PPF) and tenth pulse (post-tetanic potentiation, PTP) as a function of $t_\textit{off}$ (see also S19)\cite{zucker2002short,buonomano2000decoding,atluri1996determinants,kim2013carbon,van2017non}. 
In Fig.~\ref{fig: Fig4}a,b STP is observed, with respective maximum $P_{PPF}$ and $P_{PTP}$ values of \qty{3.8}{\percent} and \qty{40.3}{\percent}. 
Since no effective electric-double-layer (EDL) is formed in all-solid-state oxide electrolytes during millisecond pulsing in the $\qty{150}{\degreeCelsius}$ temperature range\cite{tsuchiya2013all}, the dependence of $P_{PPF}$ and $P_{PTP}$ on the pulse interval times can be described by the single-phase behavior 
\begin{align}
    P_{PPF}(t) &= C_1 \cdot \text{exp}\left(-\frac{t_\textit{off}}{\tau}\right) + P_{PPF,0} \\
    P_{PTP}(t) &= C_1 \cdot \text{exp}\left(-\frac{t_\textit{off}}{\tau}\right) + P_{PTP,0} 
\end{align}
where $t_\textit{off}$ is the pulse interval time, $C_1$ is the initial facilitation magnitude, $\tau$ is the characteristic relaxation time of the process and $P(t)$ denotes the change in plasticity right after the second (PPF), or tenth (PTP) programming pulse\cite{regehr2012short,guo2015paired}. 
For the fitted PPF we calculated $C_1 = \qty{3.8}{\percent}$ , $\tau = \qty{3.9}{\second}$, $P_{PPF,0} = \qty{-0.2}{\percent}$ and for the fitted PTP we calculated $C_1 = \qty{39.8}{\percent}$ , $\tau = \qty{1.3}{\second}$, $P_{PTP,0} = \qty{5.0}{\percent}$. 
Similar STP characteristics are also observed with shorter pulsing ($t_p = \qty{50}{\milli\second}$) with $\tau = \qty{39}{\milli\second}$ (PPF) and $\tau = \qty{136}{\milli\second}$ (PTP), as shown in S19. 
Note, that obtained relaxation time constants are consistent with those in biological synapses\cite{zucker2002short}. 

To understand the origin of the STP characteristics we performed Finite Element Model (FEM) simulations of electrochemical oxygen insertion in the ECRAM, see section S21 in supplementary information. 
The results show that the STP arises from a combination of different phenomena, such as the non-linear oxygen vacancy-conductance relationship in LSF, the resistive contribution of oxygen vacancy diffusing in the electrolyte/channel and the typical tendency of a battery-like system to reestablish the initial defect configuration with $E_W = \qty{0}{\V}$. 
For frequent pulses, oxide ions do not have enough time to migrate to the internal part of the channel, accumulating at the interface facing the reservoir. 
Due to the non-linear relation of the oxide-ion concentration/conductivity we previously observed for LSF50\cite{nizet2024optoionic}, the total variation of conductance is larger if the same oxide-ions accumulate near the interface than if they distribute homogenously across all the channel. 
Increasing $t_\textit{off}$, the system is held more time in short circuit conditions ($E_W = \qty{0}{\V}$), allowing oxygen vacancies to diffuse back to the reservoir and to restore the initial equilibrium. 
This behavior is believed to be at the origin of the STP characteristics, as qualitatively reproduced by the simulations (see Fig.~S28). 
Interestingly, these results suggest that LTP or STP behavior may be enhanced or suppressed by the design of channel’s dimensions or configuration, offering an additional tool to optimize a specific computational task’s requirement.


\subsection{Artificial Neural Network Simulation}
\label{subsec:ANN}

\begin{figure}[h!]
\centering
\includegraphics[width=\textwidth]{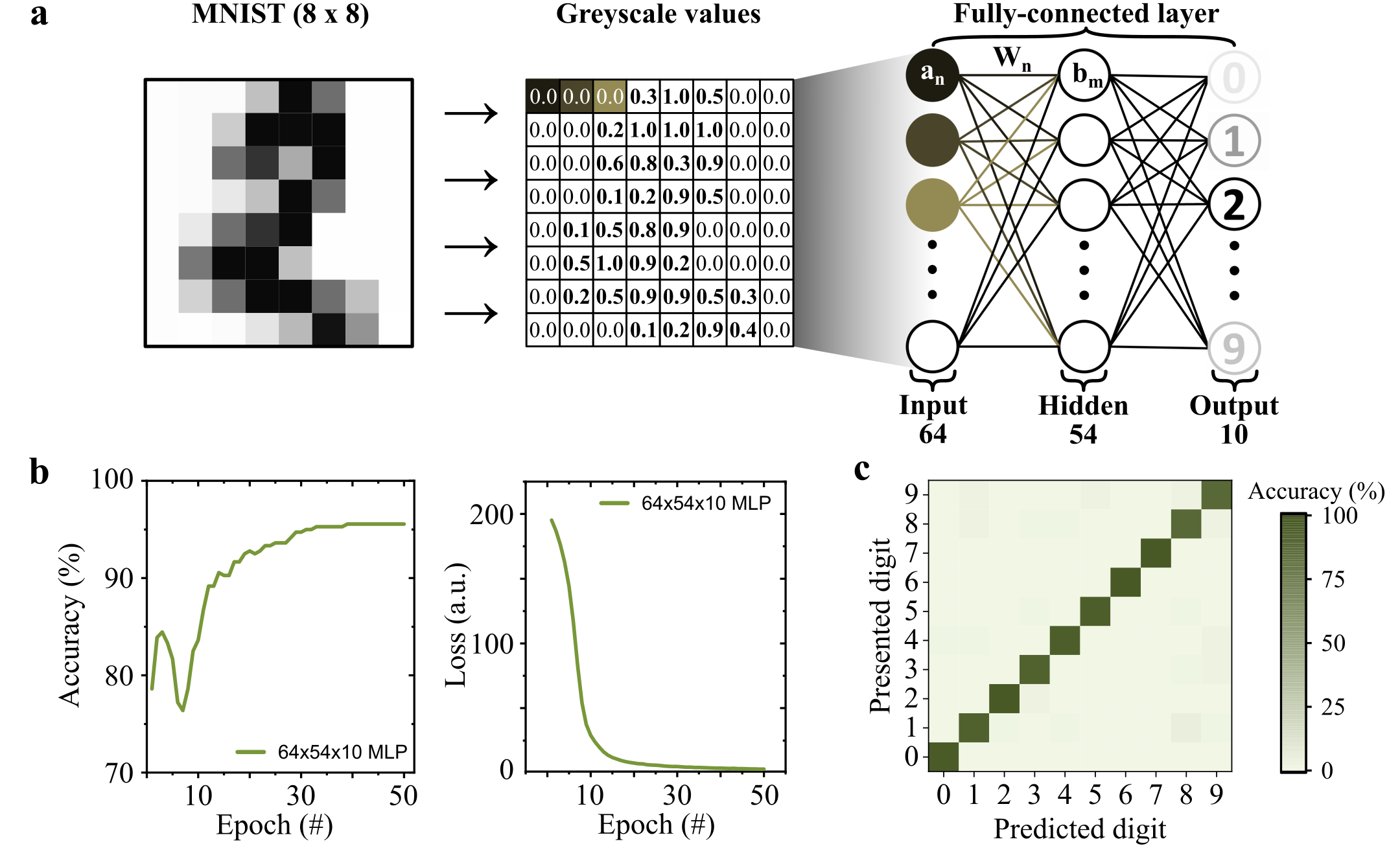}
\caption{Artificial neural network simulation of our synaptic transistor. 
\textbf{a}  \numproduct{8 x 8} pixel image of the number "2" in the training data (MNIST) is translated into grey scale values for each pixel. 
We used a three-layer (\numproduct{64 x 54 x 10}) architecture. 
Each input $a_n$, \textit{i.e.} simulating a current or voltage signal, is multiplicated with the weight-value $W_n$ stored in each neuron (simulating our synaptic transistor). 
The neuron activation $b_m$  in each layer is calculated by the activation function $tanh$, a bias $B$ and the weighted sum of the inputs.
\textbf{b} The model's accuracy on test data improving over time to \qty{96}{\percent} and loss as training progresses. 
\textbf{c} Mean confusion matrics across folds. Diagonal elements show correctly classified numbers for each class. 
Higher values on the diagonal indicate better performance. 
} 
\label{fig: Fig5}
\end{figure}

Having characterized the synaptic properties of our device and demonstrated that it meets several criteria for efficient neuromorphic computing (see Table~T\ref{tab: characteristics} for a complete list of values), we integrated the experimentally obtained synaptic properties into an artificial neural network (ANN) simulation. 
In this evaluation, the ANN was constructed such that our synaptic transistors, with their conductance values and their experimentally obtained synaptic properties, represent the weights $W$. 
These weights determine the strength of the connections between the neurons, with activation values $a,b$, in the network. 
We carried out the archetypal pattern recognition of handwritten digits employing the standard MNIST image dataset \numproduct{8 x 8} pixel.
A multi-layer perceptron neural network type was used in a three-layer architecture 
(\numproduct{64 x 54 x 10}) with backpropagation (stochastic gradient descent) optimization algorithm including 5-fold cross validation (see Fig.~\ref{fig: Fig5}a)\cite{aguirre2024hardware}. 
The simulation was implemented using the PyTorch library\cite{paszke2019pytorch}. 
After training the ANN for 50 epochs, an excellent accuracy of \qty{96}{\percent} for categorizing unseen test-data of the current fold has been achieved during the testing (see Fig.~\ref{fig: Fig5}b). 
More details on the training are given in the methods.  
Fig.~\ref{fig: Fig5}c shows the confusion matrix, a commonly used measure that illustrates an ANN’s capability to link each input pattern (an \numproduct{8 x 8} pixel handwritten digit) with its respective class (a digit from \numrange{0}{9}) and provides a graphical representation of the inference accuracy for all potential inputs\cite{aguirre2024hardware}. 
The accuracy in the confusion matrix represents the percentage of correctly predicted digits out of all predicted digits. Observed high accuracies for each predigted predicted digit illustrate the effective implementation of our device in ANNs.  
Overall, one can conclude that an ANN based on our synaptic transistor device delivers excellent fidelity results reinforcing the interest of this innovative technology for implementation in real neuromorphic hardware.  

\newpage

\section{Conclusion}\label{sec: Conclusion}

Microfabricated and fully integrable all-solid-state oxide-ion synaptic transistors for application in neuromorphic hardware were designed, fabricated and evaluated for operation as programmable resistors using an applied voltage. 
The three-terminal multi-state synaptic transistor was demonstrated in a broad operational range showing a deterministic and reversible behavior precisely controlling three orders of magnitude of the channel conductance by employing voltages as low as \qty{\pm 0.5}{\V}.
This synaptic transistor exhibits synaptic characteristics such as long- and short-term-plasticity, paired-pulse facilitation, and post-tetanic potentiation as required for mimicking biological neural networks. 
Furthermore, it demonstrates features critical for efficient neuromorphic computing including excellent linear and symmetric synaptic plasticity ($AR = \num{0.06}$, $\nu_p = \numrange{0,3}{-1.7}$, $DR = \num{5.1+-0,1}$)non-volatile states with low energy consumption ($\sim\qty{1}{\milli\joule\per\siemens}$ per pulse) and great endurance with negligible cycle-to-cycle variations. 
Finally, our study successfully integrated such experimentally derived synaptic properties into an artificial neural network simulation for handwritten digit recognition, achieving a high accuracy of 
$\qty{96}{\percent}$ in unseen data. 
These findings demonstrate the potential of our technology for effective implementation in analog neuromorphic computing based on oxide-iontronics.

\begin{table}[h!]
\caption{Synaptic transistor characteristics. }\label{tab: characteristics}
\begin{tabular}{c|c}
\toprule
Category & Characteristic  \\
\midrule
Operation window  & $\sim \qtyrange{e-1}{e2}{\siemens\per\cm}$  \\ 
Channel dimensions  & \begin{tabular}[c]{@{}c@{}}width: \qty{6}{\micro\metre} \\ length: \qty{35}{\micro\metre}\\ gap$_{Res \text{–} Ch}$: \qty{3}{\micro\metre} \end{tabular} \\ 
Pulsing  & \begin{tabular}[c]{@{}c@{}}$\qty{+- 0,5}{\V}$\\ down to $\qty{50}{\milli\second}$\end{tabular}  \\ 
Retention rate RR & \qty{2}{\percent}  \\
Asymmetric ratio AR & \qty{6 +- 1}{\percent} (200 pulses per cycle)  \\
Linearity $\nu$ & \begin{tabular}[c]{@{}c@{}}min: $\num{0,3}$ \\ max: $\num{-1,7}$ \end{tabular} \\
Dynamic range DR & max: $\num{5,1 +- 0,1}$  \\
Short-term plasticity STP & Yes \\
Long-term plasticity LTP & Yes \\
Paired-pulse facilitation PPF & Yes; $\tau = \qty{39}{\milli\second}$ \\
Post-tetanic potentiation PTP & Yes; $\tau = \qty{136}{\milli\second}$ \\
Switching accuracy  & \begin{tabular}[c]{@{}c@{}}depression: $\qty{100}{\percent}$ \\ potentiation: $\qty{98,8}{\percent}$ \end{tabular}  \\
Energy efficiency during cycling  & \qty{5.0(2.3)}{\milli\joule\per\siemens} per pulse \\
Accuracy on MNIST-data classification  & $\qty{96}{\percent}$  \\
\botrule
\end{tabular}
\end{table}

\newpage

\section{Methods}\label{sec: Methods}

\subsection{Materials}

The commercial LSF50 pellet target (\qty{99,9}{\percent}) was purchased from Codex International, France. 
For the fabrication of the home-made BICUVOX pellet target, $\text{Bi}_\text{2}\text{V}_\text{0.9}\text{Cu}_{0.1}\text{O}_\text{5.35}$ powder was first synthesized in-house by solid state synthesis, from basic oxide precursors ($\text{Bi}_\text{2}\text{O}_\text{3}$, $\text{V}_\text{2}\text{O}_\text{5}$ and CuO, \qty{99,9}{\percent} Sigma-Aldrich) mixed in the corresponding stoichiometric ratio. 
Then, a three-inch pellet was fabricated by uniaxial pressing and sintered under ambient air at \qty{850}{\degreeCelsius} for \qty{6}{\hour}.

\subsection{Device fabrication}

All BICUVOX and LSF50 thin films were deposited via pulsed laser deposition (PLD, PVD Products) at a substrate temperature of \qty{600}{\degreeCelsius} on a [100]-orientated \qtyproduct{5 x 5}{\mm} $(\text{La}\text{Al}\text{O}_\text{3})_\text{0.3}(\text{Sr}\text{Al}\text{Ta}_\text{0.6})_\text{0.7}$-substrate (LSAT). 
The films were grown with an energy fluency of \qty{0,8}{\joule\per\cm\squared} per pulse at a frequency of \qty{10}{\hertz}. 
Epitaxial BICUVOX was deposited at oxygen pressure of p$\text{O}_\text{2} = \qty{0,267}{\milli\bar}$ at a substrate-target distance of $\qty{110}{\mm}$. 
LSF50 was deposited at oxygen pressure of p$\text{O}_\text{2} = \qty{0,007}{\milli\bar}$ using a Si hard mask \qtyproduct{400 x 400}{\micro\metre} at a substrate-target distance of $\qty{90}{\mm}$. 
For microfabrication of the synaptic transistor, a mask-less photolithography was done by spin-coating a positive photoresist (S1813, Shipley) at 4000 rpm on the thin-film coated substrate. 
After exposing (MicroWriter ML3 Pro, Durham Magneto Optics) and developing, the electrode architecture for channel and reservoir was etched into the BICUVOX$\vert$LSF thin films via ion-milling ($\text{Ar}^\text{+}$) for $\qty{12}{\minute}$. 
After lift-off, a second photolithography was done in similar procedure for the Cr$\vert$Au current collector design. 
Current collectors were metallized with a $\qty{5}{\nano\metre}$ Cr adhesion-layer and $\qty{100}{\nano\metre}$ Au-layer. 
After lift-off, the microfabrication was complete.

\subsection{Thin film characterization}

Thin films microstructure was studied by Scanning Electron Microscopy (SEM, ZEISS AURIGA). 
SEM-coupled Energy-Dispersive X-ray spectroscopy (EDX) was used for the elemental distribution analysis. 
Non-contact Atomic Force Microscopy (AFM, XE 100 Park System Corp.) was additionally used for topography and top view imaging of the epitaxial BICUVOX and LSF50 films as well as the synaptic transistor channel and reservoir dimensions. 
Spectroscopic ellipsometry (UVISEL ellipsometer, Horiba scientific) measured the optical constants of BICUVOX and LSF thin films within a photon energy range from \qtyrange{0,6}{5}{\eV} with \qty{0.05}{\eV} intervals, using a \ang{70} incident light beam. 
The data acquired from ellipsometry were subjected to modeling and fitting via Horiba Scientific's DeltaPsi2 software to determine the thin films thickness. 
X-ray diffraction (XRD, Bruker D8 Advanced diffractometer) was employed for microstructural characterization and phase identification with Cu-K$_{\alpha}$ radiation in a coupled \text{$\theta$}-2\text{$\theta$} Bragg–Brentano configuration.

\subsection{Electrical and electrochemical characterization}

Electrochemical impedance spectroscopy (EIS, 4294A Precision Impedance Analyzer, Agilent) of the BICUVOX thin film was conducted using an AC bias of \qty{50}{\mV} in a range of frequency from \qty{10}{\mega\hertz} to \qty{40}{\hertz} between channel and reservoir.
The operational window of the LSF50-channel at \qty{150}{\degreeCelsius} was determined after applying a DC bias between reservoir and channel within a range of $E_W$ from \qtyrange{0,1}{-0,4}{\V} until equilibrium, followed by a cyclic voltammetry between source and drain electrode with \qty{10}{\milli\V\per\second} within a range from \qtyrange{-0,1}{0,1}{\V} at OCV ($I_W = \qty{0}{\ampere}$). 
Basic synaptic transistor operation-, long-term and short-term memory characteristics were measured in a three-electrode setup for simultaneous recording of readout- and writing data (4200 SCS parameter analyzer, Keithley Technologies Inc.). 
For pulses shorter than $\qty{200}{\milli\second}$ the PMU unit was used and for pulses longer than $\qty{200}{\milli\second}$ the SMU unit was used. 
Temperature-dependency measurements were performed using a thermal chuck with controller (S1080, Microworld) after initial calibration.
Arrhenius representation of the extrapolated ionic conductivity was obtained by applying open circuit potential between channel and reservoir at a fixed temperature between \qtyrange{100}{300}{\degreeCelsius} (in \qty{25}{\degreeCelsius}-steps) and waiting for equilibrium in ambient air. 
Once equilibrium was achieved, previously described EIS was conducted. 
The conductivity of BICUVOX was obtained from the high frequency semicircle according to our previous work\cite{garbayo2019thin}. 

\subsection{Neural network simulation}

The neural network model implemented in this study was trained and evaluated using the Optical Recognition of Handwritten Digits dataset (MNIST data, \numproduct{8 x 8} pixel) with the PyTorch library. 
The model in a three-layer fully connected architecture (\numproduct{64 x 54 x 10}) was wrapped using a scikit-learn compatible wrapper for compatibility. 
Stratified cross-validation from scikit-learn was employed for training and evaluation\cite{scikit-learn}, with the dataset split into five folds. 
For each fold, a portion of the data was used for training and a different portion was used for testing. 
During training, the model's weights were updated using a custom synaptic weight update scheme (asymmetric ratio and linear weight update S12, weight retention S17, conductance range and dynamic range S16) integrating the experimentally obtained characteristics of our synaptic transistor (Table~T\ref{tab: characteristics}). 
The optimizer used was Stochastic Gradient Descent (SGD) from the PyTorch library. 
The performance of the model was assessed by test accuracies and losses across epochs. 
The accuracy value obtained after each epoch is calculated against unseen data for the current fold. 
During each epoch, after training on the training data, the model's performance was evaluated on the test data that it has not seen during training. 
Additionally, mean confusion matrices were obtained to evaluate the model's predictive performance.

\backmatter

\bmhead{Acknowledgements}

This project received funding from the European Union's Horizon 2020 research and innovation program under grant agreement No 101066321 (TRANSIONICS) and under the Marie Skłodowska-Curie Actions Postdoctoral Fellowship grant (101107093). The authors acknowledge support from the Generalitat de Catalunya (2021-SGR-00750, NANOEN). C. B.-G. acknowledges funding from a Marie Skłodowska Curie Actions Postdoctoral Fellowship grant (101064374).

\begin{appendices}

\end{appendices}

\bibliography{ms}

\begin{thebibliography}{10}
\expandafter\ifx\csname url\endcsname\relax
  \def\url#1{\burl{#1}}\fi
\expandafter\ifx\csname urlprefix\endcsname\relax\def\urlprefix{URL }\fi
\providecommand{\bibinfo}[2]{#2}
\providecommand{\eprint}[2][]{\url{#2}}
\providecommand{\doi}[1]{\url{https://doi.org/#1}}
\bibcommenthead

\bibitem{dan2006spike}
\bibinfo{author}{Dan, Y.} \& \bibinfo{author}{Poo, M.-M.}
\newblock \bibinfo{title}{Spike timing-dependent plasticity: from synapse to
  perception}.
\newblock \emph{\bibinfo{journal}{Physiological reviews}}
  \textbf{\bibinfo{volume}{86}}, \bibinfo{pages}{1033--1048}
  (\bibinfo{year}{2006}).

\bibitem{harvey2007locally}
\bibinfo{author}{Harvey, C.~D.} \& \bibinfo{author}{Svoboda, K.}
\newblock \bibinfo{title}{Locally dynamic synaptic learning rules in pyramidal
  neuron dendrites}.
\newblock \emph{\bibinfo{journal}{Nature}} \textbf{\bibinfo{volume}{450}},
  \bibinfo{pages}{1195--1200} (\bibinfo{year}{2007}).

\bibitem{kuzum2013synaptic}
\bibinfo{author}{Kuzum, D.}, \bibinfo{author}{Yu, S.} \& \bibinfo{author}{Wong,
  H.~P.}
\newblock \bibinfo{title}{Synaptic electronics: materials, devices and
  applications}.
\newblock \emph{\bibinfo{journal}{Nanotechnology}}
  \textbf{\bibinfo{volume}{24}}, \bibinfo{pages}{382001}
  (\bibinfo{year}{2013}).

\bibitem{atluri1996determinants}
\bibinfo{author}{Atluri, P.~P.} \& \bibinfo{author}{Regehr, W.~G.}
\newblock \bibinfo{title}{Determinants of the time course of facilitation at
  the granule cell to purkinje cell synapse}.
\newblock \emph{\bibinfo{journal}{Journal of Neuroscience}}
  \textbf{\bibinfo{volume}{16}}, \bibinfo{pages}{5661--5671}
  (\bibinfo{year}{1996}).

\bibitem{zucker2002short}
\bibinfo{author}{Zucker, R.~S.} \& \bibinfo{author}{Regehr, W.~G.}
\newblock \bibinfo{title}{Short-term synaptic plasticity}.
\newblock \emph{\bibinfo{journal}{Annual review of physiology}}
  \textbf{\bibinfo{volume}{64}}, \bibinfo{pages}{355--405}
  (\bibinfo{year}{2002}).

\bibitem{debanne1996paired}
\bibinfo{author}{Debanne, D.}, \bibinfo{author}{Guerineau, N.~C.},
  \bibinfo{author}{G{\"a}hwiler, B.} \& \bibinfo{author}{Thompson, S.~M.}
\newblock \bibinfo{title}{Paired-pulse facilitation and depression at unitary
  synapses in rat hippocampus: quantal fluctuation affects subsequent release.}
\newblock \emph{\bibinfo{journal}{The Journal of physiology}}
  \textbf{\bibinfo{volume}{491}}, \bibinfo{pages}{163--176}
  (\bibinfo{year}{1996}).

\bibitem{aguirre2024hardware}
\bibinfo{author}{Aguirre, F.} \emph{et~al.}
\newblock \bibinfo{title}{Hardware implementation of memristor-based artificial
  neural networks}.
\newblock \emph{\bibinfo{journal}{Nature communications}}
  \textbf{\bibinfo{volume}{15}}, \bibinfo{pages}{1974} (\bibinfo{year}{2024}).

\bibitem{shi2013correlated}
\bibinfo{author}{Shi, J.}, \bibinfo{author}{Ha, S.~D.}, \bibinfo{author}{Zhou,
  Y.}, \bibinfo{author}{Schoofs, F.} \& \bibinfo{author}{Ramanathan, S.}
\newblock \bibinfo{title}{A correlated nickelate synaptic transistor}.
\newblock \emph{\bibinfo{journal}{Nature communications}}
  \textbf{\bibinfo{volume}{4}}, \bibinfo{pages}{2676} (\bibinfo{year}{2013}).

\bibitem{fuller2016li}
\bibinfo{author}{Fuller, E.~J.} \emph{et~al.}
\newblock \bibinfo{title}{Li-ion synaptic transistor for low power analog
  computing}.
\newblock \emph{\bibinfo{journal}{Advanced Materials}}
  \textbf{\bibinfo{volume}{29}} (\bibinfo{year}{2016}).

\bibitem{song2023recent}
\bibinfo{author}{Song, M.-K.} \emph{et~al.}
\newblock \bibinfo{title}{Recent advances and future prospects for memristive
  materials, devices, and systems}.
\newblock \emph{\bibinfo{journal}{ACS nano}} \textbf{\bibinfo{volume}{17}},
  \bibinfo{pages}{11994--12039} (\bibinfo{year}{2023}).

\bibitem{bauer2015magneto}
\bibinfo{author}{Bauer, U.} \emph{et~al.}
\newblock \bibinfo{title}{Magneto-ionic control of interfacial magnetism}.
\newblock \emph{\bibinfo{journal}{Nature materials}}
  \textbf{\bibinfo{volume}{14}}, \bibinfo{pages}{174--181}
  (\bibinfo{year}{2015}).

\bibitem{mehonic2022brain}
\bibinfo{author}{Mehonic, A.} \& \bibinfo{author}{Kenyon, A.~J.}
\newblock \bibinfo{title}{Brain-inspired computing needs a master plan}.
\newblock \emph{\bibinfo{journal}{Nature}} \textbf{\bibinfo{volume}{604}},
  \bibinfo{pages}{255--260} (\bibinfo{year}{2022}).

\bibitem{li2018efficient}
\bibinfo{author}{Li, C.} \emph{et~al.}
\newblock \bibinfo{title}{Efficient and self-adaptive in-situ learning in
  multilayer memristor neural networks}.
\newblock \emph{\bibinfo{journal}{Nature communications}}
  \textbf{\bibinfo{volume}{9}}, \bibinfo{pages}{2385} (\bibinfo{year}{2018}).

\bibitem{li2018analogue}
\bibinfo{author}{Li, C.} \emph{et~al.}
\newblock \bibinfo{title}{Analogue signal and image processing with large
  memristor crossbars}.
\newblock \emph{\bibinfo{journal}{Nature electronics}}
  \textbf{\bibinfo{volume}{1}}, \bibinfo{pages}{52--59} (\bibinfo{year}{2018}).

\bibitem{snider2007self}
\bibinfo{author}{Snider, G.~S.}
\newblock \bibinfo{title}{Self-organized computation with unreliable,
  memristive nanodevices}.
\newblock \emph{\bibinfo{journal}{Nanotechnology}}
  \textbf{\bibinfo{volume}{18}}, \bibinfo{pages}{365202}
  (\bibinfo{year}{2007}).

\bibitem{burr2015experimental}
\bibinfo{author}{Burr, G.~W.} \emph{et~al.}
\newblock \bibinfo{title}{Experimental demonstration and tolerancing of a
  large-scale neural network (165 000 synapses) using phase-change memory as
  the synaptic weight element}.
\newblock \emph{\bibinfo{journal}{IEEE Transactions on Electron Devices}}
  \textbf{\bibinfo{volume}{62}}, \bibinfo{pages}{3498--3507}
  (\bibinfo{year}{2015}).

\bibitem{li2018review}
\bibinfo{author}{Li, Y.}, \bibinfo{author}{Wang, Z.}, \bibinfo{author}{Midya,
  R.}, \bibinfo{author}{Xia, Q.} \& \bibinfo{author}{Yang, J.~J.}
\newblock \bibinfo{title}{Review of memristor devices in neuromorphic
  computing: materials sciences and device challenges}.
\newblock \emph{\bibinfo{journal}{Journal of Physics D: Applied Physics}}
  \textbf{\bibinfo{volume}{51}}, \bibinfo{pages}{503002}
  (\bibinfo{year}{2018}).

\bibitem{islam2019device}
\bibinfo{author}{Islam, R.} \emph{et~al.}
\newblock \bibinfo{title}{Device and materials requirements for neuromorphic
  computing}.
\newblock \emph{\bibinfo{journal}{Journal of Physics D: Applied Physics}}
  \textbf{\bibinfo{volume}{52}}, \bibinfo{pages}{113001}
  (\bibinfo{year}{2019}).

\bibitem{van2017non}
\bibinfo{author}{Van De~Burgt, Y.} \emph{et~al.}
\newblock \bibinfo{title}{A non-volatile organic electrochemical device as a
  low-voltage artificial synapse for neuromorphic computing}.
\newblock \emph{\bibinfo{journal}{Nature materials}}
  \textbf{\bibinfo{volume}{16}}, \bibinfo{pages}{414--418}
  (\bibinfo{year}{2017}).

\bibitem{zhang2022reconfigurable}
\bibinfo{author}{Zhang, H.-T.} \emph{et~al.}
\newblock \bibinfo{title}{Reconfigurable perovskite nickelate electronics for
  artificial intelligence}.
\newblock \emph{\bibinfo{journal}{Science}} \textbf{\bibinfo{volume}{375}},
  \bibinfo{pages}{533--539} (\bibinfo{year}{2022}).

\bibitem{yang2018all}
\bibinfo{author}{Yang, C.-S.} \emph{et~al.}
\newblock \bibinfo{title}{All-solid-state synaptic transistor with ultralow
  conductance for neuromorphic computing}.
\newblock \emph{\bibinfo{journal}{Advanced Functional Materials}}
  \textbf{\bibinfo{volume}{28}}, \bibinfo{pages}{1804170}
  (\bibinfo{year}{2018}).

\bibitem{yang2018artificial}
\bibinfo{author}{Yang, J.-T.} \emph{et~al.}
\newblock \bibinfo{title}{Artificial synapses emulated by an electrolyte-gated
  tungsten-oxide transistor}.
\newblock \emph{\bibinfo{journal}{Advanced Materials}}
  \textbf{\bibinfo{volume}{30}}, \bibinfo{pages}{1801548}
  (\bibinfo{year}{2018}).

\bibitem{tsuchiya2013all}
\bibinfo{author}{Tsuchiya, T.}, \bibinfo{author}{Terabe, K.} \&
  \bibinfo{author}{Aono, M.}
\newblock \bibinfo{title}{All-solid-state electric-double-layer transistor
  based on oxide ion migration in gd-doped ceo2 on srtio3 single crystal}.
\newblock \emph{\bibinfo{journal}{Applied Physics Letters}}
  \textbf{\bibinfo{volume}{103}} (\bibinfo{year}{2013}).

\bibitem{schwacke2024electrochemical}
\bibinfo{author}{Schwacke, M.}, \bibinfo{author}{{\v{Z}}guns, P.},
  \bibinfo{author}{del Alamo, J.}, \bibinfo{author}{Li, J.} \&
  \bibinfo{author}{Yildiz, B.}
\newblock \bibinfo{title}{Electrochemical ionic synapses with mg2+ as the
  working ion}.
\newblock \emph{\bibinfo{journal}{Advanced Electronic Materials}}
  \bibinfo{pages}{2300577} (\bibinfo{year}{2024}).

\bibitem{ling2020electrolyte}
\bibinfo{author}{Ling, H.} \emph{et~al.}
\newblock \bibinfo{title}{Electrolyte-gated transistors for synaptic
  electronics, neuromorphic computing, and adaptable biointerfacing}.
\newblock \emph{\bibinfo{journal}{Applied Physics Reviews}}
  \textbf{\bibinfo{volume}{7}} (\bibinfo{year}{2020}).

\bibitem{tang2018ecram}
\bibinfo{author}{Tang, J.} \emph{et~al.}
\newblock \bibinfo{title}{Ecram as scalable synaptic cell for high-speed,
  low-power neuromorphic computing}.
\newblock \emph{\bibinfo{journal}{2018 IEEE International Electron Devices
  Meeting (IEDM)}}  (\bibinfo{year}{2018}).

\bibitem{fuller2019parallel}
\bibinfo{author}{Fuller, E.~J.} \emph{et~al.}
\newblock \bibinfo{title}{Parallel programming of an ionic floating-gate memory
  array for scalable neuromorphic computing}.
\newblock \emph{\bibinfo{journal}{Science}} \textbf{\bibinfo{volume}{364}},
  \bibinfo{pages}{570--574} (\bibinfo{year}{2019}).

\bibitem{nikam2019near}
\bibinfo{author}{Nikam, R.~D.} \emph{et~al.}
\newblock \bibinfo{title}{Near ideal synaptic functionalities in li ion
  synaptic transistor using li3poxsex electrolyte with high ionic
  conductivity}.
\newblock \emph{\bibinfo{journal}{Scientific reports}}
  \textbf{\bibinfo{volume}{9}}, \bibinfo{pages}{18883} (\bibinfo{year}{2019}).

\bibitem{cui2023cmos}
\bibinfo{author}{Cui, J.} \emph{et~al.}
\newblock \bibinfo{title}{Cmos-compatible electrochemical synaptic transistor
  arrays for deep learning accelerators}.
\newblock \emph{\bibinfo{journal}{Nature Electronics}}
  \textbf{\bibinfo{volume}{6}}, \bibinfo{pages}{292--300}
  (\bibinfo{year}{2023}).

\bibitem{zhu2020comprehensive}
\bibinfo{author}{Zhu, J.}, \bibinfo{author}{Zhang, T.}, \bibinfo{author}{Yang,
  Y.} \& \bibinfo{author}{Huang, R.}
\newblock \bibinfo{title}{A comprehensive review on emerging artificial
  neuromorphic devices}.
\newblock \emph{\bibinfo{journal}{Applied Physics Reviews}}
  \textbf{\bibinfo{volume}{7}} (\bibinfo{year}{2020}).

\bibitem{nizet2024optoionic}
\bibinfo{author}{Nizet, P.} \emph{et~al.}
\newblock \bibinfo{title}{Optoionic impedance spectroscopy (ois): a model-less
  technique for in-situ electrochemical characterization of mixed ionic
  electronic conductors}.
\newblock \emph{\bibinfo{journal}{submitted}}  (\bibinfo{year}{2024}).

\bibitem{shi2021solid}
\bibinfo{author}{Shi, P.} \emph{et~al.}
\newblock \bibinfo{title}{Solid-state electrolyte gated synaptic transistor
  based on srfeo2. 5 film channel}.
\newblock \emph{\bibinfo{journal}{Materials \& Design}}
  \textbf{\bibinfo{volume}{210}}, \bibinfo{pages}{110022}
  (\bibinfo{year}{2021}).

\bibitem{talin2023ecram}
\bibinfo{author}{Talin, A.~A.}, \bibinfo{author}{Li, Y.},
  \bibinfo{author}{Robinson, D.~A.}, \bibinfo{author}{Fuller, E.~J.} \&
  \bibinfo{author}{Kumar, S.}
\newblock \bibinfo{title}{Ecram materials, devices, circuits and architectures:
  A perspective}.
\newblock \emph{\bibinfo{journal}{Advanced Materials}}
  \textbf{\bibinfo{volume}{35}}, \bibinfo{pages}{2204771}
  (\bibinfo{year}{2023}).

\bibitem{huang2023electrochemical}
\bibinfo{author}{Huang, M.} \emph{et~al.}
\newblock \bibinfo{title}{Electrochemical ionic synapses: progress and
  perspectives}.
\newblock \emph{\bibinfo{journal}{Advanced Materials}}
  \textbf{\bibinfo{volume}{35}}, \bibinfo{pages}{2205169}
  (\bibinfo{year}{2023}).

\bibitem{li2019low}
\bibinfo{author}{Li, Y.} \emph{et~al.}
\newblock \bibinfo{title}{Low-voltage, cmos-free synaptic memory based on li x
  tio2 redox transistors}.
\newblock \emph{\bibinfo{journal}{ACS applied materials \& interfaces}}
  \textbf{\bibinfo{volume}{11}}, \bibinfo{pages}{38982--38992}
  (\bibinfo{year}{2019}).

\bibitem{nguyen2022ultralow}
\bibinfo{author}{Nguyen, N.-a.} \emph{et~al.}
\newblock \bibinfo{title}{An ultralow power lixtio2-based synaptic transistor
  for scalable neuromorphic computing}.
\newblock \emph{\bibinfo{journal}{Advanced Electronic Materials}}
  \textbf{\bibinfo{volume}{8}}, \bibinfo{pages}{2200607}
  (\bibinfo{year}{2022}).

\bibitem{kim2019metal}
\bibinfo{author}{Kim, S.} \emph{et~al.}
\newblock \bibinfo{title}{Metal-oxide based, cmos-compatible ecram for deep
  learning accelerator}.
\newblock \emph{\bibinfo{journal}{2019 IEEE International Electron Devices
  Meeting (IEDM)}}  (\bibinfo{year}{2019}).

\bibitem{onen2021cmos}
\bibinfo{author}{Onen, M.}, \bibinfo{author}{Emond, N.}, \bibinfo{author}{Li,
  J.}, \bibinfo{author}{Yildiz, B.} \& \bibinfo{author}{Del~Alamo, J.~A.}
\newblock \bibinfo{title}{Cmos-compatible protonic programmable resistor based
  on phosphosilicate glass electrolyte for analog deep learning}.
\newblock \emph{\bibinfo{journal}{Nano Letters}} \textbf{\bibinfo{volume}{21}},
  \bibinfo{pages}{6111--6116} (\bibinfo{year}{2021}).

\bibitem{onen2022nanosecond}
\bibinfo{author}{Onen, M.} \emph{et~al.}
\newblock \bibinfo{title}{Nanosecond protonic programmable resistors for analog
  deep learning}.
\newblock \emph{\bibinfo{journal}{Science}} \textbf{\bibinfo{volume}{377}},
  \bibinfo{pages}{539--543} (\bibinfo{year}{2022}).

\bibitem{pradhan2024scalable}
\bibinfo{author}{Pradhan, D.~K.} \emph{et~al.}
\newblock \bibinfo{title}{A scalable ferroelectric non-volatile memory
  operating at 600\textdegree{} c}.
\newblock \emph{\bibinfo{journal}{Nature Electronics}} \bibinfo{pages}{1--8}
  (\bibinfo{year}{2024}).

\bibitem{melianas2020temperature}
\bibinfo{author}{Melianas, A.} \emph{et~al.}
\newblock \bibinfo{title}{Temperature-resilient solid-state organic artificial
  synapses for neuromorphic computing}.
\newblock \emph{\bibinfo{journal}{Science advances}}
  \textbf{\bibinfo{volume}{6}}, \bibinfo{pages}{eabb2958}
  (\bibinfo{year}{2020}).

\bibitem{nikam2021all}
\bibinfo{author}{Nikam, R.~D.}, \bibinfo{author}{Kwak, M.} \&
  \bibinfo{author}{Hwang, H.}
\newblock \bibinfo{title}{All-solid-state oxygen ion electrochemical
  random-access memory for neuromorphic computing}.
\newblock \emph{\bibinfo{journal}{Advanced Electronic Materials}}
  \textbf{\bibinfo{volume}{7}}, \bibinfo{pages}{2100142}
  (\bibinfo{year}{2021}).

\bibitem{lee2022strategies}
\bibinfo{author}{Lee, J.}, \bibinfo{author}{Nikam, R.~D.},
  \bibinfo{author}{Kwak, M.} \& \bibinfo{author}{Hwang, H.}
\newblock \bibinfo{title}{Strategies to improve the synaptic characteristics of
  oxygen-based electrochemical random-access memory based on material
  parameters optimization}.
\newblock \emph{\bibinfo{journal}{ACS Applied Materials \& Interfaces}}
  \textbf{\bibinfo{volume}{14}}, \bibinfo{pages}{13450--13457}
  (\bibinfo{year}{2022}).

\bibitem{jeong2021room}
\bibinfo{author}{Jeong, B.}, \bibinfo{author}{Veith, L.},
  \bibinfo{author}{Smolders, T.~J.}, \bibinfo{author}{Wolf, M.~J.} \&
  \bibinfo{author}{Asadi, K.}
\newblock \bibinfo{title}{Room-temperature halide perovskite field-effect
  transistors by ion transport mitigation}.
\newblock \emph{\bibinfo{journal}{Advanced Materials}}
  \textbf{\bibinfo{volume}{33}}, \bibinfo{pages}{2100486}
  (\bibinfo{year}{2021}).

\bibitem{garbayo2019thin}
\bibinfo{author}{Garbayo, I.} \emph{et~al.}
\newblock \bibinfo{title}{Thin film oxide-ion conducting electrolyte for near
  room temperature applications}.
\newblock \emph{\bibinfo{journal}{Journal of materials chemistry A}}
  \textbf{\bibinfo{volume}{7}}, \bibinfo{pages}{25772--25778}
  (\bibinfo{year}{2019}).

\bibitem{bae2019investigations}
\bibinfo{author}{Bae, H.} \emph{et~al.}
\newblock \bibinfo{title}{Investigations on defect equilibrium, thermodynamic
  quantities, and transport properties of la0. 5sr0. 5feo3-$\delta$}.
\newblock \emph{\bibinfo{journal}{Journal of The Electrochemical Society}}
  \textbf{\bibinfo{volume}{166}}, \bibinfo{pages}{F180} (\bibinfo{year}{2019}).

\bibitem{jana2019charge}
\bibinfo{author}{Jana, S.} \emph{et~al.}
\newblock \bibinfo{title}{Charge disproportionate antiferromagnetism at the
  verge of the insulator-metal transition in doped lafeo 3}.
\newblock \emph{\bibinfo{journal}{Physical Review B}}
  \textbf{\bibinfo{volume}{99}}, \bibinfo{pages}{075106}
  (\bibinfo{year}{2019}).

\bibitem{wang2019hole}
\bibinfo{author}{Wang, L.} \emph{et~al.}
\newblock \bibinfo{title}{Hole-induced electronic and optical transitions in l
  a 1- x s r x fe o 3 epitaxial thin films}.
\newblock \emph{\bibinfo{journal}{Physical Review Materials}}
  \textbf{\bibinfo{volume}{3}}, \bibinfo{pages}{025401} (\bibinfo{year}{2019}).

\bibitem{tang2021pushing}
\bibinfo{author}{Tang, Y.} \emph{et~al.}
\newblock \bibinfo{title}{Pushing the study of point defects in thin film
  ferrites to low temperatures using in situ ellipsometry}.
\newblock \emph{\bibinfo{journal}{Advanced Materials Interfaces}}
  \textbf{\bibinfo{volume}{8}}, \bibinfo{pages}{2001881}
  (\bibinfo{year}{2021}).

\bibitem{patrakeev2003electron}
\bibinfo{author}{Patrakeev, M.} \emph{et~al.}
\newblock \bibinfo{title}{Electron/hole and ion transport in la1- xsrxfeo3-d}.
\newblock \emph{\bibinfo{journal}{Journal of solid state chemistry}}
  \textbf{\bibinfo{volume}{172}}, \bibinfo{pages}{219--231}
  (\bibinfo{year}{2003}).

\bibitem{cheng2005thermochemistry}
\bibinfo{author}{Cheng, J.}, \bibinfo{author}{Navrotsky, A.},
  \bibinfo{author}{Zhou, X.-D.} \& \bibinfo{author}{Anderson, H.~U.}
\newblock \bibinfo{title}{Thermochemistry of la1-x sr x feo3-d solid solutions
  (0.0-x-1.0, 0.0-d-0.5)}.
\newblock \emph{\bibinfo{journal}{Chemistry of Materials}}
  \textbf{\bibinfo{volume}{17}}, \bibinfo{pages}{2197--2207}
  (\bibinfo{year}{2005}).

\bibitem{schmid2023rechargeable}
\bibinfo{author}{Schmid, A.}, \bibinfo{author}{Krammer, M.} \&
  \bibinfo{author}{Fleig, J.}
\newblock \bibinfo{title}{Rechargeable oxide ion batteries based on mixed
  conducting oxide electrodes}.
\newblock \emph{\bibinfo{journal}{Advanced Energy Materials}}
  \textbf{\bibinfo{volume}{13}}, \bibinfo{pages}{2203789}
  (\bibinfo{year}{2023}).

\bibitem{gu2023surface}
\bibinfo{author}{Gu, J.}, \bibinfo{author}{Jiang, L.}, \bibinfo{author}{Ismail,
  S.~A.}, \bibinfo{author}{Guo, H.} \& \bibinfo{author}{Han, D.}
\newblock \bibinfo{title}{Surface protonic conduction on oxide ceramics:
  Mechanism, materials, and method for characterization}.
\newblock \emph{\bibinfo{journal}{Advanced Materials Interfaces}}
  \textbf{\bibinfo{volume}{10}}, \bibinfo{pages}{2201764}
  (\bibinfo{year}{2023}).

\bibitem{norby2004hydrogen}
\bibinfo{author}{Norby, T.}, \bibinfo{author}{Wider{\o}e, M.},
  \bibinfo{author}{Gl{\"o}ckner, R.} \& \bibinfo{author}{Larring, Y.}
\newblock \bibinfo{title}{Hydrogen in oxides}.
\newblock \emph{\bibinfo{journal}{Dalton transactions}}
  \bibinfo{pages}{3012--3018} (\bibinfo{year}{2004}).

\bibitem{meng2019recent}
\bibinfo{author}{Meng, Y.} \emph{et~al.}
\newblock \bibinfo{title}{recent progress in low-temperature proton-conducting
  ceramics}.
\newblock \emph{\bibinfo{journal}{Journal of materials science}}
  \textbf{\bibinfo{volume}{54}}, \bibinfo{pages}{9291--9312}
  (\bibinfo{year}{2019}).

\bibitem{fabbri2010materials}
\bibinfo{author}{Fabbri, E.}, \bibinfo{author}{Pergolesi, D.} \&
  \bibinfo{author}{Traversa, E.}
\newblock \bibinfo{title}{Materials challenges toward proton-conducting oxide
  fuel cells: a critical review}.
\newblock \emph{\bibinfo{journal}{Chemical Society Reviews}}
  \textbf{\bibinfo{volume}{39}}, \bibinfo{pages}{4355--4369}
  (\bibinfo{year}{2010}).

\bibitem{joo2006electrical}
\bibinfo{author}{Joo, J.~H.} \& \bibinfo{author}{Choi, G.~M.}
\newblock \bibinfo{title}{Electrical conductivity of ysz film grown by pulsed
  laser deposition}.
\newblock \emph{\bibinfo{journal}{Solid State Ionics}}
  \textbf{\bibinfo{volume}{177}}, \bibinfo{pages}{1053--1057}
  (\bibinfo{year}{2006}).

\bibitem{chiodelli2005synthesis}
\bibinfo{author}{Chiodelli, G.}, \bibinfo{author}{Malavasi, L.},
  \bibinfo{author}{Massarotti, V.}, \bibinfo{author}{Mustarelli, P.} \&
  \bibinfo{author}{Quartarone, E.}
\newblock \bibinfo{title}{Synthesis and characterization of ce0. 8gd0. 2o2- y
  polycrystalline and thin film materials}.
\newblock \emph{\bibinfo{journal}{Solid State Ionics}}
  \textbf{\bibinfo{volume}{176}}, \bibinfo{pages}{1505--1512}
  (\bibinfo{year}{2005}).

\bibitem{nishitani2012three}
\bibinfo{author}{Nishitani, Y.}, \bibinfo{author}{Kaneko, Y.},
  \bibinfo{author}{Ueda, M.}, \bibinfo{author}{Morie, T.} \&
  \bibinfo{author}{Fujii, E.}
\newblock \bibinfo{title}{Three-terminal ferroelectric synapse device with
  concurrent learning function for artificial neural networks}.
\newblock \emph{\bibinfo{journal}{Journal of Applied Physics}}
  \textbf{\bibinfo{volume}{111}} (\bibinfo{year}{2012}).

\bibitem{yao2020protonic}
\bibinfo{author}{Yao, X.} \emph{et~al.}
\newblock \bibinfo{title}{Protonic solid-state electrochemical synapse for
  physical neural networks}.
\newblock \emph{\bibinfo{journal}{Nature communications}}
  \textbf{\bibinfo{volume}{11}}, \bibinfo{pages}{3134} (\bibinfo{year}{2020}).

\bibitem{kwak2021experimental}
\bibinfo{author}{Kwak, H.}, \bibinfo{author}{Lee, C.}, \bibinfo{author}{Lee,
  C.}, \bibinfo{author}{Noh, K.} \& \bibinfo{author}{Kim, S.}
\newblock \bibinfo{title}{Experimental measurement of ungated channel region
  conductance in a multi-terminal, metal oxide-based ecram}.
\newblock \emph{\bibinfo{journal}{Semiconductor Science and Technology}}
  \textbf{\bibinfo{volume}{36}}, \bibinfo{pages}{114002}
  (\bibinfo{year}{2021}).

\bibitem{li2020filament}
\bibinfo{author}{Li, Y.} \emph{et~al.}
\newblock \bibinfo{title}{Filament-free bulk resistive memory enables
  deterministic analogue switching}.
\newblock \emph{\bibinfo{journal}{Advanced Materials}}
  \textbf{\bibinfo{volume}{32}}, \bibinfo{pages}{2003984}
  (\bibinfo{year}{2020}).

\bibitem{kim2023nonvolatile}
\bibinfo{author}{Kim, D.~S.} \emph{et~al.}
\newblock \bibinfo{title}{Nonvolatile electrochemical random-access memory
  under short circuit}.
\newblock \emph{\bibinfo{journal}{Advanced Electronic Materials}}
  \textbf{\bibinfo{volume}{9}}, \bibinfo{pages}{2200958}
  (\bibinfo{year}{2023}).

\bibitem{bisri2017endeavor}
\bibinfo{author}{Bisri, S.~Z.}, \bibinfo{author}{Shimizu, S.},
  \bibinfo{author}{Nakano, M.} \& \bibinfo{author}{Iwasa, Y.}
\newblock \bibinfo{title}{Endeavor of iontronics: from fundamentals to
  applications of ion-controlled electronics}.
\newblock \emph{\bibinfo{journal}{Advanced Materials}}
  \textbf{\bibinfo{volume}{29}}, \bibinfo{pages}{1607054}
  (\bibinfo{year}{2017}).

\bibitem{lynch2004long}
\bibinfo{author}{Lynch, M.~A.}
\newblock \bibinfo{title}{Long-term potentiation and memory}.
\newblock \emph{\bibinfo{journal}{Physiological reviews}}
  \textbf{\bibinfo{volume}{84}}, \bibinfo{pages}{87--136}
  (\bibinfo{year}{2004}).

\bibitem{merolla2014million}
\bibinfo{author}{Merolla, P.~A.} \emph{et~al.}
\newblock \bibinfo{title}{A million spiking-neuron integrated circuit with a
  scalable communication network and interface}.
\newblock \emph{\bibinfo{journal}{Science}} \textbf{\bibinfo{volume}{345}},
  \bibinfo{pages}{668--673} (\bibinfo{year}{2014}).

\bibitem{seo2020recent}
\bibinfo{author}{Seo, S.} \emph{et~al.}
\newblock \bibinfo{title}{Recent progress in artificial synapses based on
  two-dimensional van der waals materials for brain-inspired computing}.
\newblock \emph{\bibinfo{journal}{ACS Applied Electronic Materials}}
  \textbf{\bibinfo{volume}{2}}, \bibinfo{pages}{371--388}
  (\bibinfo{year}{2020}).

\bibitem{jang2015optimization}
\bibinfo{author}{Jang, J.-W.}, \bibinfo{author}{Park, S.},
  \bibinfo{author}{Burr, G.~W.}, \bibinfo{author}{Hwang, H.} \&
  \bibinfo{author}{Jeong, Y.-H.}
\newblock \bibinfo{title}{Optimization of conductance change in pr 1--x ca x
  mno 3-based synaptic devices for neuromorphic systems}.
\newblock \emph{\bibinfo{journal}{IEEE Electron Device Letters}}
  \textbf{\bibinfo{volume}{36}}, \bibinfo{pages}{457--459}
  (\bibinfo{year}{2015}).

\bibitem{krogh1991simple}
\bibinfo{author}{Krogh, A.} \& \bibinfo{author}{Hertz, J.}
\newblock \bibinfo{title}{A simple weight decay can improve generalization}.
\newblock \emph{\bibinfo{journal}{Advances in neural information processing
  systems}} \textbf{\bibinfo{volume}{4}} (\bibinfo{year}{1991}).

\bibitem{buonomano2000decoding}
\bibinfo{author}{Buonomano, D.~V.}
\newblock \bibinfo{title}{Decoding temporal information: a model based on
  short-term synaptic plasticity}.
\newblock \emph{\bibinfo{journal}{Journal of Neuroscience}}
  \textbf{\bibinfo{volume}{20}}, \bibinfo{pages}{1129--1141}
  (\bibinfo{year}{2000}).

\bibitem{kim2013carbon}
\bibinfo{author}{Kim, K.}, \bibinfo{author}{Chen, C.-L.},
  \bibinfo{author}{Truong, Q.}, \bibinfo{author}{Shen, A.~M.} \&
  \bibinfo{author}{Chen, Y.}
\newblock \bibinfo{title}{A carbon nanotube synapse with dynamic logic and
  learning}.
\newblock \emph{\bibinfo{journal}{Adv. Mater}} \textbf{\bibinfo{volume}{25}},
  \bibinfo{pages}{1693--1698} (\bibinfo{year}{2013}).

\bibitem{regehr2012short}
\bibinfo{author}{Regehr, W.~G.}
\newblock \bibinfo{title}{Short-term presynaptic plasticity}.
\newblock \emph{\bibinfo{journal}{Cold Spring Harbor perspectives in biology}}
  \textbf{\bibinfo{volume}{4}}, \bibinfo{pages}{a005702}
  (\bibinfo{year}{2012}).

\bibitem{guo2015paired}
\bibinfo{author}{Guo, L.~Q.}, \bibinfo{author}{Zhu, L.~Q.},
  \bibinfo{author}{Ding, J.~N.} \& \bibinfo{author}{Huang, Y.~K.}
\newblock \bibinfo{title}{Paired-pulse facilitation achieved in
  protonic/electronic hybrid indium gallium zinc oxide synaptic transistors}.
\newblock \emph{\bibinfo{journal}{AIP Advances}} \textbf{\bibinfo{volume}{5}}
  (\bibinfo{year}{2015}).

\bibitem{paszke2019pytorch}
\bibinfo{author}{Paszke, A.} \emph{et~al.}
\newblock \bibinfo{title}{Pytorch: An imperative style, high-performance deep
  learning library}.
\newblock \emph{\bibinfo{journal}{Advances in neural information processing
  systems}} \textbf{\bibinfo{volume}{32}} (\bibinfo{year}{2019}).

\bibitem{scikit-learn}
\bibinfo{author}{Pedregosa, F.} \emph{et~al.}
\newblock \bibinfo{title}{Scikit-learn: Machine learning in {P}ython}.
\newblock \emph{\bibinfo{journal}{Journal of Machine Learning Research}}
  \textbf{\bibinfo{volume}{12}}, \bibinfo{pages}{2825--2830}
  (\bibinfo{year}{2011}).

\end{thebibliography}


\begin{thebibliography}{10}
\expandafter\ifx\csname url\endcsname\relax
  \def\url#1{\burl{#1}}\fi
\expandafter\ifx\csname urlprefix\endcsname\relax\def\urlprefix{URL }\fi
\providecommand{\bibinfo}[2]{#2}
\providecommand{\eprint}[2][]{\url{#2}}
\providecommand{\doi}[1]{\url{https://doi.org/#1}}
\bibcommenthead

\bibitem{singh2017low}
\bibinfo{author}{Singh, B.}, \bibinfo{author}{Ghosh, S.},
  \bibinfo{author}{Aich, S.} \& \bibinfo{author}{Roy, B.}
\newblock \bibinfo{title}{Low temperature solid oxide electrolytes (lt-soe): A
  review}.
\newblock \emph{\bibinfo{journal}{Journal of Power Sources}}
  \textbf{\bibinfo{volume}{339}}, \bibinfo{pages}{103--135}
  (\bibinfo{year}{2017}).

\bibitem{tarancon2009strategies}
\bibinfo{author}{Taranc{\'o}n, A.}
\newblock \bibinfo{title}{Strategies for lowering solid oxide fuel cells
  operating temperature}.
\newblock \emph{\bibinfo{journal}{Energies}} \textbf{\bibinfo{volume}{2}},
  \bibinfo{pages}{1130--1150} (\bibinfo{year}{2009}).

\bibitem{azad1992solid}
\bibinfo{author}{Azad, A.}, \bibinfo{author}{Akbar, S.},
  \bibinfo{author}{Mhaisalkar, S.}, \bibinfo{author}{Birkefeld, L.} \&
  \bibinfo{author}{Goto, K.}
\newblock \bibinfo{title}{Solid-state gas sensors: A review}.
\newblock \emph{\bibinfo{journal}{Journal of the Electrochemical Society}}
  \textbf{\bibinfo{volume}{139}}, \bibinfo{pages}{3690} (\bibinfo{year}{1992}).

\bibitem{scherrer2013proton}
\bibinfo{author}{Scherrer, B.} \emph{et~al.}
\newblock \bibinfo{title}{On proton conductivity in porous and dense yttria
  stabilized zirconia at low temperature}.
\newblock \emph{\bibinfo{journal}{Advanced Functional Materials}}
  \textbf{\bibinfo{volume}{23}}, \bibinfo{pages}{1957--1964}
  (\bibinfo{year}{2013}).

\bibitem{perez2010electrical}
\bibinfo{author}{P{\'e}rez-Coll, D.} \& \bibinfo{author}{Mather, G.~C.}
\newblock \bibinfo{title}{Electrical transport at low temperatures in dense
  nanocrystalline gd-doped ceria}.
\newblock \emph{\bibinfo{journal}{Solid State Ionics}}
  \textbf{\bibinfo{volume}{181}}, \bibinfo{pages}{20--26}
  (\bibinfo{year}{2010}).

\bibitem{meng2019recent}
\bibinfo{author}{Meng, Y.} \emph{et~al.}
\newblock \bibinfo{title}{recent progress in low-temperature proton-conducting
  ceramics}.
\newblock \emph{\bibinfo{journal}{Journal of materials science}}
  \textbf{\bibinfo{volume}{54}}, \bibinfo{pages}{9291--9312}
  (\bibinfo{year}{2019}).

\bibitem{lubben2018processes}
\bibinfo{author}{L{\"u}bben, M.}, \bibinfo{author}{Wiefels, S.},
  \bibinfo{author}{Waser, R.} \& \bibinfo{author}{Valov, I.}
\newblock \bibinfo{title}{Processes and effects of oxygen and moisture in
  resistively switching taox and hfox}.
\newblock \emph{\bibinfo{journal}{Advanced electronic materials}}
  \textbf{\bibinfo{volume}{4}}, \bibinfo{pages}{1700458}
  (\bibinfo{year}{2018}).

\bibitem{takayanagi2017thickness}
\bibinfo{author}{Takayanagi, M.} \emph{et~al.}
\newblock \bibinfo{title}{Thickness-dependent surface proton conduction in
  (111) oriented yttria-stabilized zirconia thin film}.
\newblock \emph{\bibinfo{journal}{Solid State Ionics}}
  \textbf{\bibinfo{volume}{311}}, \bibinfo{pages}{46--51}
  (\bibinfo{year}{2017}).

\bibitem{dittmann2021nanoionic}
\bibinfo{author}{Dittmann, R.}, \bibinfo{author}{Menzel, S.} \&
  \bibinfo{author}{Waser, R.}
\newblock \bibinfo{title}{Nanoionic memristive phenomena in metal oxides: the
  valence change mechanism}.
\newblock \emph{\bibinfo{journal}{Advances in Physics}}
  \textbf{\bibinfo{volume}{70}}, \bibinfo{pages}{155--349}
  (\bibinfo{year}{2021}).

\bibitem{li2020filament}
\bibinfo{author}{Li, Y.} \emph{et~al.}
\newblock \bibinfo{title}{Filament-free bulk resistive memory enables
  deterministic analogue switching}.
\newblock \emph{\bibinfo{journal}{Advanced Materials}}
  \textbf{\bibinfo{volume}{32}}, \bibinfo{pages}{2003984}
  (\bibinfo{year}{2020}).

\bibitem{nikam2021all}
\bibinfo{author}{Nikam, R.~D.}, \bibinfo{author}{Kwak, M.} \&
  \bibinfo{author}{Hwang, H.}
\newblock \bibinfo{title}{All-solid-state oxygen ion electrochemical
  random-access memory for neuromorphic computing}.
\newblock \emph{\bibinfo{journal}{Advanced Electronic Materials}}
  \textbf{\bibinfo{volume}{7}}, \bibinfo{pages}{2100142}
  (\bibinfo{year}{2021}).

\bibitem{huang2023electrochemical}
\bibinfo{author}{Huang, M.} \emph{et~al.}
\newblock \bibinfo{title}{Electrochemical ionic synapses: progress and
  perspectives}.
\newblock \emph{\bibinfo{journal}{Advanced Materials}}
  \textbf{\bibinfo{volume}{35}}, \bibinfo{pages}{2205169}
  (\bibinfo{year}{2023}).

\bibitem{kim2019metal}
\bibinfo{author}{Kim, S.} \emph{et~al.}
\newblock \bibinfo{title}{Metal-oxide based, cmos-compatible ecram for deep
  learning accelerator}.
\newblock \emph{\bibinfo{journal}{2019 IEEE International Electron Devices
  Meeting (IEDM)}}  (\bibinfo{year}{2019}).

\bibitem{lee2022strategies}
\bibinfo{author}{Lee, J.}, \bibinfo{author}{Nikam, R.~D.},
  \bibinfo{author}{Kwak, M.} \& \bibinfo{author}{Hwang, H.}
\newblock \bibinfo{title}{Strategies to improve the synaptic characteristics of
  oxygen-based electrochemical random-access memory based on material
  parameters optimization}.
\newblock \emph{\bibinfo{journal}{ACS Applied Materials \& Interfaces}}
  \textbf{\bibinfo{volume}{14}}, \bibinfo{pages}{13450--13457}
  (\bibinfo{year}{2022}).

\bibitem{lee2020pr}
\bibinfo{author}{Lee, C.} \emph{et~al.}
\newblock \bibinfo{title}{Pr 0.7 ca 0.3 mno 3-based three-terminal synapse for
  neuromorphic computing}.
\newblock \emph{\bibinfo{journal}{IEEE Electron Device Letters}}
  \textbf{\bibinfo{volume}{41}}, \bibinfo{pages}{1500--1503}
  (\bibinfo{year}{2020}).

\bibitem{kwak2021experimental}
\bibinfo{author}{Kwak, H.}, \bibinfo{author}{Lee, C.}, \bibinfo{author}{Lee,
  C.}, \bibinfo{author}{Noh, K.} \& \bibinfo{author}{Kim, S.}
\newblock \bibinfo{title}{Experimental measurement of ungated channel region
  conductance in a multi-terminal, metal oxide-based ecram}.
\newblock \emph{\bibinfo{journal}{Semiconductor Science and Technology}}
  \textbf{\bibinfo{volume}{36}}, \bibinfo{pages}{114002}
  (\bibinfo{year}{2021}).

\bibitem{kim2023nonvolatile}
\bibinfo{author}{Kim, D.~S.} \emph{et~al.}
\newblock \bibinfo{title}{Nonvolatile electrochemical random-access memory
  under short circuit}.
\newblock \emph{\bibinfo{journal}{Advanced Electronic Materials}}
  \textbf{\bibinfo{volume}{9}}, \bibinfo{pages}{2200958}
  (\bibinfo{year}{2023}).

\bibitem{agarwal2016resistive}
\bibinfo{author}{Agarwal, S.} \emph{et~al.}
\newblock \bibinfo{title}{Resistive memory device requirements for a neural
  algorithm accelerator}.
\newblock \emph{\bibinfo{journal}{2016 International Joint Conference on Neural
  Networks (IJCNN)}} \bibinfo{pages}{929--938} (\bibinfo{year}{2016}).

\bibitem{nikam2019near}
\bibinfo{author}{Nikam, R.~D.} \emph{et~al.}
\newblock \bibinfo{title}{Near ideal synaptic functionalities in li ion
  synaptic transistor using li3poxsex electrolyte with high ionic
  conductivity}.
\newblock \emph{\bibinfo{journal}{Scientific reports}}
  \textbf{\bibinfo{volume}{9}}, \bibinfo{pages}{18883} (\bibinfo{year}{2019}).

\bibitem{schwacke2024electrochemical}
\bibinfo{author}{Schwacke, M.}, \bibinfo{author}{{\v{Z}}guns, P.},
  \bibinfo{author}{del Alamo, J.}, \bibinfo{author}{Li, J.} \&
  \bibinfo{author}{Yildiz, B.}
\newblock \bibinfo{title}{Electrochemical ionic synapses with mg2+ as the
  working ion}.
\newblock \emph{\bibinfo{journal}{Advanced Electronic Materials}}
  \bibinfo{pages}{2300577} (\bibinfo{year}{2024}).

\bibitem{onen2021cmos}
\bibinfo{author}{Onen, M.}, \bibinfo{author}{Emond, N.}, \bibinfo{author}{Li,
  J.}, \bibinfo{author}{Yildiz, B.} \& \bibinfo{author}{Del~Alamo, J.~A.}
\newblock \bibinfo{title}{Cmos-compatible protonic programmable resistor based
  on phosphosilicate glass electrolyte for analog deep learning}.
\newblock \emph{\bibinfo{journal}{Nano Letters}} \textbf{\bibinfo{volume}{21}},
  \bibinfo{pages}{6111--6116} (\bibinfo{year}{2021}).

\bibitem{cui2023cmos}
\bibinfo{author}{Cui, J.} \emph{et~al.}
\newblock \bibinfo{title}{Cmos-compatible electrochemical synaptic transistor
  arrays for deep learning accelerators}.
\newblock \emph{\bibinfo{journal}{Nature Electronics}}
  \textbf{\bibinfo{volume}{6}}, \bibinfo{pages}{292--300}
  (\bibinfo{year}{2023}).

\bibitem{yang2018artificial}
\bibinfo{author}{Yang, J.-T.} \emph{et~al.}
\newblock \bibinfo{title}{Artificial synapses emulated by an electrolyte-gated
  tungsten-oxide transistor}.
\newblock \emph{\bibinfo{journal}{Advanced Materials}}
  \textbf{\bibinfo{volume}{30}}, \bibinfo{pages}{1801548}
  (\bibinfo{year}{2018}).

\bibitem{tang2021pushing}
\bibinfo{author}{Tang, Y.} \emph{et~al.}
\newblock \bibinfo{title}{Pushing the study of point defects in thin film
  ferrites to low temperatures using in situ ellipsometry}.
\newblock \emph{\bibinfo{journal}{Advanced Materials Interfaces}}
  \textbf{\bibinfo{volume}{8}}, \bibinfo{pages}{2001881}
  (\bibinfo{year}{2021}).

\bibitem{schmid2018voltage}
\bibinfo{author}{Schmid, A.}, \bibinfo{author}{Rupp, G.~M.} \&
  \bibinfo{author}{Fleig, J.}
\newblock \bibinfo{title}{Voltage and partial pressure dependent defect
  chemistry in (la, sr) feo 3- $\delta$ thin films investigated by chemical
  capacitance measurements}.
\newblock \emph{\bibinfo{journal}{Physical Chemistry Chemical Physics}}
  \textbf{\bibinfo{volume}{20}}, \bibinfo{pages}{12016--12026}
  (\bibinfo{year}{2018}).

\bibitem{ishigaki1988diffusion}
\bibinfo{author}{Ishigaki, T.}, \bibinfo{author}{Yamauchi, S.},
  \bibinfo{author}{Kishio, K.}, \bibinfo{author}{Mizusaki, J.} \&
  \bibinfo{author}{Fueki, K.}
\newblock \bibinfo{title}{Diffusion of oxide ion vacancies in perovskite-type
  oxides}.
\newblock \emph{\bibinfo{journal}{Journal of Solid State Chemistry}}
  \textbf{\bibinfo{volume}{73}}, \bibinfo{pages}{179--187}
  (\bibinfo{year}{1988}).

\end{thebibliography}

\end{document}


\title[Solid-State Oxide-Ion Synaptic Transistor]{Solid-State Oxide-Ion Synaptic Transistor for Neuromorphic Computing. \linebreak \textbf{- Supplementary Information -}}


\author[1]{\fnm{Philipp} \sur{Langner}}

\author*[1]{\fnm{Francesco} \sur{Chiabrera}}\email{fchiabrera@irec.cat}

\author[1]{\fnm{Paul} \sur{Nizet}}

\author[1]{\fnm{Nerea} \sur{Alayo}}

\author[2]{\fnm{Luigi} \sur{Morrone}}

\author[1]{\fnm{Carlota} \sur{Bozal-Ginesta}}

\author[1]{\fnm{Alex} \sur{Morata}}

\author*[1,3]{\fnm{Albert} \sur{Taranc\`on}}\email{atarancon@irec.cat}

\affil[1]{\orgdiv{Department of Advanced Materials for Energy}, \orgname{Catalonia Institute for Energy Research (IREC)}, \orgaddress{\street{Jardins de les Dones de Negre 1, 2}, \city{Sant Adrià de Besós}, \postcode{08930}, \state{Barcelona}, \country{Spain}}}

\affil[2]{\orgdiv{Scientific Services NANOQUIM}, \orgname{Institut de Ci\`encia de Materials de Barcelona (CSIC-ICMAB)}, \orgaddress{\street{Campus UAB}, \city{Bellaterra}, \postcode{08193}, \state{Barcelona}, \country{Spain}}}

\affil[3]{\orgname{Catalan Institution for Research and Advanced Studies (ICREA)}, \orgaddress{\street{Passeig Lluis Companys 23}, \postcode{08010}, \city{Barcelona}, \country{Spain}}}

\maketitle

\newpage

\tableofcontents

\newpage


\subsection{Device Architecture and Operation Principle}
\label{subsec:architecture,principle}


\subsubsection[S1]{S1 XRD of BICUVOX}

\begin{figure}[h]
\centering
\includegraphics[width=\textwidth]{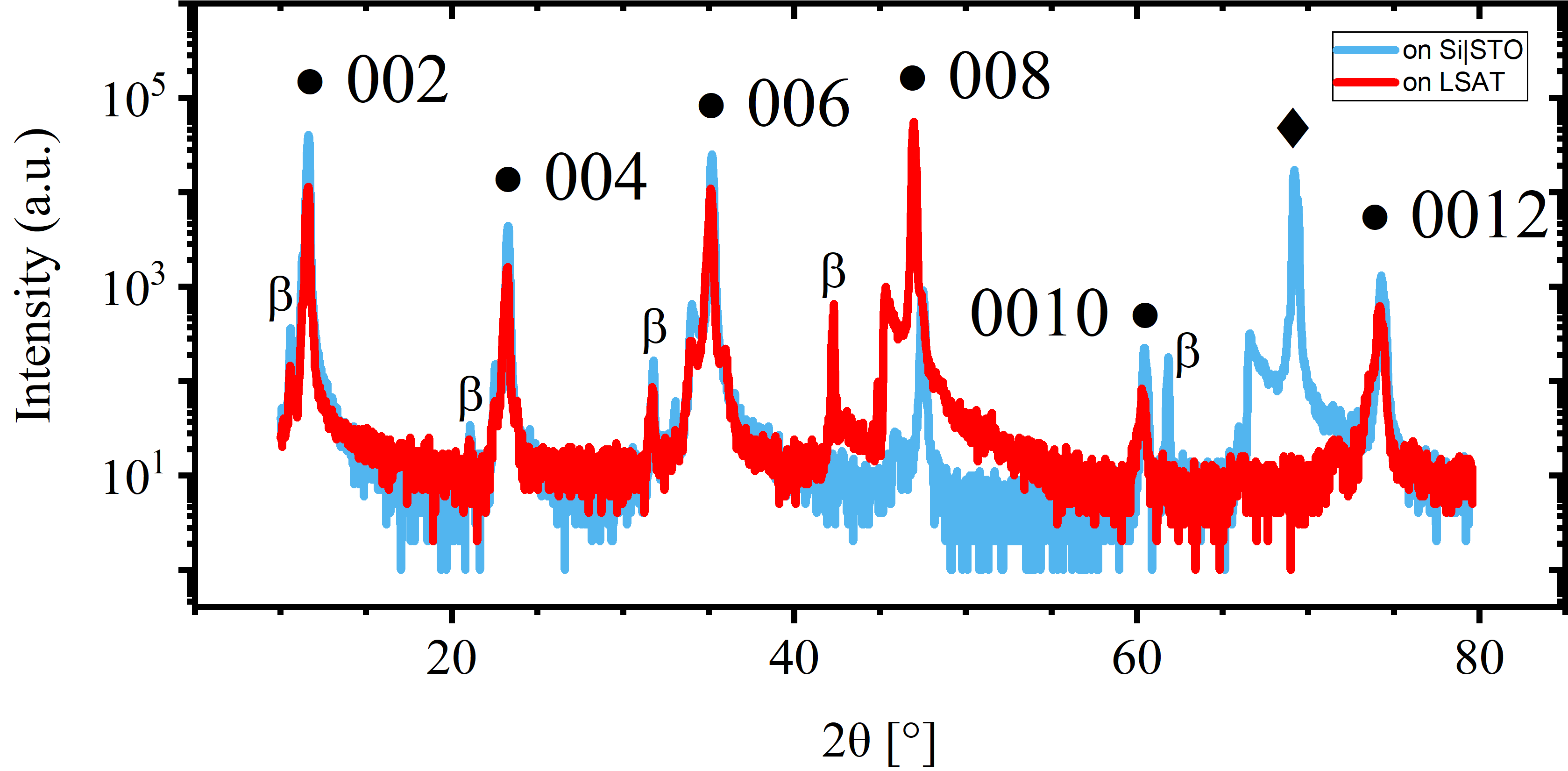}
\caption[S1]{
Scanning XRD of BICUVOX thin film on single crystal LSAT (red) and Si$\vert$STO-substrate (blue) before microfabrication of the synaptic transistors shows a highly epitaxial orientation in the [00l] plane conforming the great quality of produced thin films. Diamond-indicated peak is Si$_\text{100}$-substrate.
}
\label{fig: S1}
\end{figure}

\newpage

\subsubsection[S2]{S2 Microfabrication route}

\begin{figure}[h]
\centering
\includegraphics[width=\textwidth]{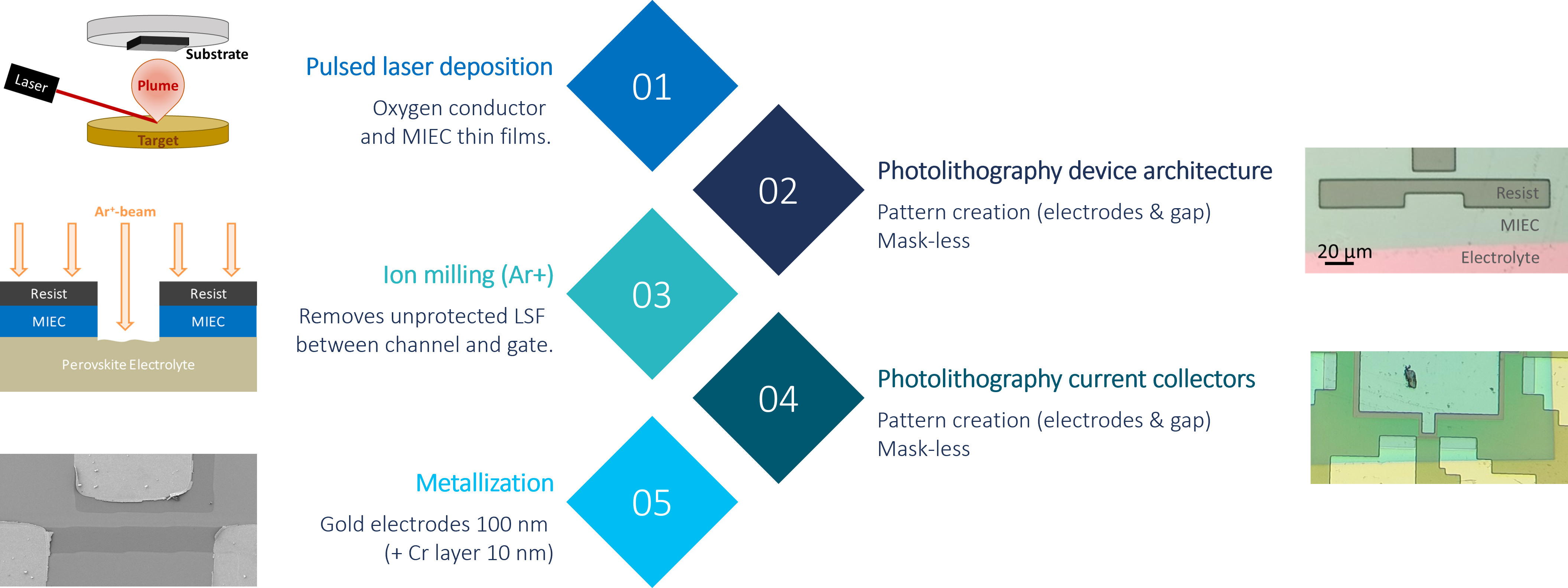}
\caption[S2]{
Synaptic transistor microfabrication route. See details in methods section.
}
\label{fig: S2}
\end{figure}

\newpage


\subsubsection[S3]{S3 Synaptic transistor architecture}

\begin{figure}[h]
\centering
\includegraphics[width=\textwidth]{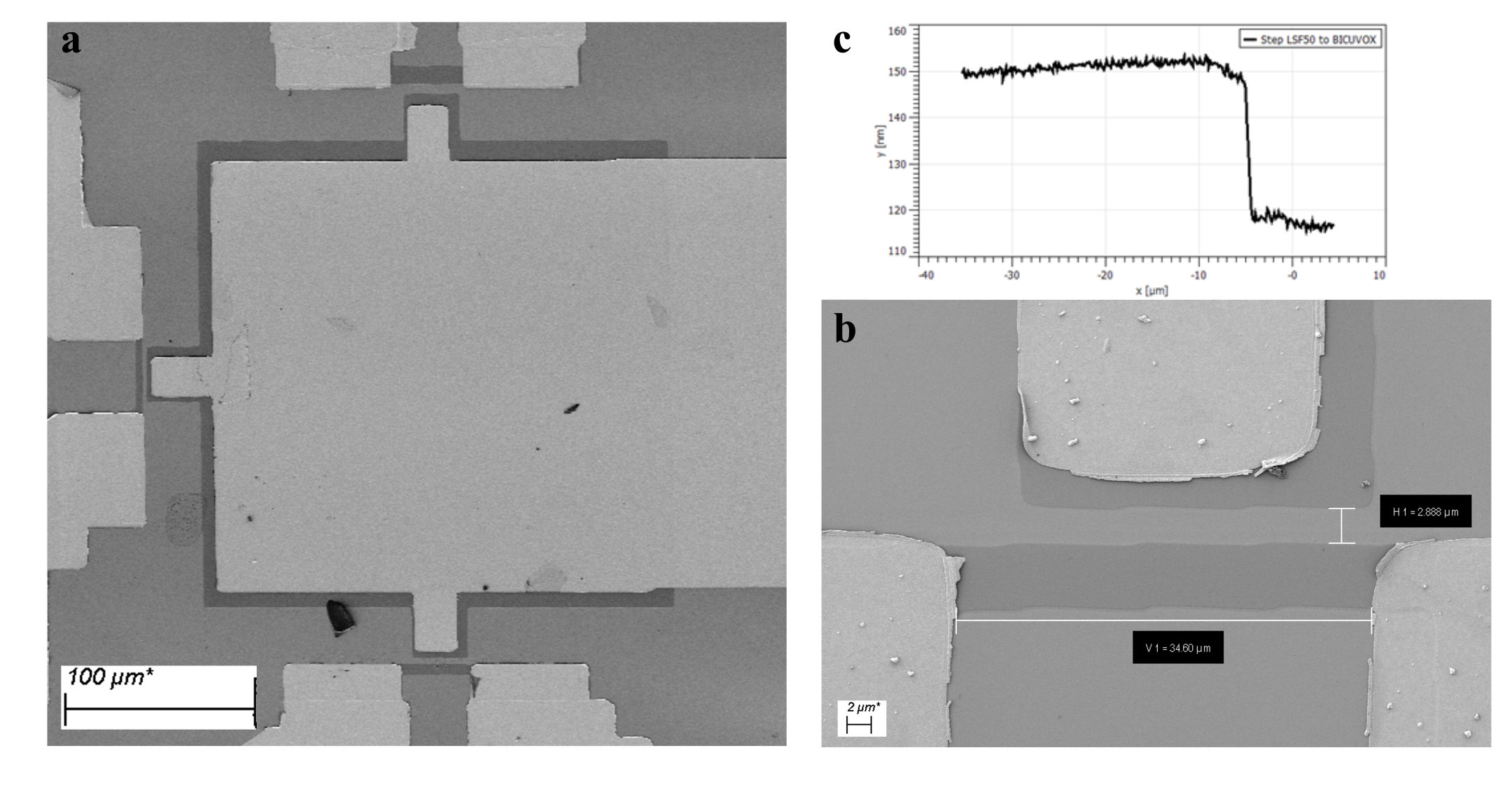}
\caption[S3]{
Device Architecture. 
\textbf{A)} Three microfabricated synaptic transistor devices on the same LSAT-BICUVOX-LSF50 substrate with \qtylist{2;3;5}{\micro\metre} separation between channel and reservoir respectively. 
\textbf{B)} SEM zoom of reported synaptic transistors (\qty{3}{\micro\metre} gap). 
\textbf{C)} calculation channel dimensions via thickness of deposited LSF50 thin film on BICUVOX before microfabrication (\qty{35}{\nano\metre} thickness). 
}
\label{fig: S3}
\end{figure}

\newpage


\subsubsection[S4]{S4 Atomic Force Microscopy}

\begin{figure}[h]
\centering
\includegraphics[width=\textwidth]{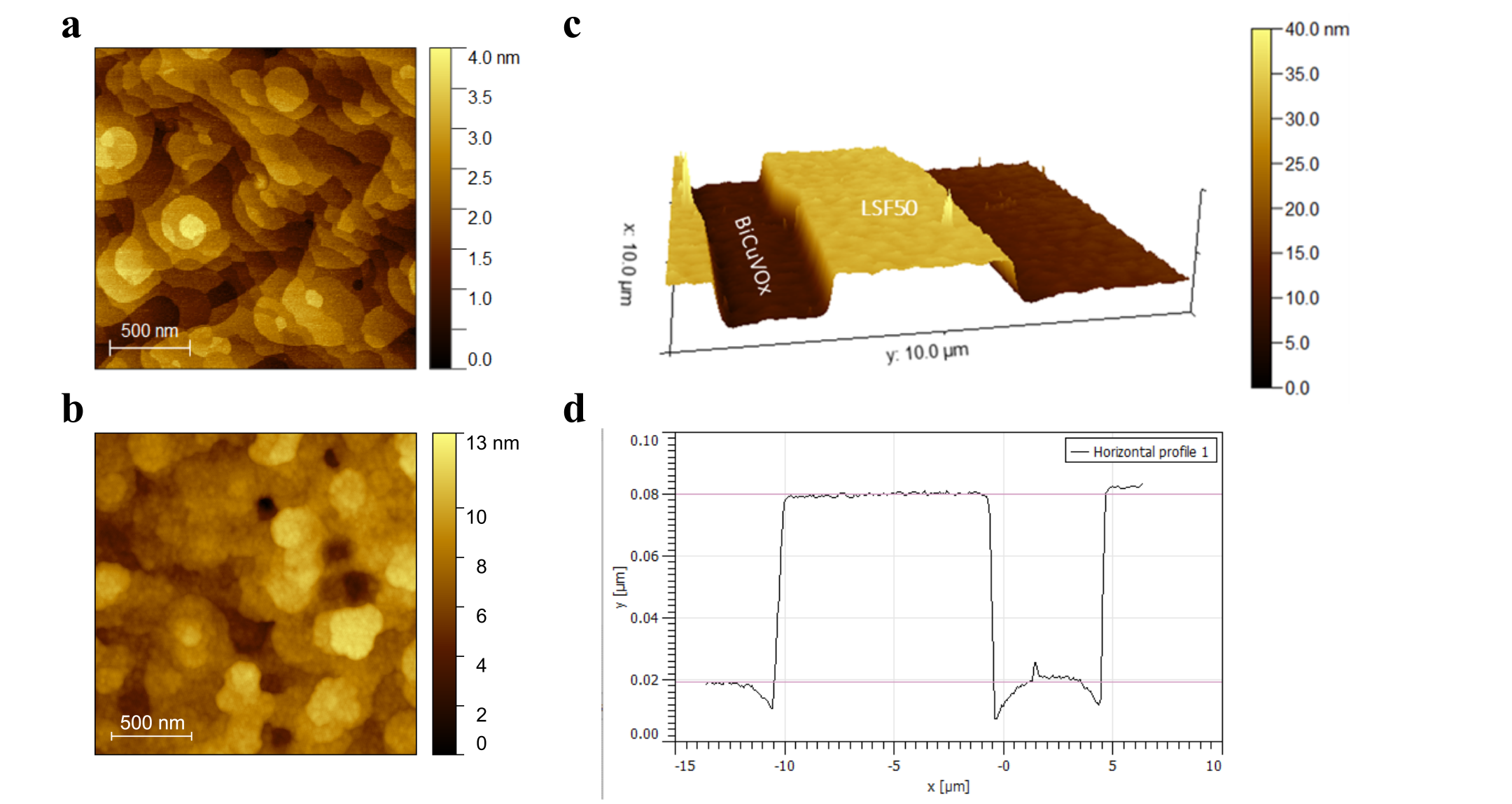}
\caption[S4]{
AFM images of layered BICUVOX surface before microfabrication confirm XRD results showing highly uniform and epitaxial orientation of the BICUVOX thin film. 
\textbf{a)} Layered, epitaxial BICUVOX surface. 
\textbf{b)} LSF50 surface showing great compatibility of LSF50 with epitaxial BICUVOX.   
\textbf{c)} 3D AFM image of channel and reservoir dimensions. 2px mean value filter was applied. 
\textbf{d)} AFM height profile of channel and reservoir after microfabrication. It is clear that during the ion-milling, the BICUVOX in the gap between channel and reservoir has been $\approx\qty{25}{\nano\metre}$ been etched into, leaving $\approx\qty{115}{\nano\metre}$ of BICUVOX thin film between channel and reservoir for our synaptic transistor.
}
\label{fig: S4}
\end{figure}

\newpage


\subsubsection[S5]{S5 Electrochemical impedance spectroscopy}

\begin{figure}[h]
\centering
\includegraphics[width=\textwidth]{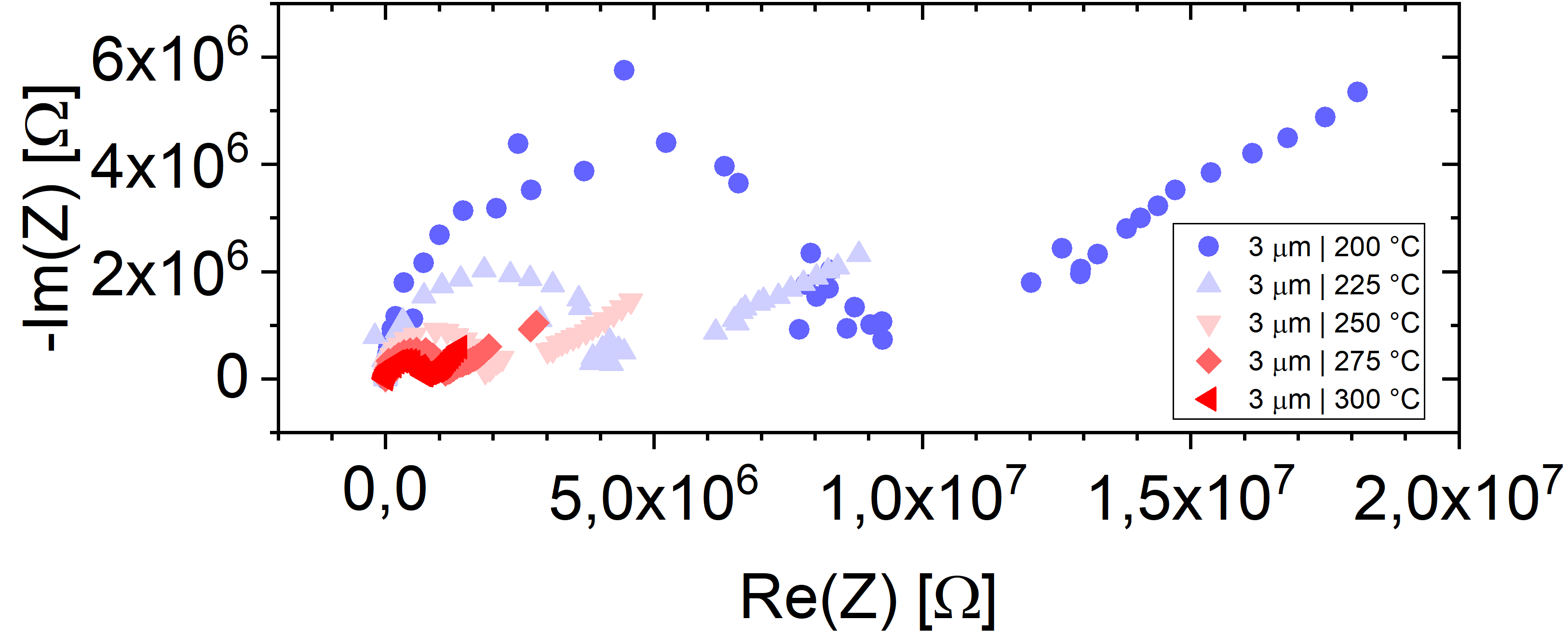}
\caption[S5]{
EIS of the BICUVOX thin film after microfabrication measured between channel and reservoir electrode referring to Figure 1b showing great ionic conduction assigned to oxide ions.
}
\label{fig: S5}
\end{figure}

\newpage


\subsubsection[S6]{S6 Discussion}

The search of high oxide ion ceramic conductors aims to lower the operating temperature of various oxide ion-driven solid-state devices, such as solid oxide fuel cells (SOFCs) or oxygen sensors\cite{singh2017low,tarancon2009strategies}. 
Promising electrolyte materials like yttria-stabilized zirconia (YSZ), Hafnia (HfO$_\text{2-$\delta$}$), or gadolinium-doped ceria (GDC) have been identified\cite{singh2017low,tarancon2009strategies}. 
By utilizing these materials in thin films the operation temperature of oxide ion-driven solid-state devices has been reduced to as low as $\qty{450}{\degreeCelsius}$\cite{azad1992solid}. 
However, recently reported oxide ion-based synaptic transistors utilizing these solid-state electrolyte materials have demonstrated operation below $\qty{450}{\degreeCelsius}$, reaching even room temperature (see table~TS\ref{tab: S6}). 
At these lower temperatures, YSZ, HfO$_\text{2-$\delta$}$, and GDC are also well known as proton conductors and resistive switching is highly moisture depending\cite{scherrer2013proton,perez2010electrical,meng2019recent,lubben2018processes}. 
This behavior arises from the significantly lower energy barrier for proton incorporation and diffusion compared to oxygen vacancies\cite{takayanagi2017thickness,scherrer2013proton}. 
Hence, the ionic conductivity of these materials decreases as temperature decreases down to $\qty{200}{\degreeCelsius}$, attributed to hindered oxide ion conductivity. 
Below $\qty{200}{\degreeCelsius}$, conductivity increases due to the growing influence of protonic conduction\cite{scherrer2013proton,meng2019recent}. 
The influence of protonic conduction cannot be excluded a priori and measurements as a function of the atmosphere are necessary to claim predominant oxide ion transport, especially at room temperature. 
Moreover, it is important to comment that a synaptic transistor relies on a pure ionic electrolyte to permit the redox insertion of ions into the channel. 
When using lightly or undoped materials such as HfO$_\text{2}$ or ZrO$_\text{2}$, redox reactions can also take place within the electrolyte, modifying its stoichiometry in a memristive fashion\cite{dittmann2021nanoionic}.
This behavior may be indeed effective to modulate the channel’s conductance but may suffers from disadvantages of resistive switching, such as stochasticity or non-linearity. 
On the contrary, a synaptic transistor based on a pure ionic conductor offers the advantage of a more reliable operation, since (considering a Coulombic efficiency equal to unit, \textit{i.e.} no electronic transport) each voltage pulse is expected to move the same ionic current. 
In the following, previously reported oxide ion based synaptic transistors from table~TS\ref{tab: S6} are briefly revised:  

\begin{itemize}
    
\item Li \textit{et al.} (2020) operated with both bulk and thin film YSZ ($\qty{400}{\nm}$), demonstrating synaptic plasticity with pulses down to $\qty{2}{\micro\second}$\cite{li2020filament}. 
Minimum temperature was \qty{160}{\degreeCelsius}, compatible with oxide-ion conductivity in YSZ.

\item Nikam \textit{et al.} (2021) reduced the electrolyte thickness to $\qty{20}{\nm}$ YSZ in their device, operating with \qty{10}{\milli\second} pulses at room temperature\cite{nikam2021all}. 
Yet, the atmosphere used was not reported, and considering the governing protonic surface conductivity in YSZ thin films below \qty{120}{\nm} thickness at low temperatures, it suggests that the operation of this device may not be governed by oxide ion diffusion\cite{takayanagi2017thickness}. 
Further discussion is provided in Huang \textit{et al.} (2022)\cite{huang2023electrochemical}. 

\item Kim \textit{et al.} (2019) utilized HfO$_\text{x}$ as an electrolyte, achieving nanosecond pulsing and demonstrating synaptic plasticity in N$_\text{2}$-atmosphere\cite{kim2019metal}. 
However, the device shows high asymmetry and nonlinearity with small dynamic range ($DR\approx 1$) and parasitic circuits, as previously discussed by Huang \textit{et al.} (2022)\cite{huang2023electrochemical}. 

\item Lee \textit{et al.} (2022) employed both zirconia and HfO$_\text{x}$ as electrolytes in their synaptic transistors at room temperature in a \qty{20}{\percent} humidity atmosphere\cite{lee2022strategies}. 
Considering measured activation energy $E_{ion} = \qty{0,29}{\eV}$ in thin film HfO$_\text{x}$, which is significantly lower than the energy barrier for creating oxygen vacancies ($\approx\qty{1}{\eV}$), surface protonics for these materials are indicated. 
This is also reported in the limitations of their study and is also applicable for synaptic transistor devices reported in Lee \textit{et al.} (2020) which are based on thin film HfO$_\text{x}$ electrolyte ($\qty{20}{\nm}$)\cite{lee2020pr}. 
Furthermore, the device is electrochemically not fully characterized (no electrochemical impedance spectroscopy reported). 

\item Kwak \textit{et al.} (2021) reports pulsing times down to $\qty{100}{\ns}$ at room temperature for their neuromorphic device based on HfO$_\text{x}$ electrolyte but doesn’t report endurance and retention characteristics\cite{kwak2021experimental}. 

\item Kim \textit{et al.} (2023) developed a YSZ-based cross-plane ECRAM device with great retention characteristics (up to 10 years at $\qty{85}{\degreeCelsius}$) but operates at $\qty{200}{\degreeCelsius}$ with $\qty{30}{\second}$ pulses ($\qty{2}{\V}$) in Ar-atmosphere due to the large thickness of the electrolyte\cite{kim2023nonvolatile}. 
\end{itemize}

Consequently, BICUVOX in our synaptic transistor presents a significant improvement in terms of oxide-ionic conductivity (Fig. 1b) compared to other oxygen based synaptic transistors in combination with potential CMOS-compatibility.
In order to fully exploit the potential of BICUVOX in our two-dimensional synaptic transistor compared to cross-plain YSZ synaptic transistors, simulations of temperature dependent pulsing times are given in S\ref{fig: S1911}. 

\setcounter{table}{5}
\begin{table}[h!]
\caption{
Overview of previously reported oxide-ion synaptic transistors.
}\label{tab: S6}
\begin{tabular}{@{}llll@{}}
\toprule
Reference                               & Electrolyte (thickness)                              & Pulses                 & Atmosphere          \\ \midrule
\cite{li2020filament}  & Bulk and thin film YSZ ($\qty{400}{\nm}$)            & $\qty{2}{\ms}$ - $\qty{2}{\micro\second}$   & $\qty{160}{\degreeCelsius}$; Ar-glove box  \\
\cite{kim2019metal} & YSZ (n.a.)                                           & $\qty{100}{\ns}$        & RT; n.a.             \\
\cite{nikam2021all}       & YSZ ($\qty{20}{\nm}$)                                & $\qty{10}{\ms}$         & RT; n.a.            \\
\cite{lee2022strategies}         & HfO$_\text{x}$ ($\qty{100}{\nm}$)                              & $\qty{100}{\ms}$        & RT; $\qty{20}{\percent}$ humidity    \\
\cite{lee2022strategies}         & ZrO$_\text{x}$ ($\qty{100}{\nm}$)                              & $\qty{100}{\ms}$        & RT; $\qty{20}{\percent}$ humidity    \\
\cite{kwak2021experimental}           & HfO$_\text{x}$ ($\qty{20}{\nm}$)                               & Down to $\qty{100}{\ns}$ & RT; atmosphere       \\
\cite{kim2023nonvolatile}                                & YSZ ($\qty{300}{\micro\metre}$)                      &  $\qty{30}{\second}$           & $\qty{200}{\degreeCelsius}$; Ar          \\ \bottomrule
\end{tabular}
\end{table}

\newpage


\subsubsection[S7]{S7 LSF50 channel operational window}

\setcounter{figure}{6}
\begin{figure}[h]
\centering
\includegraphics[width=0.6\textwidth]{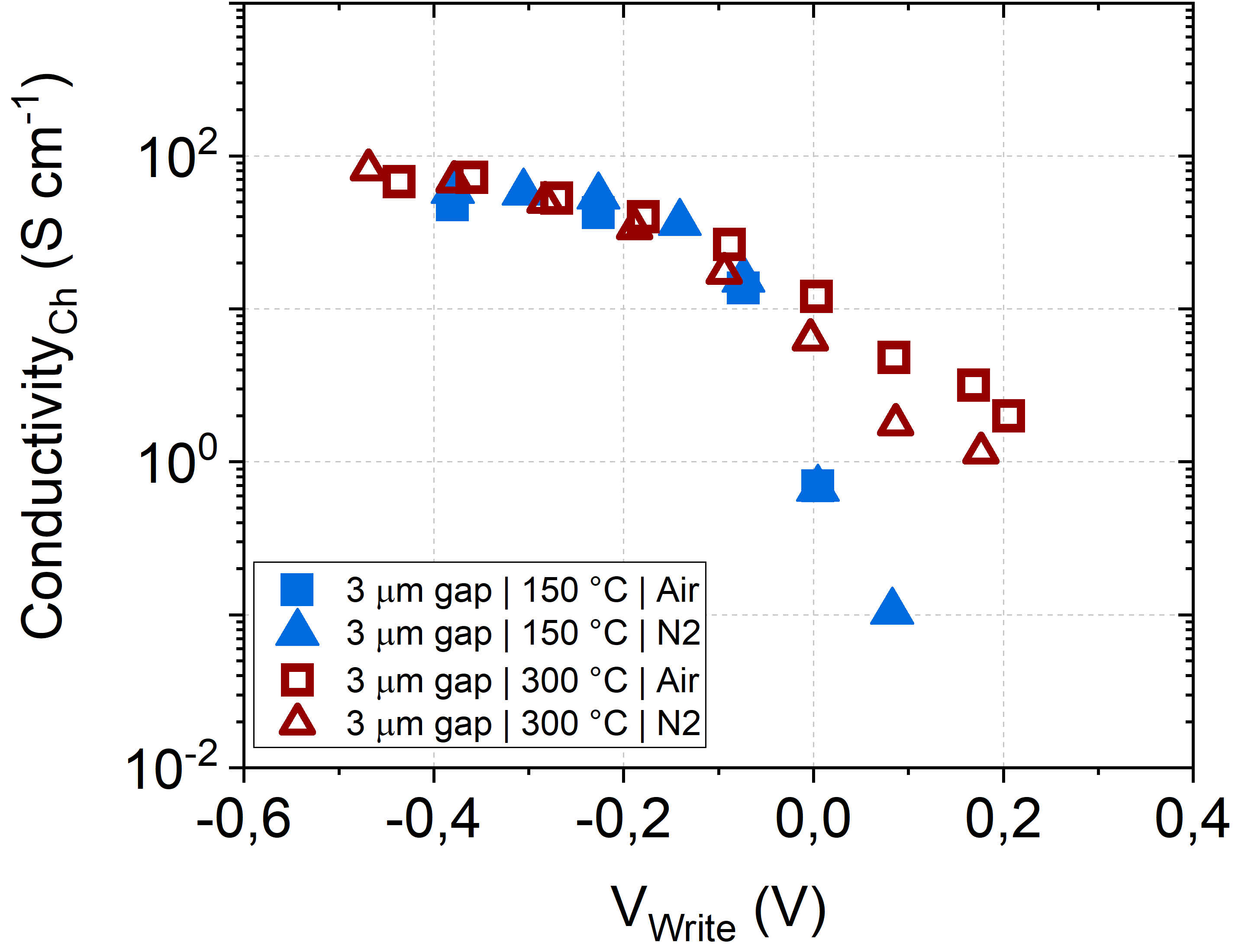}
\caption[S7]{
LSF50 channel operational window at $\qty{150}{\degreeCelsius}$ (blue) and $\qty{300}{\degreeCelsius}$ (red) in dry nitrogen atmosphere (triangle) and ambient air (squares) showing no significant differences between both atmospheres, supporting previous results excluding significant protonic conduction in our synaptic transistor device. 
}
\label{fig: S7}
\end{figure}

\newpage

\subsubsection[S8]{S8 Operation at \qty{300}{\degreeCelsius}}

\begin{figure}[h]
\centering
\includegraphics[width=\textwidth]{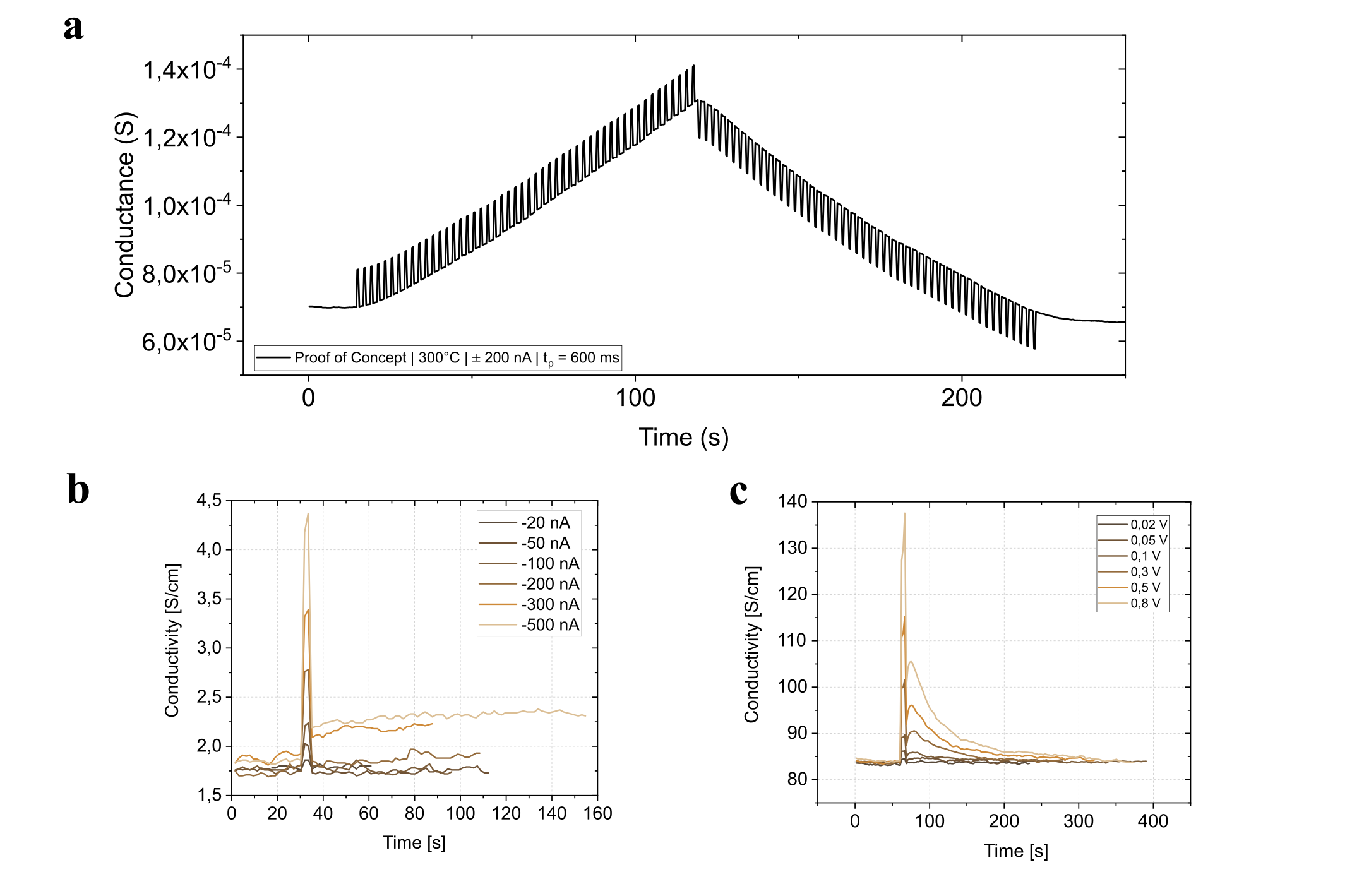}
\caption[S8]{
\textbf{a)} Channel Conductance of a synaptic transistor with dimensions $w_{Ch} = \qty{500}{\micro\metre}$, length $l_{Ch} = \qty{500}{\milli\metre}$ and channel-reservoir gap of \qty{200}{\micro\metre} and pulsing time $t_p = \qty{600}{\milli\second}$.
\textbf{b)} Current-controlled pulsing with $t_p = \qty{5}{\second}$ and varying pulse amplitude.
\textbf{b)} Voltage-controlled pulsing with $t_p = \qty{5}{\second}$ and varying pulse amplitude.
}
\label{fig: S8}
\end{figure}

\newpage


\subsubsection[S9]{S9 Cycling at \qty{100}{\degreeCelsius}}

\begin{figure}[h]
\centering
\includegraphics[width=0.6\textwidth]{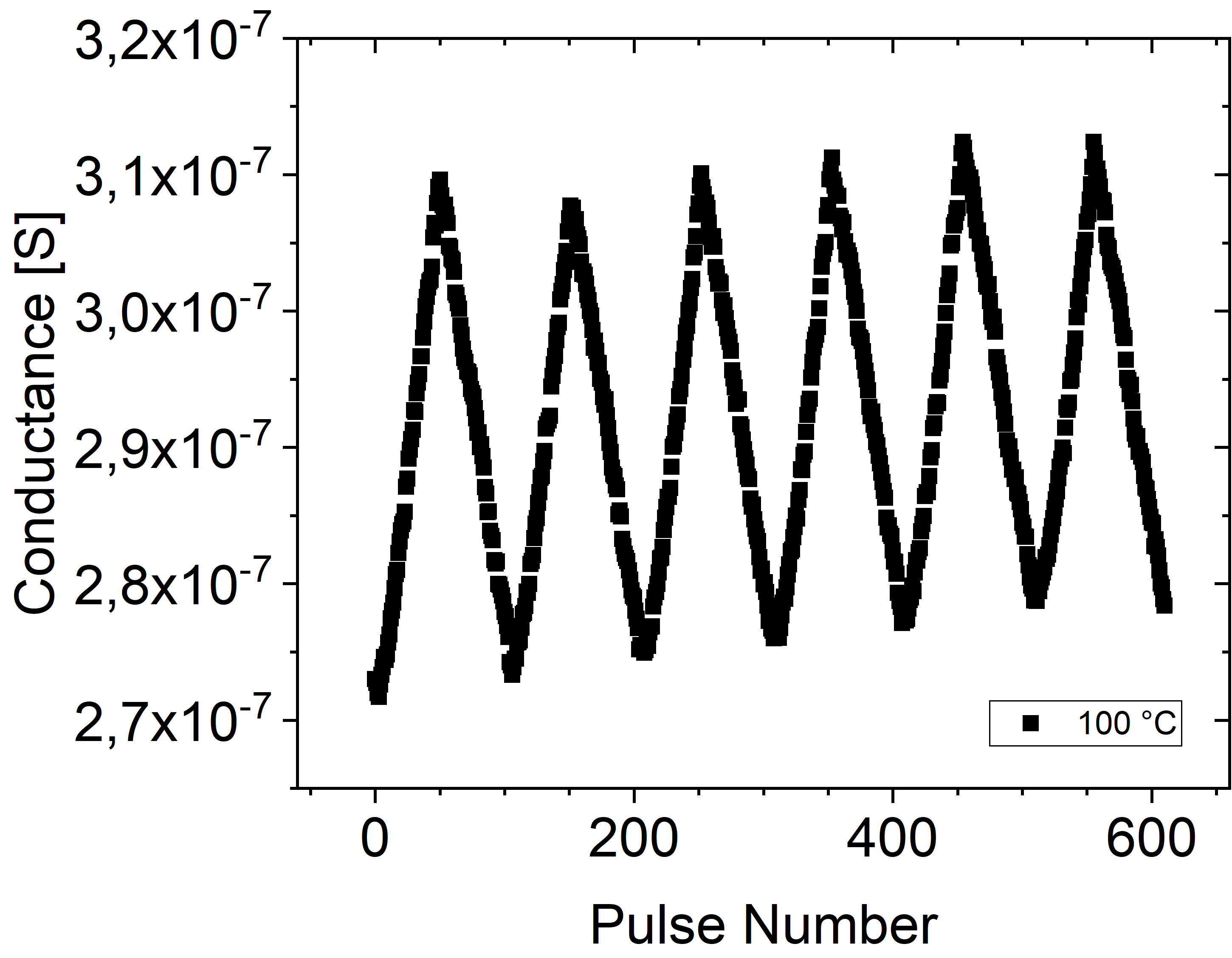}
\caption[S9]{
Channel Conductance of the reported synaptic transistor during cycling at $\qty{100}{\degreeCelsius}$ (50 pulses) with $t_p = \qty{220}{\milli\second}$ proofing operation at CMOS compatible temperatures.
}
\label{fig: S9}
\end{figure}

\newpage

\subsection{Synaptic Properties of the Synaptic Transistor}
\label{subsec:syn properties}



\subsubsection[S10]{S10 Pulse dependency of \qty{5}{\micro\metre}-gap device}

\begin{figure}[h]
\centering
\includegraphics[width=\textwidth]{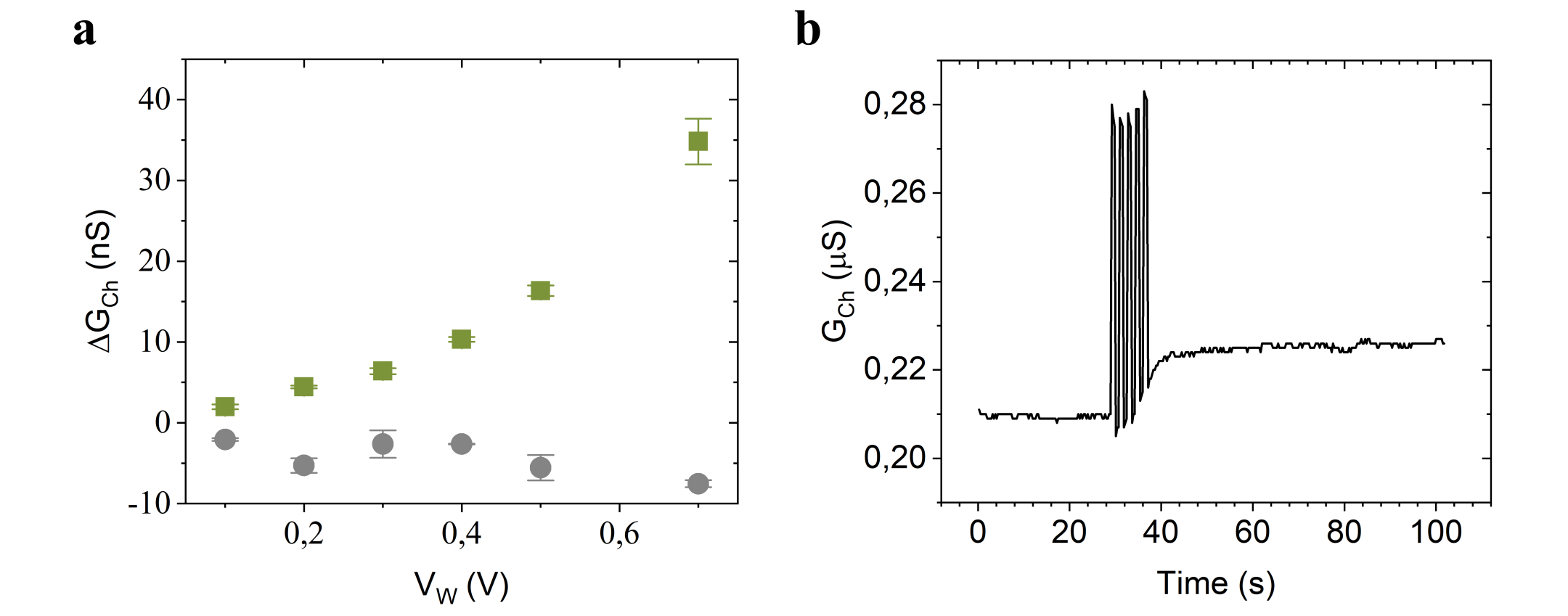}
\caption[S10]{
Pulse dependencies of a second device from same microfabricated chip (gap between channel and reservoir is $\qty{5}{\micro\metre}$.
\textbf{A)} voltage controlled dependency of conductivity changer per pulse on pulsing amplitude ($t_p = \qty{600}{\ms}$). 
\textbf{B)} voltage controlled retention characteristic after pulsing ($t_p = \qty{600}{\ms}$). 
}
\label{fig: S11}
\end{figure}

\newpage


\subsubsection[S11]{S11 $\Delta G$ dependency on $I_W$ and $t_p$ }

\begin{figure}[h]
\centering
\includegraphics[width=\textwidth]{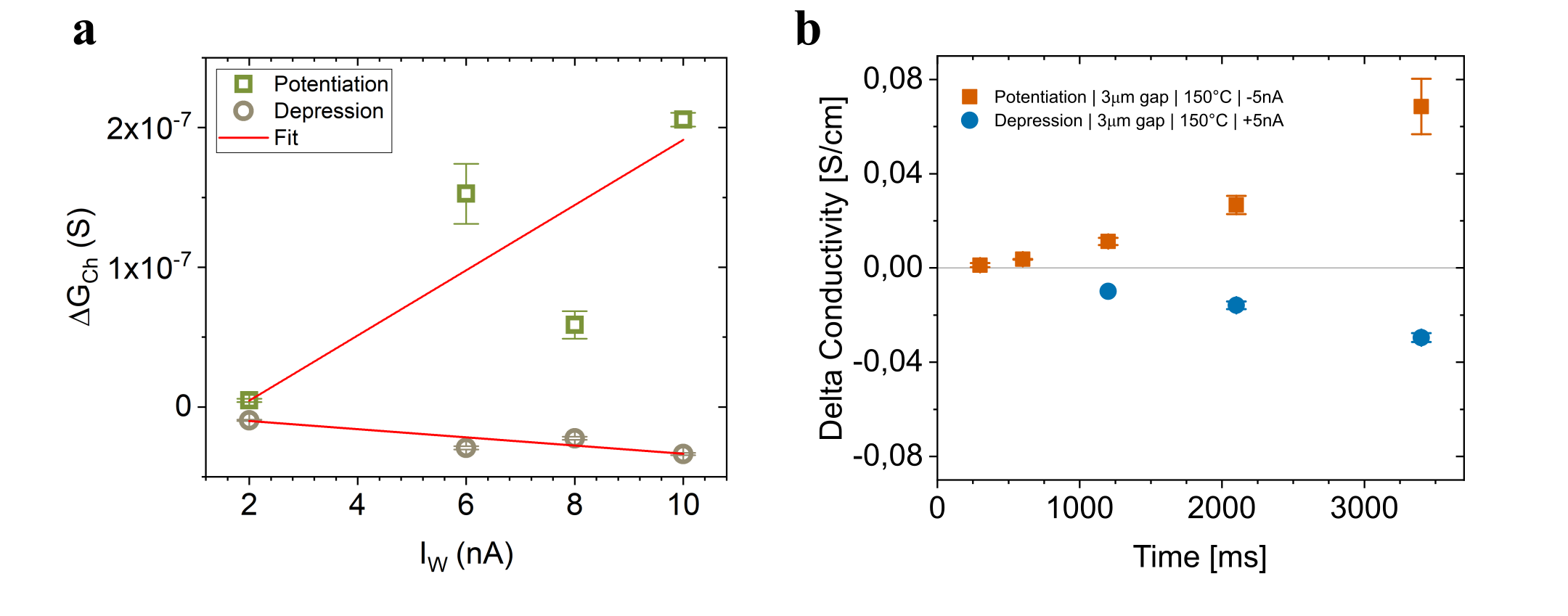}
\caption[S11]{
\textbf{A)} current controlled dependency of conductivity change $\Delta\sigma$ per pulse on pulsing time ($I_W=\qty{+-5}{\nA}$).
\textbf{B)} current controlled dependency of conductance change $\Delta G$ per pulse on pulsing amplitude ($t_p = \qty{220}{\ms}$)
}
\label{fig: S10}
\end{figure}

\newpage


\subsubsection[S12]{S12 Linearity and asymmetric ratio}

The linearity of synaptic weight update during potentiation and depression is determined by the non-linearity factor $\nu$ referring to either potentiation ($\nu_p$) or depression ($\nu_d$)~\cite{agarwal2016resistive}. 
For weight updating, the conductance is modeled by
\begin{equation}
   G = G_1 \cdot \left( 1-e^{-\nu P} \right) + G_{min} \\
\end{equation}
where $G_1 = \frac{G_{max} - G_{min}}{1 - e^{-\nu}}$, and  $G_{max}$, $G_{min}$ represent the conductance of the channel in the high-conductance and low-conductance state, respectively.
$P$ is the normalized pulse number. 
In a perfectly symmetric and linear device, $\nu$ equals zero.

Symmetry in synaptic weight update during potentiation and depression can be quantified by the asymmetric ratio which is defined as 
\begin{equation}
    AR = \frac{\text{max} \left\vert G_p(n) - G_d(n) \right\vert}{G_p(P_{max}) - G_d(P_{max})} \text{ for } n = 1 \text{ to } P_{max}
\end{equation}
where $G_p$ and $G_d$ are the channel conductance values after the $n^\text{th}$-potentiation and depression pulse, respectively, and $G_p(P_{max})$ and $G_d(P_{max})$ are the conductance values of the last pulse in the potentiation and depression pulse-train with $P_{max}$ pulses~\cite{nikam2019near}. 
$AR$ should be zero for an ideal symmetric case.

\newpage

\subsubsection[S13]{S13 Pulse variation}

\setcounter{figure}{12}
\begin{figure}[h]
\centering
\includegraphics[width=\textwidth]{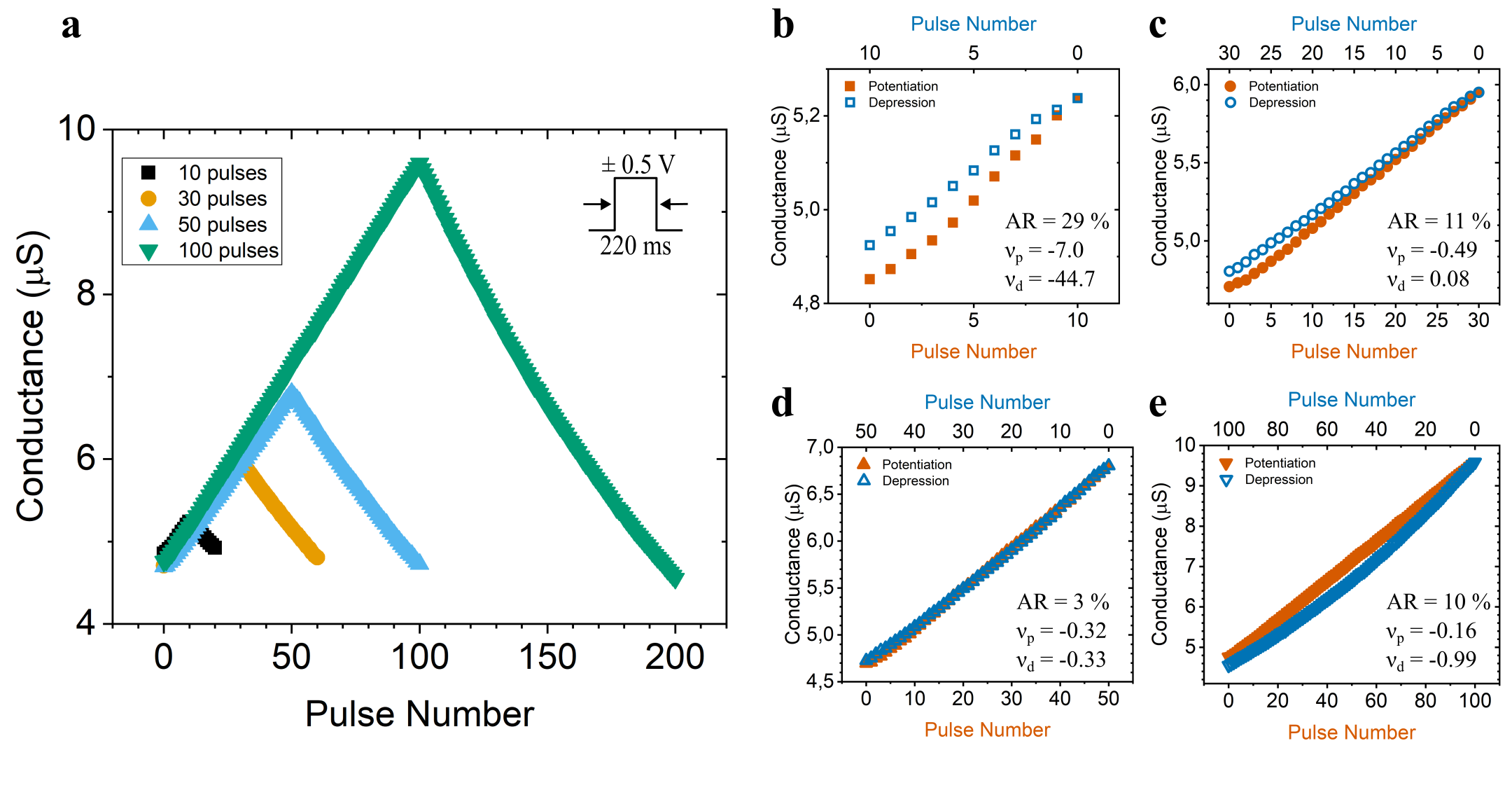}
\caption[S13]{
Pulse variation in the medium conductance range at \qty{150}{\degreeCelsius} with $P_{max} = \numlist{10;30;50;100}$ potentiation and depression pulses (voltage gating). \numlist{30;50;100} pulse-trains show great linearity and symmetry. 
The pulse train with \num{50} potentiation- and depression pulses (\textbf{B}) outperform the benchmarks in symmetry and linearity for oxide-ionic synaptic transistors as reported\cite{huang2023electrochemical}. 
}
\label{fig: S13}
\end{figure}

\newpage

\subsubsection[S14]{S14 $\Delta G$ and $I_W$ per pulse}

\begin{figure}[h]
\centering
\includegraphics[width=0.6\textwidth]{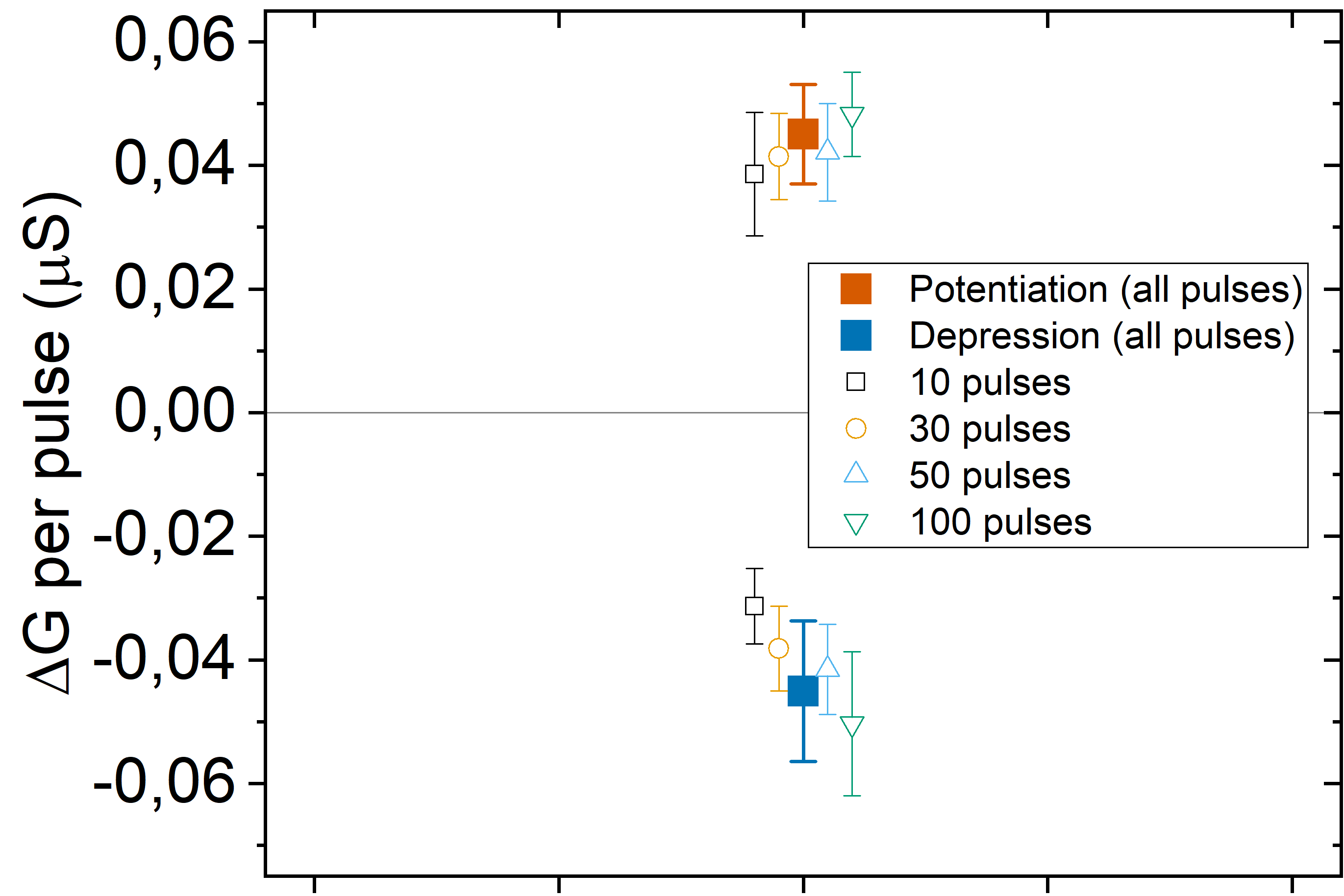}
\caption[S14]{
\textbf{A)} averaged change in conductance per pulse for varying pulse number experiments in Fig.~2e,f,g including averaged $\Delta G_p = \qty{45 +- 8}{\nano\siemens}$ for potentiation and averaged $\Delta G_d = \qty{45 +- 11}{\nano\siemens}$ for depression pulses. 
\textbf{B)} averaged writing current referring to $E_W = \qty{\pm 0.5}{\V}$ applied during varying pulse number experiments in Fig.~2e,f,g.
}
\label{fig: S14}
\end{figure}

\newpage

\subsubsection[S15]{S15 Energy calculation}

As described in Schwacke \textit{et al.} (2024), the energy consumption $\Delta E$ per programming pulse is calculated to
\begin{equation}
\Delta E = I_W \cdot \Delta t_p \cdot E_W = \Delta Q_W \cdot E_W
\end{equation}
where $E_W$ is the writing voltage, and $I_W$ and $Q_W$ are the current and the charge, respectively, that flow between channel and reservoir during the programming pulse with duration $t_p$\cite{schwacke2024electrochemical}.
For constant pulse duration, the programming charge is constant, too.
For only small changes, the change in conductivity $\Delta\sigma$ is proportional to the change in oxygen vacancies $\left[V_{O}^{\bullet\bullet}\right]$ and inverse proportional to the channel volume $V_{tot}$,
\begin{equation}
\Delta\sigma \propto \Delta\left[V_{O}^{\bullet\bullet}\right] \propto \frac{\Delta Q}{V_{tot}}
\end{equation} 
In the case of homogeneous charge carrier distribution inside the channel volume during programming, the change in conductance $\Delta G$ depends on the channel dimensions width $w$, lenght $l$ and thickness $th$,
\begin{equation}
\Delta G = \frac{th \cdot w}{l} \cdot \Delta\sigma = \frac{th \cdot w}{l} \cdot \frac{1}{th \cdot w \cdot l} \cdot \Delta Q = \frac{1}{l^2} \cdot \Delta Q
\end{equation}
Thus, the energy consumption during programming is inverse proportional to $l^2$ 
\begin{equation}
\Delta G \propto \Delta\sigma \propto \frac{1}{l^2} \cdot \Delta Q
\end{equation}
and the energy efficiency for our in-plane synaptic transistor is proportional to $l^2$
\begin{equation}
\frac{\Delta E}{\Delta G} \propto E_W \cdot l^2
\end{equation}
This proportionality to $l^2$ is also observed for out-of-plane devices, as discussed in refs.~\cite{schwacke2024electrochemical,onen2021cmos}.

For our synaptic transistor, with $l = \qty{35}{\micro\metre}$, the energy efficiency per pulse of pulse-variation experiments in Fig.~2~e,f,g is calculated to
\begin{align}
E &= \frac{\qty{4,6(0,1)}{\nano\ampere} \cdot \qty{220}{\ms} \cdot \qty{0,5}{\V}}{\qty{45(8)}{\nano\siemens}} = \frac{\qty{0.51(0.01)}{\nano\joule}}{\qty{45(8)}{\nano\siemens}} = \qty{1.1(0.2)}{\milli\joule\per\siemens} \\
E &= \frac{\qty{6,2(0,1)}{\nano\ampere} \cdot \qty{220}{\ms} \cdot \qty{0,5}{\V}} {\qty{45(11)}{\nano\siemens}} = \frac{\qty{0.68(0.01)}{\nano\joule}}{\qty{45(11)}{\nano\siemens}} =\qty{1.5(0.4)}{\milli\joule\per\siemens}
\end{align}
and for the cycling experiment in Fig.~3a to
\begin{equation}
 E = \frac{\qty{13(1)}{\nano\ampere} \cdot \qty{440}{\ms} \cdot \qty{0,7}{\V}}{\qty{80(36)}{\nano\siemens}} =\frac{\qty{4.0(0.3)}{\nano\joule}}{\qty{80(36)}{\nano\siemens}} = \qty{5.0(2.3)}{\milli\joule\per\siemens}
\end{equation}
These values scale linear with the device energy efficiency discussed in ref.~\cite{onen2021cmos}, Fig.~4, and represent an improvement in energy efficiency of $\sim \num{e2}$ with respect to other CMOS-compatible oxide-ion intercalation devices\cite{kim2019metal}.
Accordingly, we estimate the energy consumption per programming pulse of a 100~$\times$~100 \si{\nm\squared} oxide-ion synaptic transistor to $\sim \qty{10}{\femto\joule}$~\cite{onen2021cmos}, 
considering the expected increased homogeneity of charge carrier distribution when reducing the channel width.

\newpage


\subsubsection[S16]{S16 Dynamic range during cycling}

\setcounter{figure}{15}
\begin{figure}[h]
\centering
\includegraphics[width=0.6\textwidth]{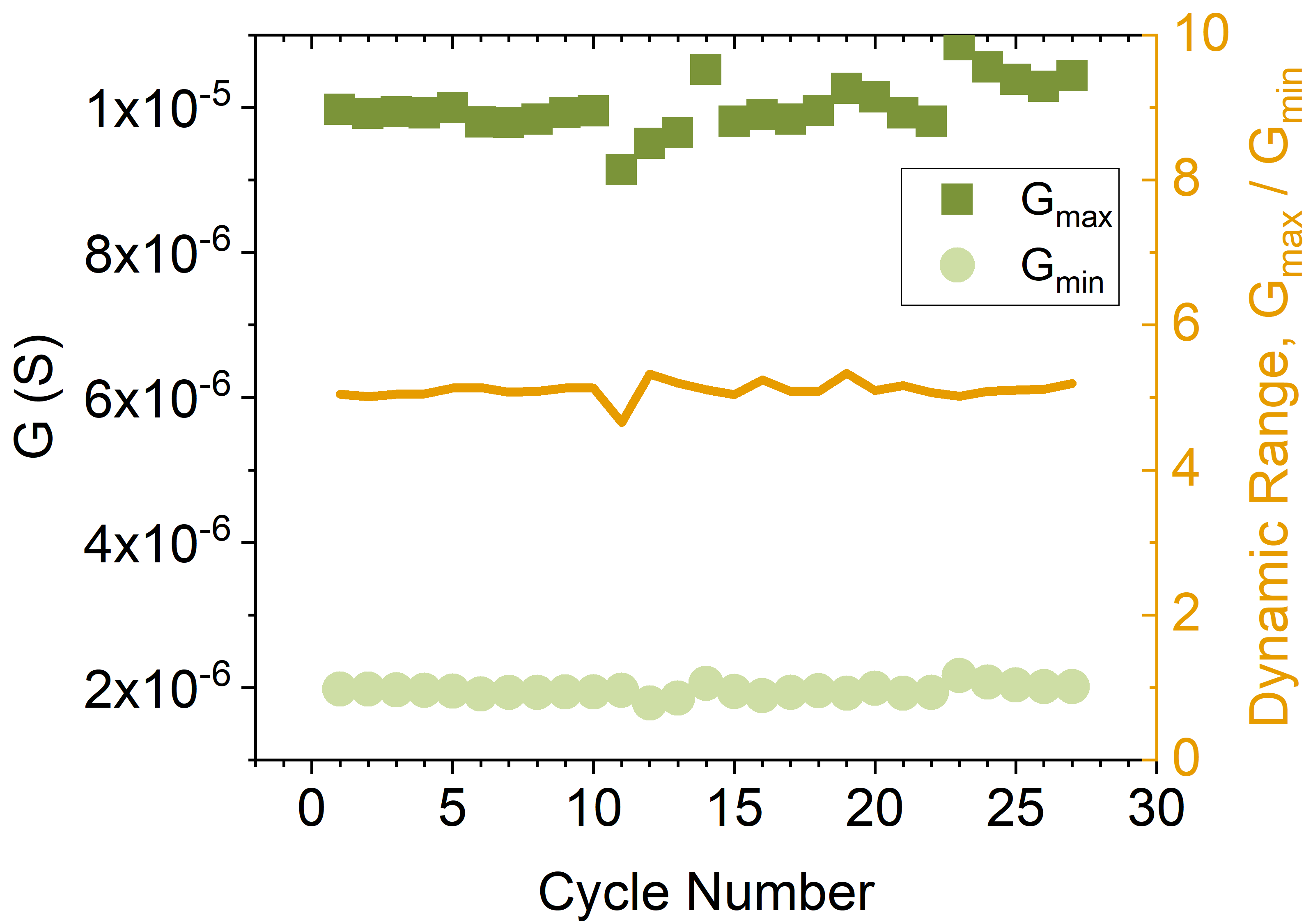}
\caption[S16]{
Endurance characteristics during cycling in Fig. 3a.
Minimum conductance ($G_{min}$) as well as maximum conductance ($G_{max}$) after 100 pulses stay constant during cycling with slight fluctuations in $G_{max}$. 
The $G_{max}\text{/}G_{min}$-ratio (dynamic range) stays very constant at \num{5.1 +-0,1}. 
}
\label{fig: S16}
\end{figure}

\newpage

\subsubsection[S17]{S17 Retention characteristics}

\begin{figure}[h]
\centering
\includegraphics[width=0.6\textwidth]{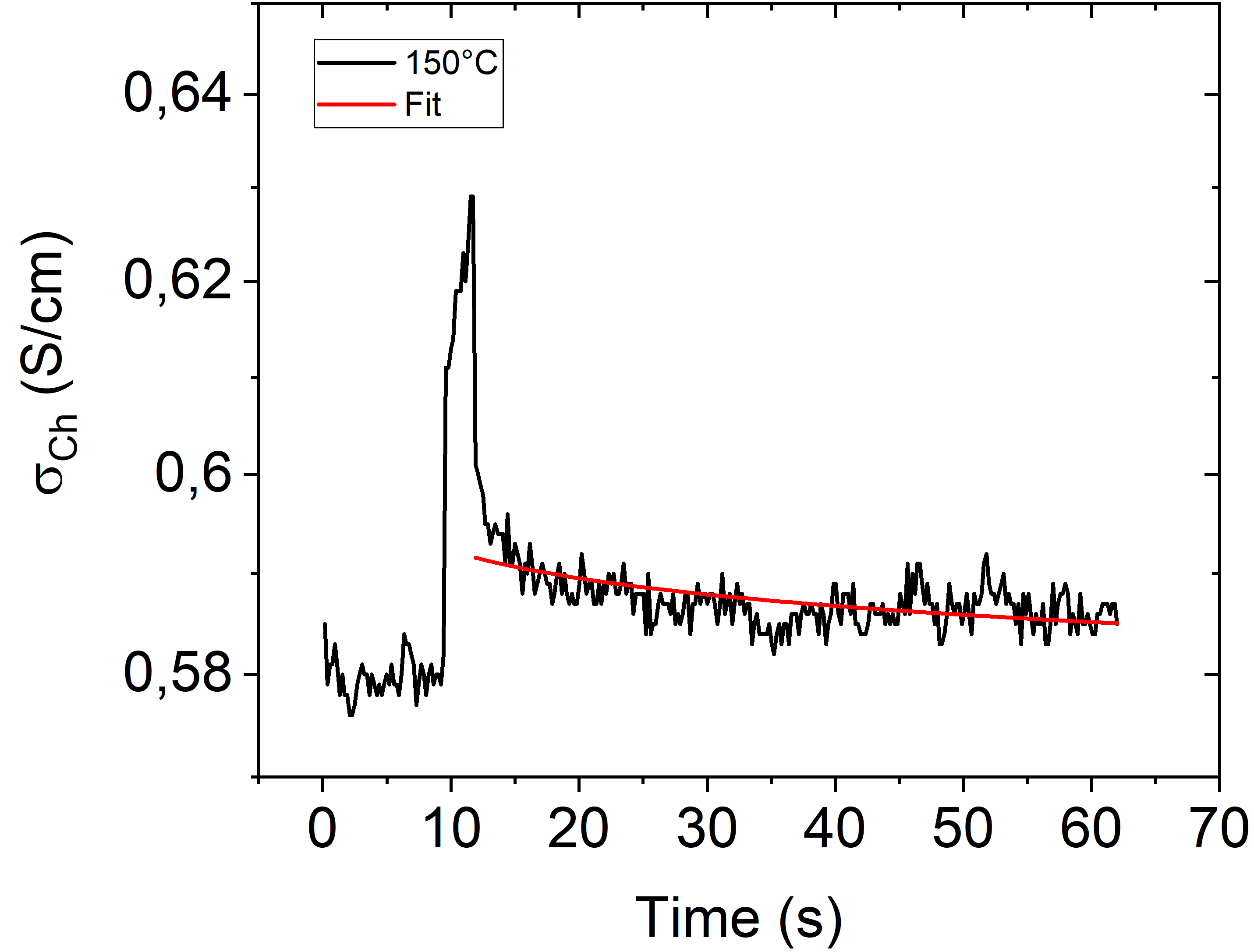}
\caption[S17]{
Retention characteristics of our synaptic transistor after voltage-controlled pulsing ($t_p = \qty{2,1}{\second}$, $E_W = \qty{-0.5}{\V}$) at $\qty{150}{\degreeCelsius}$ showing distinguishable change in state after applied pulse. 
A rention rate of \qty{2}{\percent} is calculated with a drift coefficient of \num{0,007}. The power-law has been applied for calculation\cite{cui2023cmos}.
}
\label{fig: S17}
\end{figure}

\newpage

\subsubsection[S18]{S18 Switching accuracy}

\begin{figure}[h]
\centering
\includegraphics[width=0.6\textwidth]{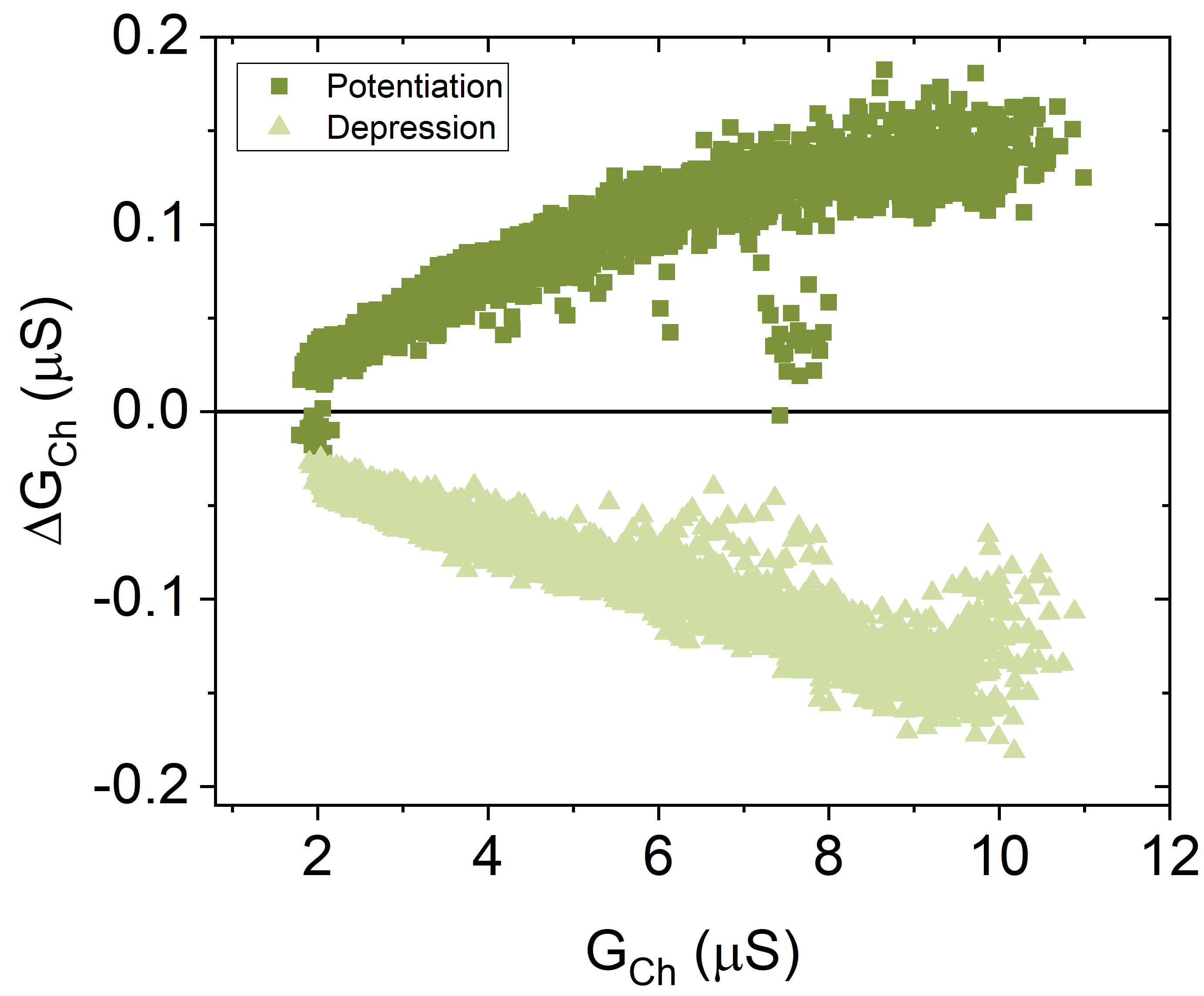}
\caption[S18]{
Mean change in conductance during cycling in Fig. 3a. 
An averaged change in conductance for potentiation ($\Delta G_p = \qty{81 +- 39}{\nano\siemens}$) and depression ($\Delta G_D = \qty{-80 +- 32}{\nano\siemens}$) is calculated. 
}
\label{fig: S18}
\end{figure}

\newpage

\subsubsection[S19]{S19 Short-term memory}

\begin{figure}[h]
\centering
\includegraphics[width=\textwidth]{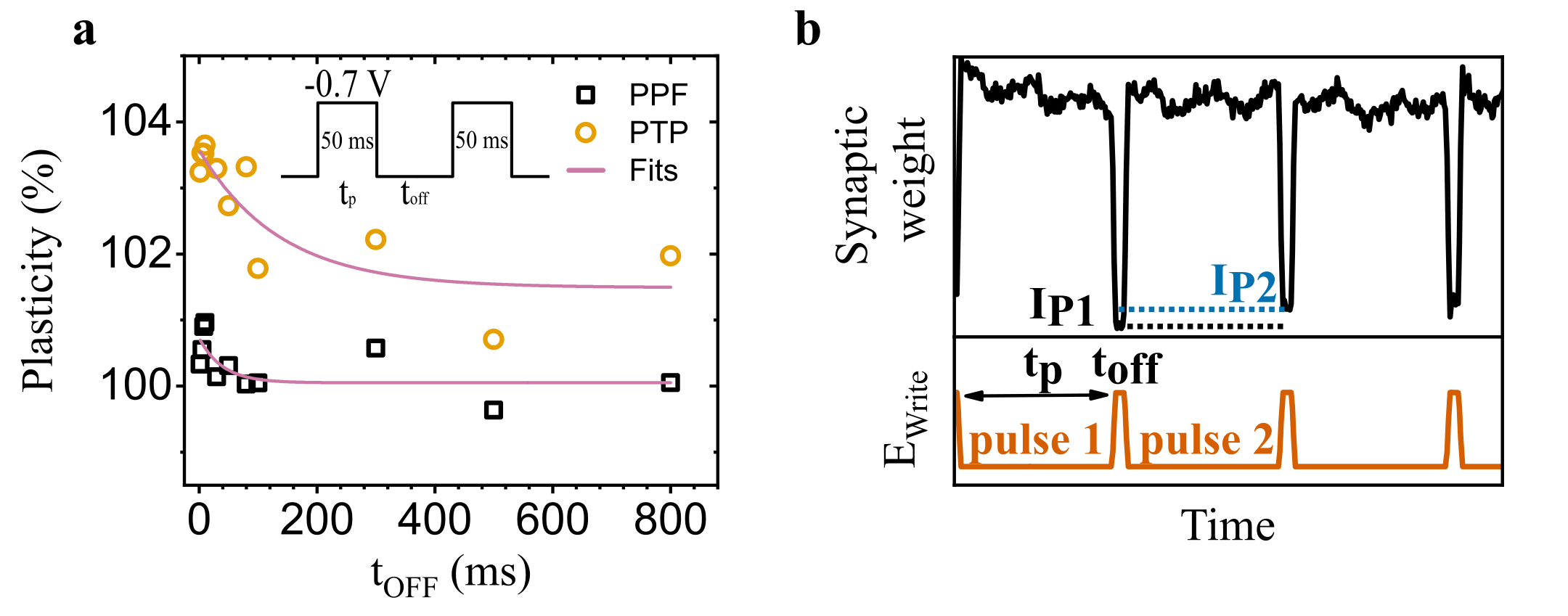}
\caption[S19]{
\textbf{a} STP dynamics under the application of voltage trains of 10 pulses with duration $t_p = \qty{50}{\milli\second}$ and off-time $t_{off}$ varying from $\qtyrange{2}{800}{\milli\second}$. 
\textbf{b} 
Pulse characteristics of short-term memory experiments. 
}
\label{fig: S19}
\end{figure}

The PPF and PTP of our synaptic transistor are defined as 

\begin{align}
    P_{PPF} &= \frac{I_{R,P2} - I_{R,P1}}{I_{W,P1}} \times \qty{100}{\percent} \\
    P_{PTP} &= \frac{I_{R,P10} - I_{R,P1}}{I_{W,P1}} \times \qty{100}{\percent}  
\end{align}
Where $P$ is the change in plasticity in percentage, $I_{R,P1-10}$ is the channel current recorded right after the first, second, and tenth pulse, respectively (see S\ref{fig: S19})\cite{yang2018artificial}. 

\newpage

\subsection{Simulations}
\label{subsec:simulation}


\subsubsection[S20]{S20 Comparison of time ion insertion speed for in-plane and out-of-plane configuration}
\label{S191}

\begin{figure}[h]
\centering
\includegraphics[width=0.7\textwidth]{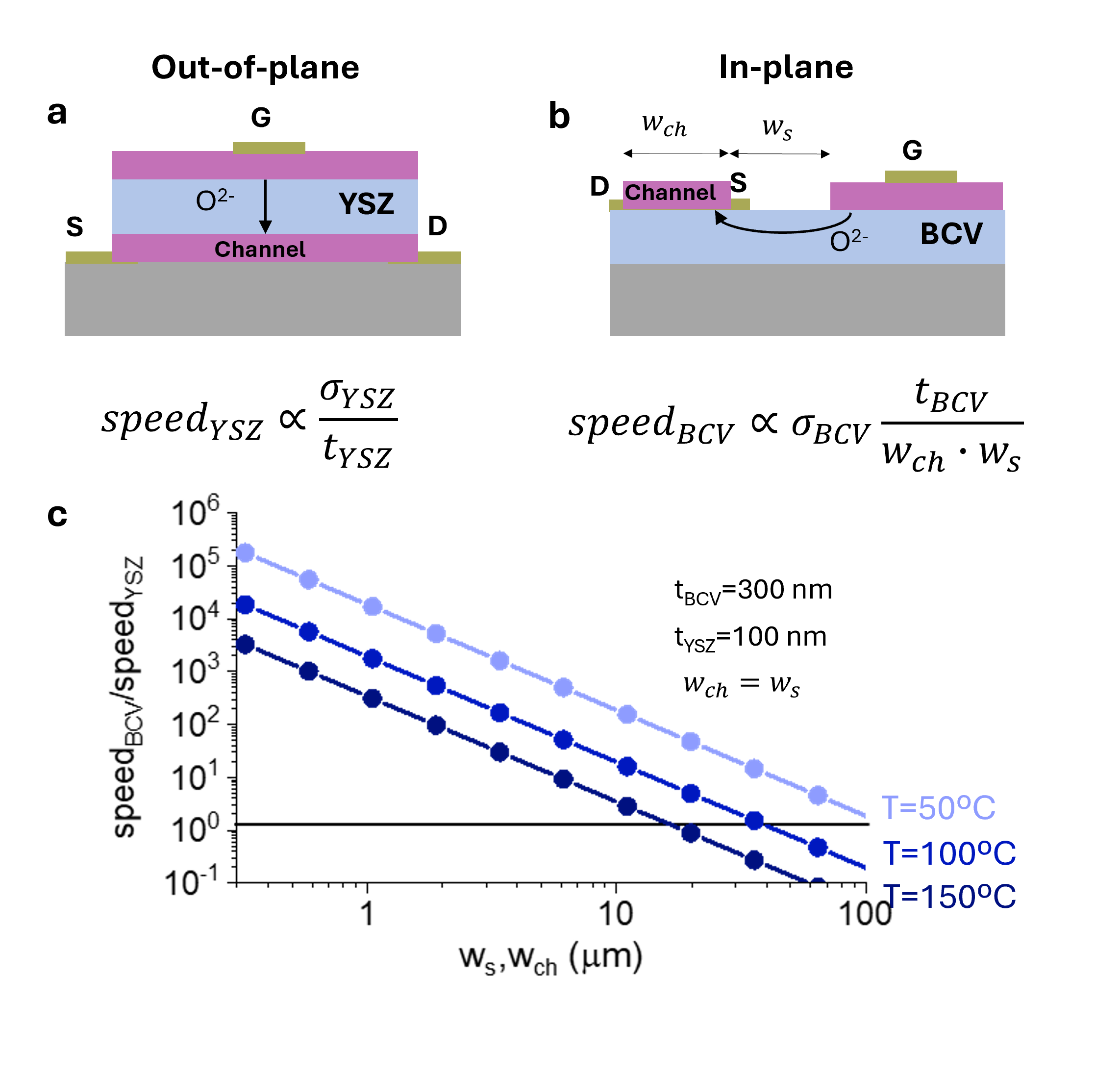}
\caption[S19]{
Schematic representation of out-of-plane configuration with YSZ electrolyte~\textbf{(a)}. A in-plane configuration with BICUVOX electrolyte~\textbf{(b)}. 
Calculated ratio of speed necessary for inserting the same amount of charge in the channel for in-plane BICVUOX and out-of-plane YSZ. 
BICUVOX is observed to outperform YSZ of several orders of magnitude decreasing the channel’s sizes in the \unit{\um} range or decreasing the T.
}
\label{fig: S1911}
\end{figure}

In this section we computed the expected ion insertion velocity for an out-of-plane configuration based on YSZ electrolyte and an in-plane configuration based on BICUVOX(001). We chose YSZ as it represent the state-of-art oxygen ion conductors for solid-oxide fuel cell application and due to the large amount of previous works that used this material\cite{kim2019metal,lee2022strategies,li2020filament,nikam2021all,kwak2021experimental,kim2023nonvolatile}. 
We consider a system where all the resistive losses are coming from the electrolyte ionic diffusion, see Fig.~S\ref{fig: S1911}. 
In this case, we aim at calculating the speed required for inserting in the channel a fixed amount of oxygen vacancies $\left[V_{O}^{\bullet\bullet}\right]_0$ (Kr\"oger-Vink notation), moving them through the electrolyte. 
First, we transform $\left[V_{O}^{\bullet\bullet}\right]_0$ from per-unit-cell concentration to the amount of charge that it holds in~$\left[\unit{\coulomb\per\cubic\centi\metre}\right]$: 
\begin{equation}
q_0 \left[\unit{\coulomb\per\cubic\centi\metre}\right] = \frac{2\left[V_{O}^{\bullet\bullet}\right]_0 F\rho}{M}
\end{equation}
where $F$ is the Faraday constant, $\rho$ the density and $M$ the molar mass of the channel’s material. 
The factor 2 considers the double charge of an oxygen vacancy. 
The total amount of charge can be then obtained by multiplying the relative charge by the dimensions of the channel:
\begin{equation}
Q\left[ \unit{\coulomb}\right] = q_0 \cdot A_{Ch} \cdot t_{Ch}
\end{equation}
where $A_{Ch}$ and $t_{Ch}$ are the contact area of the channel on the electrolyte and the electrolyte’s thickness, respectively. 
Considering that the system presents a single resistance ($R_{ion}$) of the electrolyte, the time ($time$) needed to transport the charge $Q$ upon the application of a potential $E_{set}$ is:
\begin{equation}
time = \frac{Q}{E_{set}}\cdot R_{ion} = \frac{q_0 \cdot t_{Ch}}{E_{set}} \cdot A_{Ch} \cdot R_{ion}
\end{equation}
The $R_{ion}$ depends on the configuration considered. 
For the YSZ in out-of-plane configuration, we can write:
\begin{equation}
R_{ion,YSZ} = \frac{1}{\sigma_{ion,YSZ}} \cdot \frac{t_{YSZ}}{A_{Ch}} 
\end{equation}
where $\sigma_{ion,YSZ}$ is the conductivity of the YSZ and $t_{YSZ}$ the thickness of the electrolyte. 
The insertion time in the out-of-plane direction with YSZ is therefore:
\begin{equation}
time_{YSZ} = \frac{q_0 \cdot t_{Ch}}{E_{set}}\cdot \frac{t_{YSZ}}{\sigma_{ion,YSZ}}
\end{equation}
Independent on the Area of the channel. In the case of BICUVOX in-plane, the resistance ($R_{ion,BCV}$) is:
\begin{equation}
R_{ion,BCV} = \frac{1}{\sigma_{ion,BCV}} \cdot \frac{w_s + w_Ch}{t_{BCV} \cdot l_{Ch}}
\end{equation}
Where $w_s$ and $w_{Ch}$ are the channel$\vert$gate distance and channel’s width, $l_{Ch}$ is the channel’s length and $t_{BCV}$ is the electrolyte thickness. 
Considering $w_s=w_{Ch}$ for simplification purpose and that $A_{Ch}=l_{Ch}\cdot w_{Ch}$, we can write:
\begin{equation}
time_{BCV} = \frac{q_0 \cdot t_{Ch}}{E_{set}}\cdot \frac{1}{\sigma_{ion,BCV}} \cdot \frac{2w_{Ch}^2}{t_{BCV}}
\end{equation}

Since the speed will be proportional to the inverse of insertion time, the ratio between the speed of insertion in BCV and YSZ is (considering the same potential applied and the same channel’s dimensions): 
\begin{equation}
\frac{speed_{BCV}}{speed_{YSZ}} = \frac{\sigma_{ion,BCV}}{\sigma_{ion,YSZ}} \cdot \frac{t_{YSZ} \cdot t_{BCV}}{2w_{Ch}^2}
\end{equation}
Therefore, considering the conductivity of YSZ from literature\cite{kim2023nonvolatile} and the in-plane conductivity of BICVUOX we can compute the ratio of the speed of the two devices as a function of the $w_{Ch}$(or $w_s$), see Fig.~S\ref{fig: S1911}c. 
In-plane configuration with BICUVOX is expected to present better performances than YSZ in out-of-plane configuration already for channel’s size of tens of \unit{\um} at \qty{150}{\degreeCelsius}. 
Lowering the T or decreasing the channel’s size in the \unit{\um} or sub-~\unit{\um}~range, in-plane BICUVOX is expected to outperform out-of-plane YSZ by several orders of magnitude.

\newpage

\subsubsection[S21]{S21 Finite Element Model (FEM) Simulations}
\label{S19S1}

\begin{figure}[h]
\centering
\includegraphics[width=\textwidth]{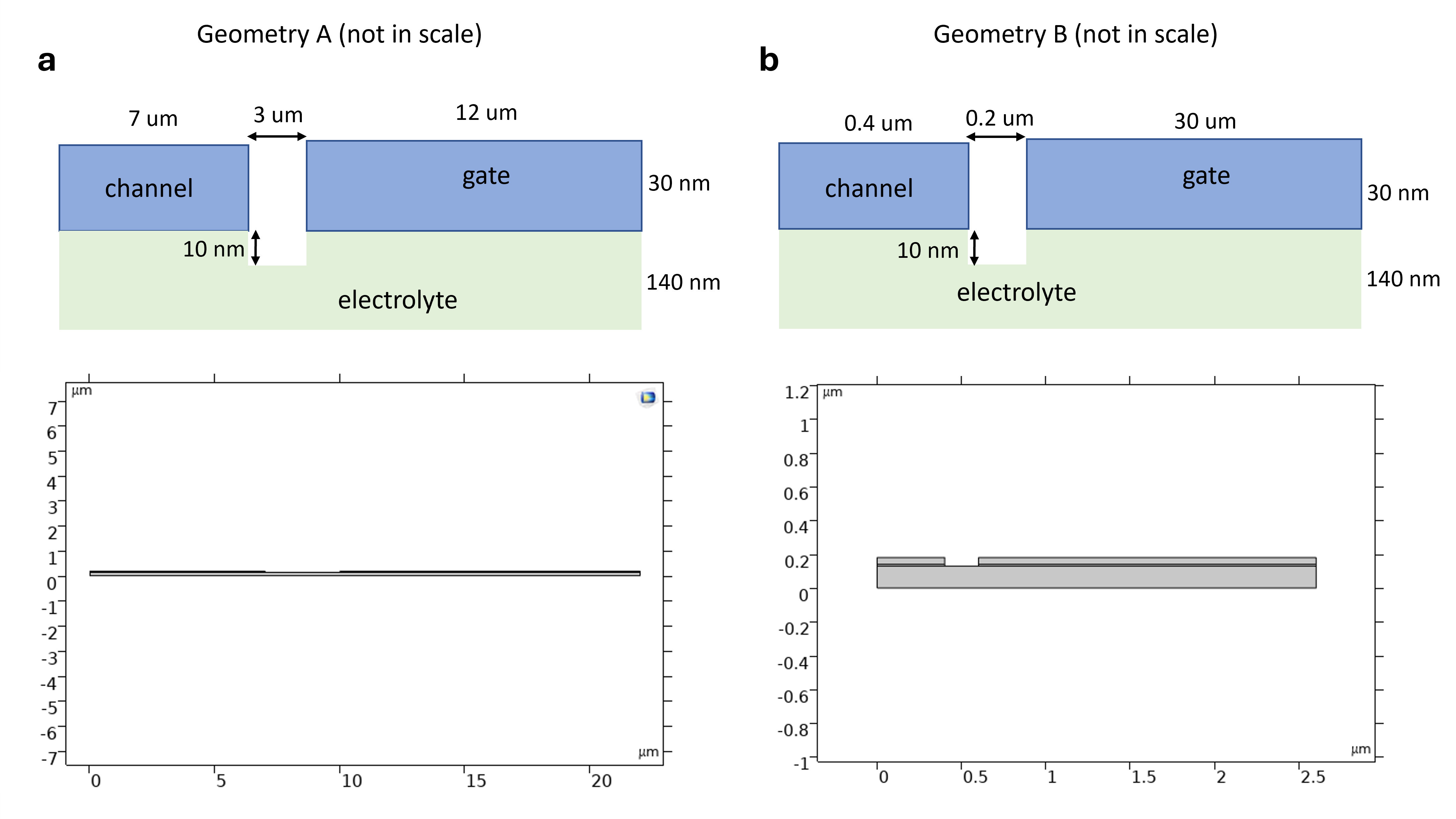}
\caption[S19]{
Sketch and scaled geometries used in the simulation. 
Geometry A \textbf{(a)} corresponds to the geometry of our microfabricated synaptic transistor, Geometry B \textbf{(b)} is used to test the low temperature limitations of BICUVOX based devices.
}
\label{fig: S19S1}
\end{figure}

FEM simulations were performed by COMSOL Multiphysics to get qualitative insights into operation of our synaptic transistor. 
The 2D model, based on previous ECRAM 1D and 2D simulations\cite{huang2023electrochemical},  is directly based on the cross section of the microfabricated in-plane configuration shown in Figure 1 of the main text. 
We tested two different geometries, as depicted in Fig.~S\ref{fig: S19S1}: 
Geometry A presents the same dimensions of the synaptic transistor microfabricated device of Figure 1, while Geometry B presents smaller channel and channel-gate lateral dimensions. 
This allows us to investigate the possibility of lowering the operation temperature of the device. 
The geometry presents a \qty{10}{\nm} gap in the BICUVOX electrolyte to take in account possible over-etching during the microfabrication of the channel$\vert$gate.

\begin{figure}[h]
\centering
\includegraphics[width=\textwidth]{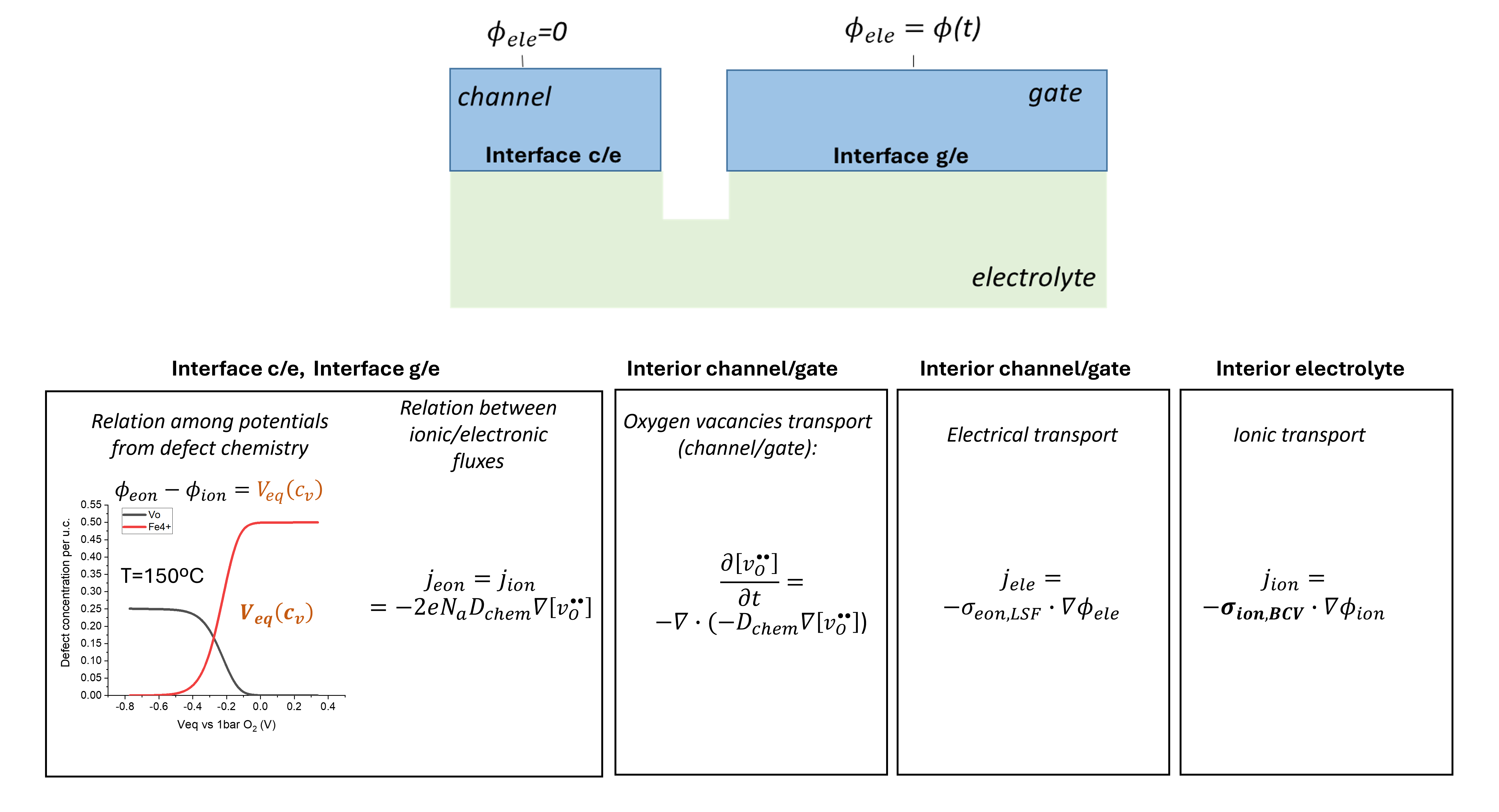}
\caption[S19]{
Schematic representation of the simulation’s geometry with the main physics considered.
}
\label{fig: S19S2}
\end{figure}

The FEM simulations requires the simultaneous solution of different physics, see Fig.~S\ref{fig: S19S2}. 
First, we considered gate$\vert$channel with predominant electronic conduction, meaning that the electronic conductivity will be always higher that the ionic one. 
Under these assumptions (which will be commented later), the current flow is dominated by electronic contribution ($j_{eon}$), as simulated by the equation: 
\begin{equation}
j_{eon} = -\sigma_{eon,LSF}\cdot\nabla \phi_{eon}
\label{eq: S1}
\end{equation}
where $\sigma_{eon,LSF}$ is the electronic conductivity of the LSF and $\phi_{eon}$ is the electrical potential. 
Channel is set to ground ($\phi_{eon}=0$) while potential pulses are applied to the gate ($\phi_{eon}=f(t)$). 
On the other hand, the BICUVOX electrolyte is considered as a pure ionic conductor, meaning that the current flow can be described as:
\begin{equation}
j_{ion} = -\sigma_{ion,BCV}\cdot\nabla \phi_{ion}
\end{equation}
where $\sigma_{ion,BCV}$ is the ionic conductivity of the BICUVOX (written as a matrix to take in account the anisotropic in and out-of-plane conduction) and $\phi_{ion}$ its internal ionic potential. 
Application of an electrical potential through an electrolyte shifts the equilibrium of point defects in the LSF channel$\vert$gate, similarly to what observed in our previous study on LSF$\vert$YSZ(001) system\cite{tang2021pushing},  or to what takes place in a battery. 
Considering a small resistance at the LSF$\vert$BICUVOX interface (as usually considered in oxygen electrolytes/electrode interface in solid oxide fuel cells), we can describe the relation between electrical and ionic potential at the gate/channel and electrolyte boundary as:
\begin{equation}
\phi_{ele} - \phi_{ion} = V_{eq}(c_v)
\label{eq: potential quilibrium}
\end{equation}
where $V_{eq}(c_v)$ represent the equilibrium between the oxygen vacancies and the applied potential. 
This relation can be understood as the shift of oxygen chemical potential taking place in an MIEC electrode under the application of a Nernst potential\cite{tang2021pushing,schmid2018voltage}. 
In this model, only electronic holes are considered as charge carriers, in agreement with the low temperature and potentials explored in this work. 
We also considered spatial electroneutrality, meaning that in any point of channel$\vert$gate it is valid the equilibrium:
\begin{equation}
[Sr_{La}^{'}] = 2\left[V_{O}^{\bullet\bullet}\right] + \left[h^{\bullet}\right]
\end{equation}
where $[Sr_{La}^{'}]$,  $\left[V_{O}^{\bullet\bullet}\right]$, and $\left[h^{\bullet}\right]$ are the concentrations of aliovalent Sr doping, oxygen vacancies and electronic holes written according to the Kr\"oger–Vink notation.
As no redistribution of cations are expected at such low temperatures, this relation implies that the insertion/extraction of oxygen vacancies will determine a opposite local decrease/increase of electronic holes. 
Indeed, we can consider the potential equilibrium of eq.~\ref{eq: potential quilibrium} as the driving force of ionic current, shifting the equilibrium of oxygen vacancies and creating an electrical flow of:
\begin{equation}
j_{ele} = j_{ion} = -2eN_aD_{chem} \nabla \left[V_{O}^{\bullet\bullet}\right]
\label{eq: boundary ChG}
\end{equation}
where $e$, $N_a$, and $D_{chem}$ are the electron charge, the Avogadro number, and the oxygen chemical diffusivity. 
Eq.~\ref{eq: boundary ChG} represent the boundary equation of the channel/gate and electrolyte interface. Importantly, it also defines the continuity between the electrical and the ionic current in the gate$\vert$channel and electrolyte, which must be equal for maintaining electroneutrality. 

Finally, ion transport in the gate$\vert$channel is described by the time-dependent Fick’s law:
\begin{equation}
\frac{\partial{\left[V_{O}^{\bullet\bullet}\right]}}{\partial{t}} = - \nabla \cdot \left( D_{chem}\nabla \left[V_{O}^{\bullet\bullet}\right]\right)
\label{eq: Fick}
\end{equation}

The set of equations~\ref{eq: S1}-\ref{eq: Fick} determines a system that can be solved to find the 2D time-distribution of oxygen vacancies inside the channel as a function of voltage pulses applied to the gate. 

\begin{figure}[h]
\centering
\includegraphics[width=0.7\textwidth]{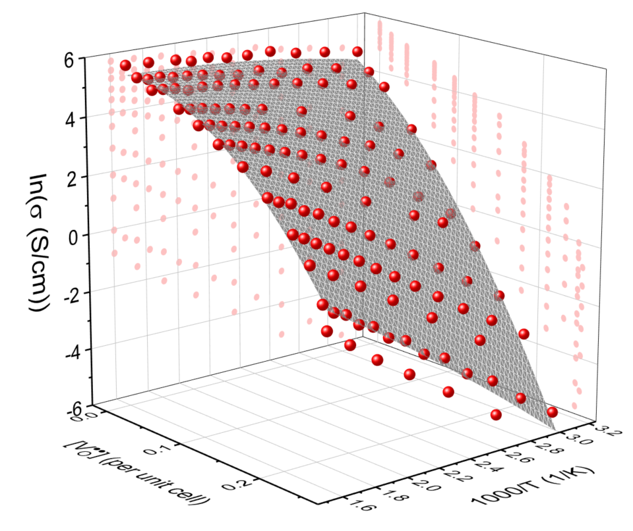}
\caption[S19]{
Experimental electrical conductivity measured for LSF50 as a function of oxygen vacancy concentration and temperature by Niezet et al. and 2D polynomial fitting used in the simulations.
}
\label{fig: S19S3}
\end{figure}

Regarding the simulation’s parameters, we first considered the T-dependent equilibrium between the oxygen vacancies and the applied potential which was derived from our previous work, where we measured the defect chemistry of LSF50 thin films on YSZ(001) electrolytes\cite{tang2021pushing}. 
We recently investigated the variation of electronic conductivity of LSF50 as a function of oxygen vacancy concentration and temperature.
We found that, due to a magnetic transition, the electronic conductivity decreases orders of magnitude when reducing the material[NIZET]. 
Therefore, we fitted a 2D quadratic function to the experimental data measure as a function of T and $\left[V_{O}^{\bullet\bullet}\right]$, valid for $0.01 < \left[V_{O}^{\bullet\bullet}\right] < 0.24$ (per unit cell) and $50 < \left[V_{O}^{\bullet\bullet}\right] < \qty{350}{\degreeCelsius}$, yielding to (see Fig.~S\ref{fig: S19S3}):
\begin{equation}
{\ln(\sigma_{eon}) = 4.44 + 23.08 \cdot \left[V_{O}^{\bullet\bullet}\right] + \cdot \frac{1190}{T} - 89.33 \cdot \left[V_{O}^{\bullet\bullet}\right]^2 - \cdot \frac{370000}{T^2} - \cdot \left[V_{O}^{\bullet\bullet}\right] \cdot \frac{14380}{T}}
\label{eq: Fit} 
\end{equation}
with $\left[V_{O}^{\bullet\bullet}\right]$ in conc. per unit cell and T in \unit{\kelvin}. 
Then, eq.~\ref{eq: Fit} was inserted in the FEM model to compute the electrical conductivity of each point of the channel$\vert$gate. 
The average electrical conductivity of the channel was calculated by integrating the local conductivity over its cross section. 

\begin{figure}[h]
\centering
\includegraphics[width=0.8\textwidth]{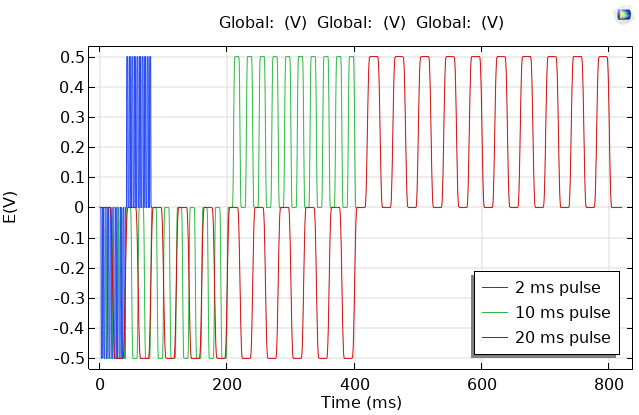}
\caption[S19]{
Explementary pulses sequences used in the first simulation.
}
\label{fig: S19S4}
\end{figure}

Oxygen vacancy diffusivity ($D_v$) was extracted from literature bulk measurement and extrapolated to the temperatures used in the simulations\cite{ishigaki1988diffusion}.  
The T-dependent equation used is:
\begin{equation}
D_v = 0.02 \left[ \unit{\cm\squared\per\second} \right] \cdot \exp \left( \frac{\num{-7500} \left[ \unit{\joule\per\mol} \right]}{8.314 \left[ \unit{\joule\per\kelvin\per\mol} \right]} \cdot T \right) 
\end{equation}
Nernst-Einstein equation was used to calculate the oxygen conductivity of LSF:
\begin{equation}
\sigma_{ion,LSF} = D_v \cdot \left( 4e \cdot \frac{e}{k_bT} \right) \cdot \left[V_{O}^{\bullet\bullet}\right] \cdot N_a
\end{equation}
The $\sigma_{ion,LSF}$ was computed to confirm that at any point and time the assumption of predominantly electronic transport in LSF was true ($\sigma_{ion,LSF} << \sigma_{eon,LSF}$), which allowed the use of eq.~\ref{eq: S1} to compute the electrical potential inside gate/channel. 
Finally, $D_{chem}$ was calculated from:
\begin{equation}
D_{chem} = \Gamma_v \cdot D_v
\end{equation}
Where $\Gamma_v$ is the thermodynamic factor of oxygen vacancies $\left( \Gamma_v = -\frac{1}{2} \frac{\partial{\ln pO_2}}{\partial{\ln\left[V_{O}^{\bullet\bullet}\right]}}\right)$ , which was calculated in the simulations as a function of x,y point and time from the defect chemical equilibrium, resulting in a $D_{chem}$ that vary in time and space to accommodate for different ion concentration. 
Finally, the temperature dependent ionic conductivity of BICUVOX was obtained from the EIS measurements in the in-plane direction, while extracted from our previous work for the out of-plane direction. 
In this way, the simulations are expected to properly catch the variation of temperature and dynamic conditions (e.g. spatial and temporal vacancy concentration) taking place in real devices.
The potential $\Phi_{eon}$ at the gate was set as an evolution of voltage pulses, whose amplitude ($E_{set}$), on/off time ($t_{on}$ and $t_{off}$) and sequence type depends on the different simulations performed. 
Fig.~S\ref{fig: S19S4} shows the three pulse sequences used in the first simulations (10 pulses $E_{set}=\qty{-0.5}{\volt}$, 10 pulses $E_{set}=\qty{+0.5}{\volt}$, $t_{on}=\qtylist{2;10;20}{\ms}$ and $t_{on}=t_{off}$). 
The initial concentration of oxygen vacancies was set to \num{0.1} per unit cell, nearly half of the available oxygen vacancies ($\left[V_{O}^{\bullet\bullet}\right] = \frac{1}{2}[Sr_{La}^{'}]=0.25$). 
Finally, rectangular elements were used for the adaptive mesh and FEM simulations were performed in time dependent mode.

\subsubsection*{Comparison between different geometries and temperature operation limits}
\label{S19S5}

\begin{figure}[h]
\centering
\includegraphics[width=\textwidth]{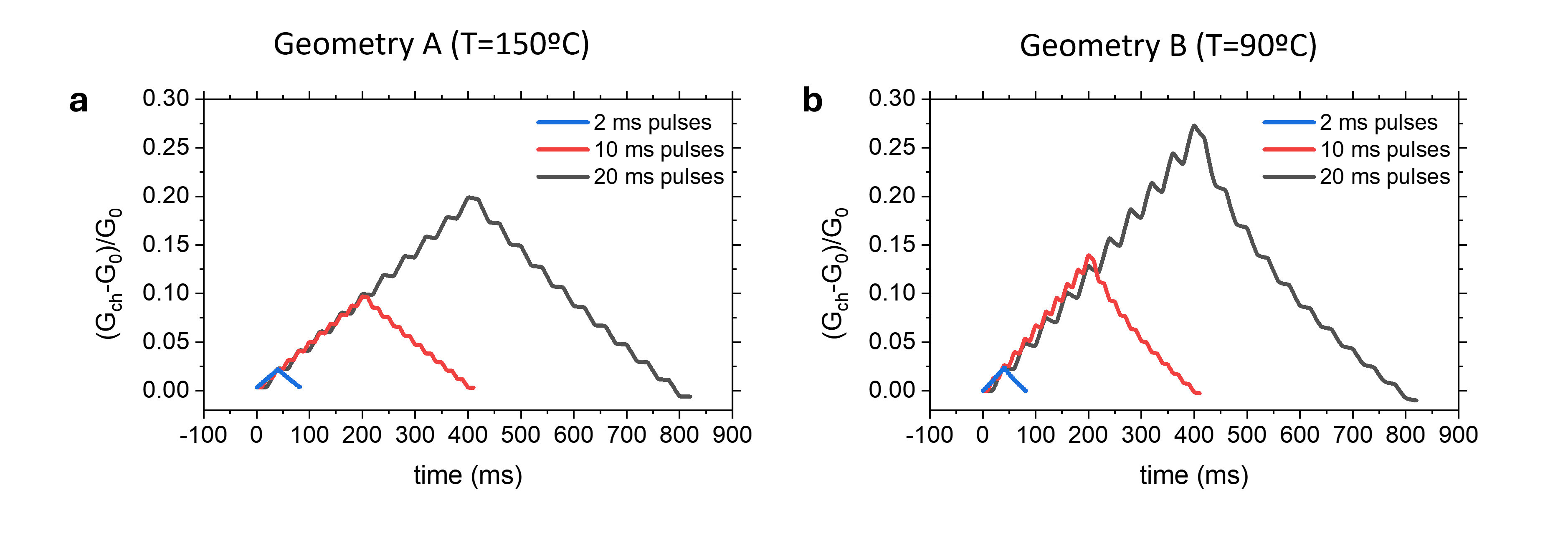}
\caption[S19]{
Simulated potentiation/depression curves obtained for Gemotry A \textbf{(a)} and B \textbf{(b)} for the sequences of train pulses shown in Figure S4
}
\label{fig: S19S5}
\end{figure}

We first investigated the possibility of decreasing the operation temperature and the limitations of in-plane synaptic transistors based on BICUVOX thin films. 
Fig.~S\ref{fig: S19S5} shows the normalized increased of channel’s conductance $\frac{(G_{Ch}-G_0)}{G_0}$ obtained in the simulations for the application of the train’s pulses shown in Fig.~S\ref{fig: S19S4} for the geometry A (see Fig.~S\ref{fig: S19S1}a), with dimensions comparable with the microfabricated synaptic transistor at \qty{150}{\degreeCelsius}. 
The simulations are able to qualitatively reproduce the potentiation/depression measured in the experiments (Figure 2). 
Although the exact fitting of the experimental data is out of the scope of the simulations, it is worth to discuss the similarities and discrepancies between simulations and experiments. 
First, we note that the writing current ($I_W$, the current flowing from channel to gate) that we obtained in the simulations for pulses of \qty{0.5}{\volt} is around \qty{2}{\nA}, comparable to the \qty{5}{\nA} obtained in the experimental measurements. 
This suggests that the resistances to the current flow that we considered in our simulations are in agreement with the experimental ones ($I_W=\frac{E_{set}}{R_tot}$, where $R_{tot}$ is the sum of the electrolyte, diffusivity and geometrical resistive contributions), essential to explore the behavior for different geometries and different T. 
The increase of $\frac{(G_{Ch}-G_0)}{G_0}$ after 10 pulses is equal to \num{0.2} in the simulations for \qty{20}{\ms} pulses, while the experimental data present an average $\frac{(G_{Ch}-G_0)}{G_0}$ of $\approx \num{0.1}$ for the mid-conductivity range with \qty{220}{\ms} pulses (see Figure 2). 
The larger increase of conductance in the simulations compared to the experiments may have the following origins:
\begin{itemize}
\item 	Different initial oxygen vacancy concentration. The $\left[V_{O}^{\bullet\bullet}\right]-T-\sigma_{eon}$ relation (Fig.~S\ref{fig: S19S3}) is very sensitive to the initial oxygen vacancies concentration and for more oxidized samples the variation tends to be less exponential. 
Simulations performed with initial $\left[V_{O}^{\bullet\bullet}\right] = \num{0.05}$ shows a decrease in the relative increase of channel’s conductance ($\frac{(G_{Ch}-G_0)}{G_0}$ of \num{0.07} after 10 pulses of \qty{20}{\ms}). 
\item 	The $\left[V_{O}^{\bullet\bullet}\right] - T-\sigma_{eon}$ relation is different for LSF50 measured on YSZ (Fig.~S\ref{fig: S19S3}) and on BICUVOX due to different crystalline microstructure or interfaces, which may modify the effect of an oxygen content increase/decrease on the channel’s properties.
\item 	The experimental system presents small current leakages (e.g. oxygen incorporation/evolution taking place at the triple phase boundary BICUVOX/LSF50/atmosphere). 
This would mean that a part of the current flowing in the system does not directly convert in an increase/decrease of  $\left[V_{O}^{\bullet\bullet}\right]$ in the channel, partially explaining the slightly higher $I_W$ and the lower $\frac{(G_{Ch}-G_0)}{G_0}$ of the experiments.
\end{itemize}

\begin{figure}[h]
\centering
\includegraphics[width=\textwidth]{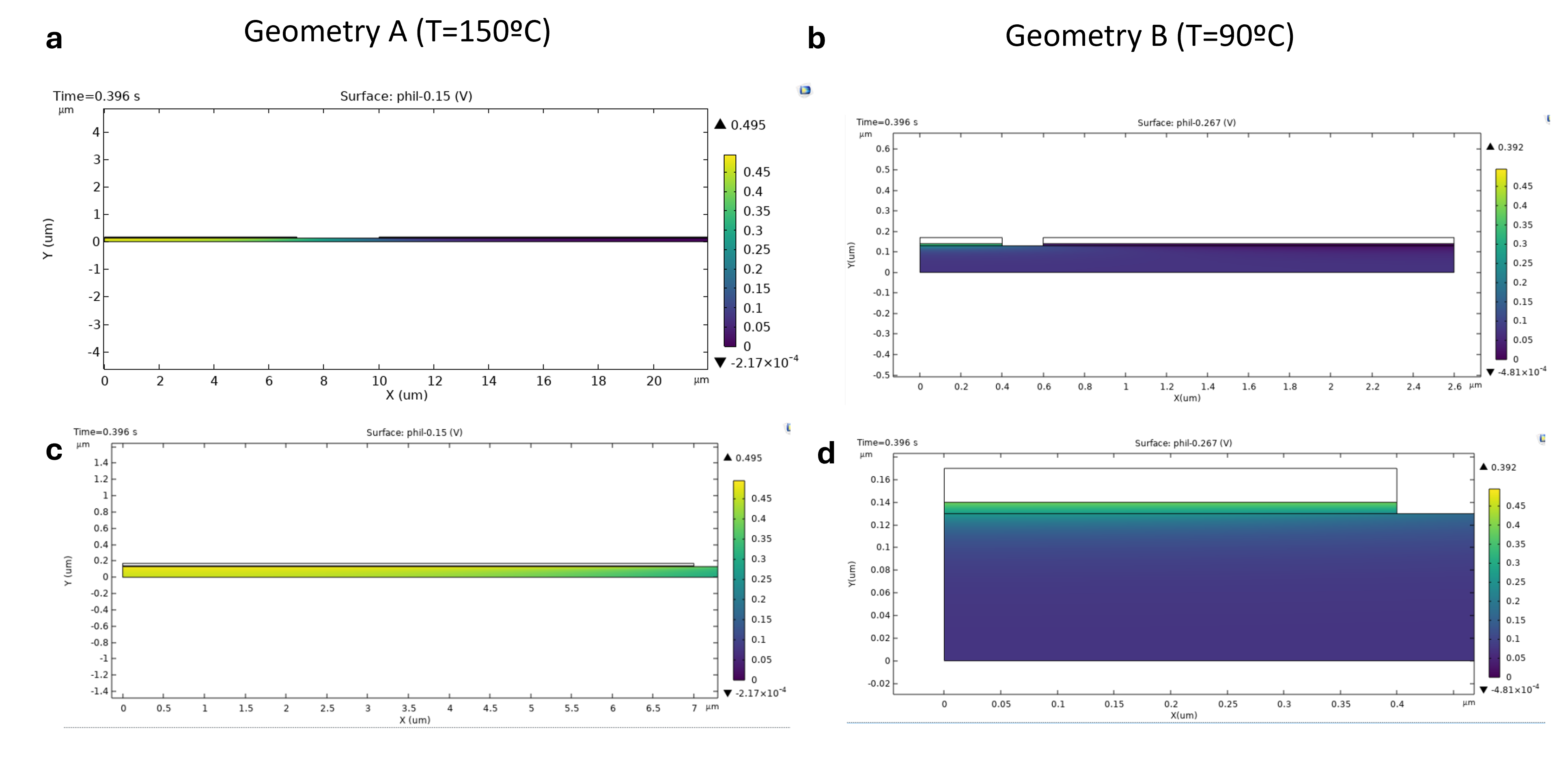}
\caption[S19]{
Distribution of internal ionic potential $\Phi_{ion}$ simulated for geometry A \textbf{(a,c)} and B \textbf{(b,d)} obtained for the \qty{20}{\ms}-pulse train after \qty{0.396}{\second} (last potentiation pulse). 
A smaller potential drop and a more homogenous potential distribution is observed for the reduced dimensions of Geometry B.
}
\label{fig: S19S6}
\end{figure}

We then explored the effects of a smaller sizes of the synaptic transistor on the potentiation/depression curve. 
Fig.~S\ref{fig: S19S5}b shows the normalized increased of channel’s conductance $\frac{(G_{Ch}-G_0)}{G_0}$ obtained for Geometry B (Fig.~S\ref{fig: S19S1}b) at \qty{90}{\degreeCelsius}. 
The total variation of conductance is similar to the one obtained with Geometry A at \qty{150}{\degreeCelsius}, suggesting that by decreasing of lateral sizes of the channel the operation temperature can be substantially reduced. 
Interestingly, although the total variation of conductance is similar, Geometry B at \qty{90}{\degreeCelsius} presents less linearity and more conductance decay during the off time. 
To understand the origin of this feature is useful to consider the distribution of internal ionic potential and $\Phi_{ion}$ for geometry A (Fig.~S\ref{fig: S19S6}a) and B (Fig.~S\ref{fig: S19S6}b) obtained for the \qty{20}{\ms}-pulse train after \qty{0.396}{\second} (last potentiation pulse). 
The total potential drop is \qty{0.495}{\volt} and \qty{0.392}{\volt} for Geometry A and B, respectively. This means that, although the \qty{60}{\degreeCelsius} decrease of temperature, the BICUVOX electrolyte presents a smaller drop of potential (i.e. less resistive contribution) in the simulations with reduced dimensions. 
Moreover, a clear decrease of $\Phi_{ion}$ is visible for Geometry A along the channel’s width, meaning that a lateral gradient of intercalation potentials will develop in this device. 
On the contrary, Geometry B presents a quite homogenous $\Phi_{ion}$ in the lateral direction, implying that the smaller dimension of the channel helps in homogenizing the potential and providing a similar ionic flux in all channel’s points. 
These results suggest that a reduced channel/gate gap and small channel’s dimensions are key for decreasing the operation temperature and providing a homogenous variation of channel’s conductance. 
Nevertheless, the results do not explain the decrease in linearity observed in Fig.~S\ref{fig: S19S5} for geometry 2.

\begin{figure}[h]
\centering
\includegraphics[width=\textwidth]{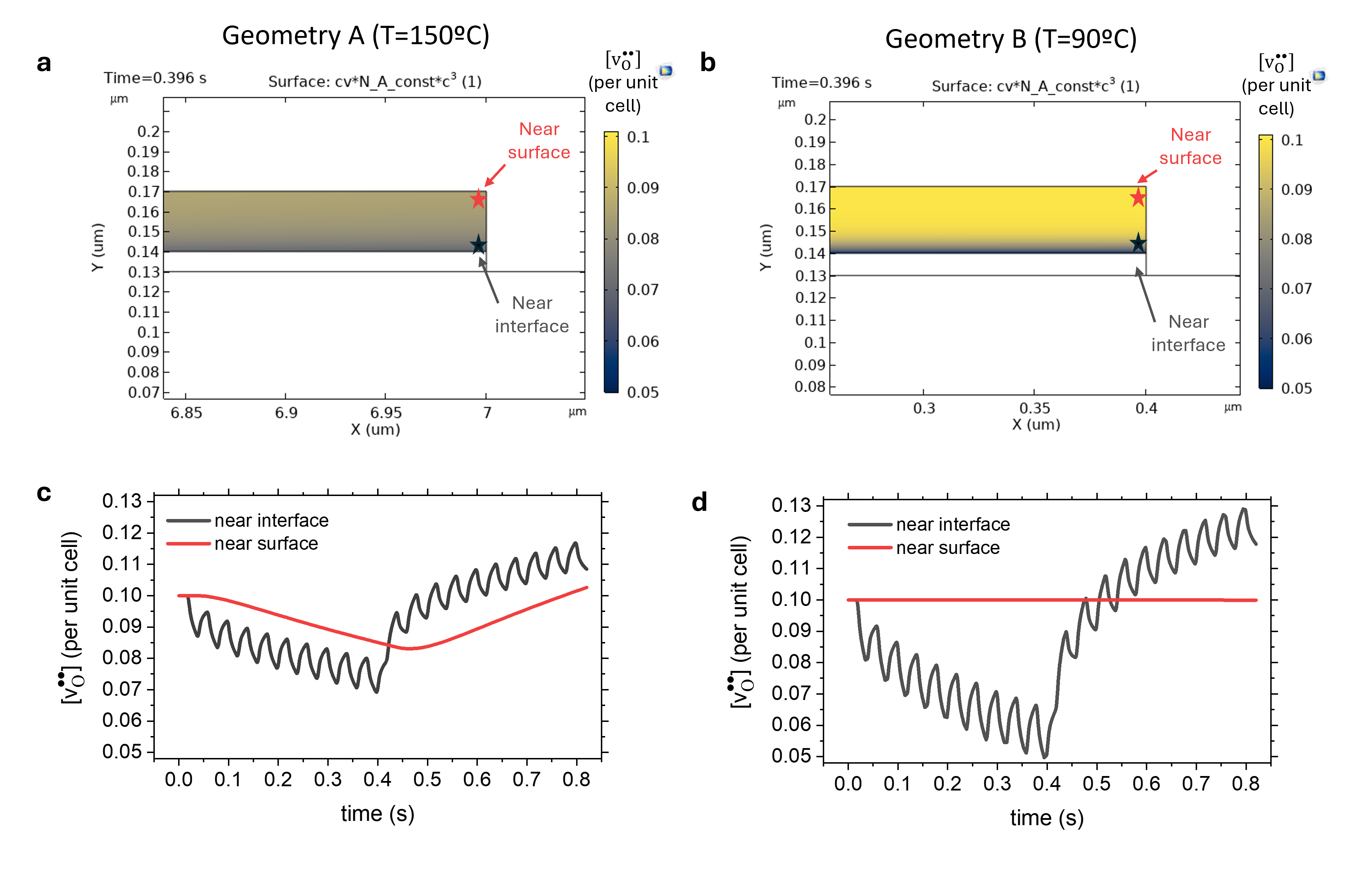}
\caption[S19]{
Distribution of oxygen vacancies obtained in the channel for Geometry A \textbf{(a)} and B \textbf{(b)} for the \qty{20}{\ms}-pulse train after \qty{0.396}{\second} (last potentiation pulse). 
Temporal evolution of the oxygen vacancy concentration obtained for Geometry A (a) and B(b) in the near-surface and near-interface regions shown in a and b
}
\label{fig: S19S7}
\end{figure}

We then considered the other major resistive contribution of the device: the oxygen vacancy diffusion in the channel. 
Fig.~S\ref{fig: S19S7}a and b show the distribution of oxygen vacancies obtained in the channel for Geometry A and B, respectively, for the \qty{20}{\ms}-pulse train after \qty{0.396}{\second} (last potentiation pulse). 
While a quite homogenous out-of-plane distribution is observed at \qty{150}{\degreeCelsius} (Geometry A), a substantial gradient is visible at \qty{90}{\degreeCelsius} (Geometry B). 
Fig.~S\ref{fig: S19S7}a and b present the temporal variation of the local variation of $\left[V_{O}^{\bullet\bullet}\right]$ obtained in the channel’s surface and electrolyte-interface regions for Geometry A and B, respectively. 
For Geometry A, potential pulses generate a sawtooth-like variation of $\left[V_{O}^{\bullet\bullet}\right]$ near the interface, which is followed by a smoother variation of $\left[V_{O}^{\bullet\bullet}\right]$ in the surface. 
Combined with the substantial ionic drop observed in Fig.~S\ref{fig: S19S6}a,c this suggests that at \qty{150}{\degreeCelsius} ionic diffusion in LSF is not substantially limiting the device. 
On the contrary, Geometry B presents a larger sawtooth-like variation of $\left[V_{O}^{\bullet\bullet}\right]$ near the interface and no variation in the surface, suggesting that at \qty{90}{\degreeCelsius} and for the reduced lateral dimensions of Geometry B, oxygen vacancy diffusivity starts to limit the device’s properties. 

Therefore, the potentiation/depression observed in Fig.~S\ref{fig: S19S5}b mostly comes from the variation of $\left[V_{O}^{\bullet\bullet}\right]$ in the interface region, explaining why during the off time the channel’s conductance presents a more volatile behavior. 
Overall, the simulations suggest that reducing the lateral dimensions should allow the use of BICUVOX at \qty{90}{\degreeCelsius} or even below. 
However, to reach room temperature operation, strategies to improve the oxygen diffusivity in the channel must be pursued.

\subsubsection*{Investigation of the volatile and PPF behavior}
\label{S19S8}

\begin{figure}[h]
\centering
\includegraphics[width=\textwidth]{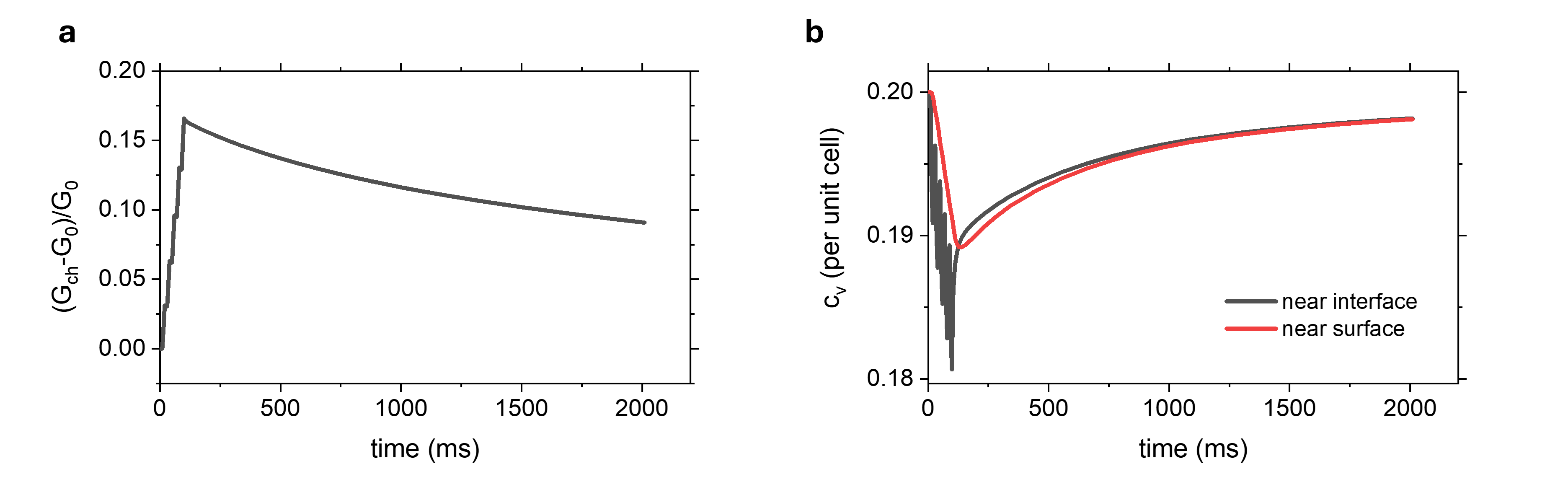}
\caption[S19]{
Evolution of conductance \textbf{(a)} and near-surface/near-interface oxygen vacancies (\textbf{b}, probed at the same location shown in Fig.~S\ref{fig: S19S7}a) obtained for Geometry A at $T=\qty{150}{\degreeCelsius}$ for 1) 5 initial potentiation pulses 2) forcing the same potential in gate and channel ($E_{set}=0$).
}
\label{fig: S19S8}
\end{figure}

We first investigated the effect of forcing the same potential in gate and channel ($E_{set}=0$) on the channel’s properties. 
For these simulations, we used Geometry A ($T=\qty{90}{\degreeCelsius}$) and an initial $\left[V_{O}^{\bullet\bullet}\right]=\num{0.2}$. 
Fig.~S\ref{fig: S19S8}a shows the evolution of channel’s conductance $\frac{(G_{Ch}-G_0)}{G_0}$ obtained for: 
1) Application of 5 pulses of (\qty{+0.5}{\volt}, $t_{on}=t_{off}=\qty{10}{\degreeCelsius}$) 
2) Application of $E_{set}=0$ for \qty{2}{\second}. 
Channel’s conductance potentiates with the application of the train’s pulses and then slowly decays during the application of $E_{set}=0$. 
This behavior can be explained observing the evolution of the $\left[V_{O}^{\bullet\bullet}\right]$ probed at the near-surface and near-interface regions depicted in Fig.~S\ref{fig: S19S7}a, see Fig.~S\ref{fig: S19S8}b. 
The application of $E_{set}=0$ tends to reestablish the initial oxygen vacancy concentration, as it is expected to take place in a battery-like closed system. 
The difference in time constant of the exponential decay observed in the conductance (Fig.~S\ref{fig: S19S8}a) and oxygen vacancies (Fig.~S\ref{fig: S19S8}b) is caused by the heterogeneous ionic intercalation across the channel, as commented in Fig.~S\ref{fig: S19S6}. 

\begin{figure}[h]
\centering
\includegraphics[width=\textwidth]{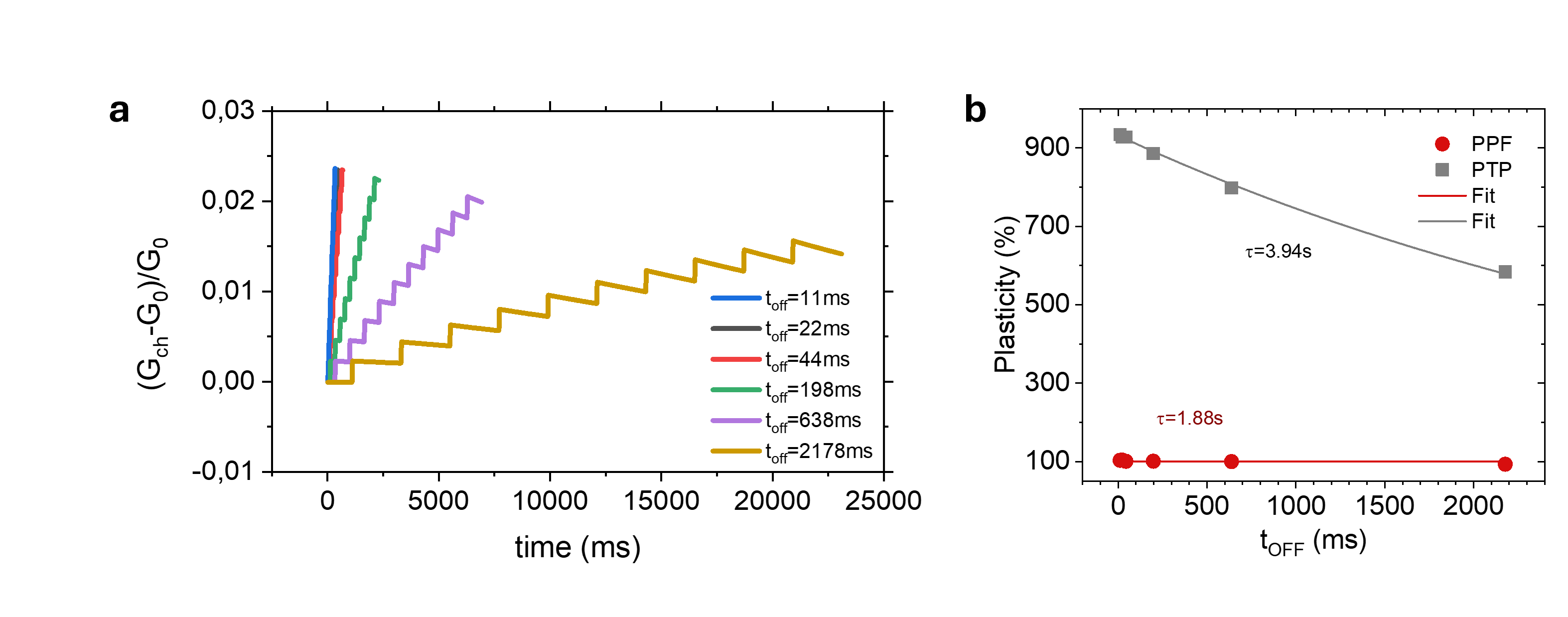}
\caption[S19]{
\textbf{a}, Evolution of normalized conductance of Geometry A at \qty{150}{\degreeCelsius} after the application of 10 pulses with  $t_{on}=\qty{20}{\ms}$ and different $t_{off}$. 
\textbf{b}, PPT and PPF extracted from the simulations and fitted with an exponential function.
}
\label{fig: S19S9}
\end{figure}

Then, we investigated the PPF behavior by simulating the dynamic response of our synaptic transistor under the application of a train of pulses with $t_{on}=\qty{22}{\ms}$ and different $t_{off}$ (between \qtylist{11;2178}{\ms}). 
For these simulations, we used Geometry A ($T=\qty{90}{\degreeCelsius}$), an initial $\left[V_{O}^{\bullet\bullet}\right]=\num{0.1}$ and a $E_{set}=\qty{-0.15}{\volt}$. 
Fig.~S\ref{fig: S19S9}a shows the potentiation of channel’s conductance $\frac{(G_{Ch}-G_0)}{G_0}$ obtained after 10 pulses with different $t_{off}$. 
Clearly, higher potentiation is observed for shorter $t_{off}$, similarly to the experimental data (see Fig.~4). 
Moreover, the PPF and PPT extracted from the simulations (Fig.~S\ref{fig: S19S9}b) can be fitted with an exponential decay function that shows time constants comparable with the experiments (see Fig.~4). 

\begin{figure}[h]
\centering
\includegraphics[width=\textwidth]{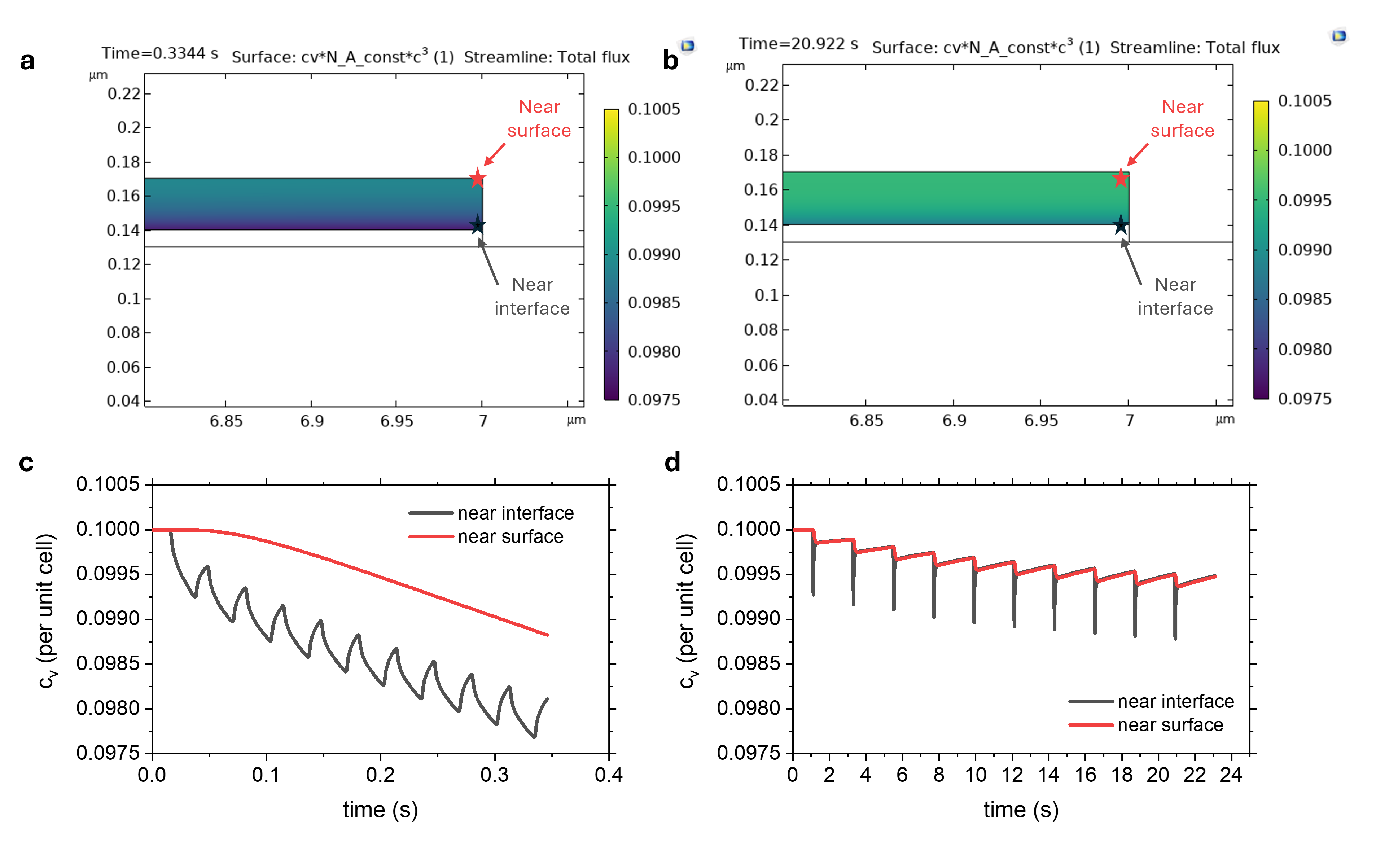}
\caption[S19]{
Distribution of oxygen vacancies obtained in the channel for $t_{off}=\qty{11}{\ms}$ \textbf{(a)} and $t_{off}=\qty{2178}{\ms}$ \textbf{(b)} for the \qty{12}{\ms}-pulse train after \qty{0.334}{\second} and \qty{20.92}{\second} (last potentiation pulse). 
Temporal evolution of the oxygen vacancy concentration obtained for $t_{off}=\qty{11}{\ms}$ \textbf{(c)} and $t_{off}=\qty{2178}{\ms}$ \textbf{(d)} in the near-surface and near-interface regions shown in c and d.
}
\label{fig: S19S10}
\end{figure}

The simulations show that the PPF behavior is mainly originated from a combination of different phenomena:
\begin{itemize}
\item When $E_{set}=0$ (i.e. during $t_{off}$) the channel tends to reestablish its initial $\left[V_{O}^{\bullet\bullet}\right]$, giving rise to a decay of channel’s conductance (as commented before, see Fig.~S\ref{fig: S19S8}). 
Therefore, the larger the $t_{off}$ the more time the system has to restore equilibrium. 
This is clearly visible in temporal evolution the $\left[V_{O}^{\bullet\bullet}\right]$ probed near the interface and near the surface for $t_{off}=\qty{11}{\ms}$ ms and $t_{off}=\qty{2178}{\ms}$, see Fig.~S\ref{fig: S19S10}a,c and b,d, respectively. 
A lower concentration of oxygen vacancies is obtained after 10 pulses with shorter off time compared to longer ones, both in the near surface and interface regions. 
\item Resistive phenomena such as out of plane oxygen vacancy diffusion in the channel or in-plane oxygen-ion conductance in the electrolyte generate a gradient of $\left[V_{O}^{\bullet\bullet}\right]$ in different directions that enhances the PPF. 
This is for instance observed in the higher gradient forming in the out-of-plane direction of the channel with shorter $t_{off}$ times see Fig.~S\ref{fig: S19S9}. 
However, we note that for this mechanism to give rise to an increase of PPF in the conductance the relationship between $\left[V_{O}^{\bullet\bullet}\right]$ and $\sigma_{eon}$ must be non-linear. 
If this takes place, a local decrease of $\left[V_{O}^{\bullet\bullet}\right]$ in the near electrolyte interface (diffusion limitations) or in the region near the channel$\vert$gate gap (electrolyte limitation) may give rise to a higher increase  of channel’s conductance compared to a distributed increase of the same amount of $\left[V_{O}^{\bullet\bullet}\right]$. 
The limiting mechanism will determine the time constant of the conductance decay. 
In our simulation, the time constant is mainly related to the BICUVOX, as shown in Fig.~S\ref{fig: S19S6}. 
By varying the geometry, the temperature or the channel’s material different PPF time constant can be obtained.
\end{itemize}

\newpage

\bibliography{zsupplementary_Langner}